%% file: draft0930.tex
\documentclass[11pt,a4paper]{article}

\usepackage{geometry,amsmath,amsfonts}
\usepackage{slashed}
\usepackage{epsfig}
\usepackage{latexsym}
\usepackage{graphicx}
\usepackage{amssymb}
\usepackage{verbatim}
\usepackage{bm}
\usepackage{dsfont}
\usepackage[footnotesize]{caption}
\usepackage{cancel}
\usepackage{cite}
\usepackage{graphicx}

\usepackage{epsfig}

\usepackage{multirow}
\usepackage{tikz}
\usepackage{subfig}
\usepackage{color}
\usepackage{multirow}
\usepackage{color}
\usepackage{rotating}
\usepackage{ifthen}

\newcommand{\non}[0]{\nonumber \\}
\newcommand{\bee}[0]{\begin{eqnarray}}
\newcommand{\eee}[0]{\end{eqnarray}}
\newcommand{\Tr}[1]{\mbox{Tr}\left[ #1 \right]}

\usepackage{color}
\definecolor{cblue}{RGB}{100,5,255}
\definecolor{cred}{RGB}{180,50,40} 
\definecolor{cgreen}{RGB}{40,255,40} 
\definecolor{corange}{RGB}{120,120,40} 
\definecolor{mathblue}{RGB}{100,100,250} 
     
\begin{document}

\begin{minipage}{0.4\textwidth}
 \begin{flushleft}
SISSA 30/2015 FISI\\
LPT-Orsay-15-50
\end{flushleft}
\end{minipage} \hfill
\begin{minipage}{0.4\textwidth}
\begin{flushright}
 July 2015
\end{flushright}
 \end{minipage}

\vskip 1.5cm

\begin{center}
{\LARGE\bf Lepton number violation as a key to}

\vspace{2mm}

{\LARGE\bf low-scale leptogenesis}

\vskip 2cm

{\large  Asmaa Abada$^a$, Giorgio Arcadi$^a$, Valerie~Domcke$^{b,c}$ and  Michele Lucente$^{a,b}$ }\\[3mm]
{\it{
\vspace*{.5cm} 
$^{a}$ Laboratoire de Physique Th\'eorique, CNRS -- UMR 8627, \\
Universit\'e de Paris-Sud, F-91405 Orsay Cedex, France
\vspace*{.2cm} 

$^{b}$ SISSA, Via Bonomea 265, 34136 Trieste, Italy
}

$^{c}$ INFN-Sezione di Trieste, 34136 Trieste, Italy
}
\end{center}

\vskip 1cm

\begin{abstract}
We explore the possibility of having 
a successful leptogenesis through oscillations between new sterile fermion states added to the Standard Model field content in a well motivated framework,  naturally giving rise to the 
required mass splitting between the sterile states through a small total lepton number violation. 
We propose a framework with only two sterile states forming a pseudo-Dirac state,  in which their mass difference as well as the smallness of the neutrino masses are due to two sources of lepton number violation with $\Delta L=2$,  corresponding to an Inverse Seesaw framework extended by a Linear Seesaw mass term. 
We also explore the pure Inverse Seesaw mechanism in its minimal version, requiring at least four new sterile states in order to comply with neutrino data. 
Our analytical and numerical studies  reveal that one can have a successful leptogenesis at the temperature of the electroweak scale through oscillations between the two sterile states with a ``natural'' origin of the strong degeneracy in their mass spectrum. 
We also revisit the analytical expression of the baryon asymmetry of the Universe in the weak washout regime of this framework.
\end{abstract}

\thispagestyle{empty}

\newpage

\setcounter{page}{1}


\input{introduction0720}

\input{analytic0720}

\input{parametrization0720}

\input{numerics0720}

\input{numerics_results0805}

\input{strongwashout_0720}

\input{iss0805}

\input{conclusion0720}

\vspace{1cm}
\subsubsection*{Acknowledgements}

We thank M.~Drewes,  M.~Kekic, J.~Lopez-Pavon and S.T~Petcov
 for very helpful discussions. 
This work has been supported in part by the European Union FP7-ITN INVISIBLES (Marie Curie Action PITAN-GA-2011-289442-INVISIBLES). G.A. acknowledges support from the ERC Advanced Grant Higgs@LHC.


\appendix

\input{app_leptogenesis_0720_v2}

\input{tables0720}



\bibliography{draftbiblio}{}
\bibliographystyle{JHEP}

\end{document}

%% file: introduction0720.tex
\section{Introduction}\label{introduction}

The origin of neutrino masses and the leptonic mixing,  the Dark Matter (DM) nature and the origin of the baryon asymmetry of the Universe (BAU) are three of
the most pressing open questions of particle and  astroparticle physics. 

In order to account for neutrino masses and mixings, many extensions
of the Standard Model (SM) call upon the introduction of right-handed  neutrinos - which are sterile states - giving
rise to a Dirac mass term for the neutral leptons. One of the most economical possibility is the embedding of the (standard) seesaw mechanism~\cite{Minkowski:1977sc,GellMann:1980vs,Yanagida:1979as,Glashow:1979nm,Mohapatra:1979ia} into the Standard Model (SM). 
The problem with these scenarios is that they cannot be probed directly: in
order to have natural neutrino Yukawa couplings,  
the typical scale of the extra particles
such as right-handed (RH) neutrinos (or even for scalar (fermion)  isospin triplets in the case of a type II (III) seesaw) is in general very high, 
potentially very close to the gauge coupling unification (GUT) scale, thus implying that direct experimental tests of
the seesaw hypothesis might be  impossible.  
In contrast, low-scale seesaw mechanisms~\cite{Mohapatra:1986bd,GonzalezGarcia:1988rw,Pilaftsis:1991ug,Deppisch:2004fa,Asaka:2005an,Gavela:2009cd,Ibarra:2010xw}, in which sterile fermions with masses
around the electroweak (EW) scale or even much lower are added to the SM, are very attractive since the new states can be
produced in collider and/or low-energy experiments, and their
contributions to physical processes can be sizeable, provided a not excessively suppressed mixing to the light (mostly active) neutrinos is present (cf.\, e.g.~\cite{Drewes:2015iva} for a recent overview of laboratory and cosmological constraints). 
For instance, this is the case 
for the $\nu$MSM~\cite{Asaka:2005an}, the Inverse Seesaw
(ISS)~\cite{Mohapatra:1986bd}, the Linear Seesaw (LSS)~\cite{Barr:2003nn,Malinsky:2005bi} and the low-scale
type-I seesaw~\cite{Gavela:2009cd,Ibarra:2010xw}. 
In general, within the seesaw mechanism it is possible to account for the BAU via thermal leptogenesis, see for instance~\cite{JosseMichaux:2008ix,Davidson:2008bu}. This kind of mechanism normally requires very high seesaw  scales, above $10^8\,\mbox{GeV}$. An efficient thermal leptogenesis can be nonetheless achieved at a seesaw  scale $\sim \,\mbox{TeV}$ (see for instance~\cite{Dev:2009aw,Blanchet:2010kw})  in the presence of a resonant amplification~\cite{Pilaftsis:2003gt}. At even lower seesaw  scales thermal leptogenesis is no longer at work and one must consider different mechanisms for the generation of any lepton asymmetry.
A viable possibility is provided by the mechanism first proposed in~\cite{Akhmedov:1998qx}, in which a lepton asymmetry is produced by the CP-violating oscillations of a pair of heavy neutrinos. This kind of mechanism has been successfully implemented in the so-called $\nu$MSM. Aiming at simultaneously
addressing the problems of neutrino mass generation, BAU and providing a viable DM
candidate, the $\nu$MSM is a truly minimal extension of the SM through the
inclusion of three RH neutrinos ($N_R^{1,2,3}$)~\cite{Asaka:2005an,Asaka:2005pn,Shaposhnikov:2008pf,Canetti:2012kh}. The lightest of these new states, with mass at the keV scale, is substantially sterile, i.e.\ with highly suppressed couplings both to active neutrinos and to the two other new states, and represents the Dark Matter candidate. The latter two heavy neutrinos are instead responsible for the light neutrino mass generation, as well as for lepton asymmetries both at early times, giving rise to the BAU, and at later times, accommodating the production of the correct amount of DM~\cite{Laine:2008pg}.  
For this to work, the spectrum in the additional sterile states requires a certain pattern, the two heaviest states $N_R^{2}$ and $N_R^{3}$ being almost degenerate. 
Notice however that this requirement can be relaxed by considering, relaxing the DM hypothesis, all the three right-handed neutrinos involved in the leptogenesis process, as it was shown in~\cite{Drewes:2012ma}.

The degeneracy between the heavy neutrinos, which is phenomenologically imposed in the $\nu$MSM, can be however naturally explained in frameworks in which the  smallness of (active) neutrino masses is directly linked to  a small violation of the total lepton number~\cite{Shaposhnikov:2006nn}. 
This can be achieved when, for instance, the Inverse Seesaw~\cite{Wyler:1982dd, Mohapatra:1986bd} mechanism  is embedded into the SM.  Indeed here, 
contrary to most of (type I)
low-energy seesaw realisations, 
neutrino masses and mixings  are accommodated with natural values of the Yukawa couplings for a
comparatively low seesaw scale. The possibility of having sizeable 
mixings between the active and sterile states renders the model
phenomenologically rich, with a potential impact for a number of
observables. 
 Moreover, depending on its actual realisation, the ISS  allows to accommodate
the observed DM relic abundance and (potential) indirect DM
detection hints (see for instance~\cite{Abada:2014vea,Abada:2014zra}). 
The mechanism consists in the addition of at least  two sets of additional neutrinos 
with opposite lepton numbers, 
allowing for a small $\Delta L=2$ lepton number violating (LNV) mass parameter $\mu$ corresponding to a Majorana mass in the sterile sector. The masses of the mostly active neutrinos (light neutrinos) are proportional to $\mu$, while the remaining mostly sterile states are coupled into  pseudo-Dirac pairs with mass differences of the order of the LNV parameter $\mu$. 
In the limit where $\mu\rightarrow 0$,
lepton number conservation  is restored. However, in order to be phenomenologically viable, this mechanism requires at least four extra fermions~\cite{Abada:2014vea}. 
Another mechanism based on a small violation of the total lepton number is  the so called Linear Seesaw~\cite{Barr:2003nn,Malinsky:2005bi}.  It is similar to the ISS,  in the sense that it requires  the introduction of two types of fermion singlets with opposite assignment for the total lepton number, and the smallness of the neutrino masses is linked to the small violation of the total lepton number conservation  (by two units). The difference with respect to the ISS is that the lepton number violation arises from the new LNV  Yukawa couplings of steriles with the left-handed neutrinos.

The present study focuses on the possibility of simultaneously having a very low  scale seesaw mechanism - typically at $1 - 10$ GeV - at work for generating neutrino masses and mixings as well as an efficient leptogenesis at the electroweak scale, by considering a ``natural'' and minimal framework (with only two additional neutrinos) giving rise to the needed degeneracy in the spectrum of the sterile states.
For this, we consider the ISS and the LSS frameworks and revisit the mechanism of leptogenesis through oscillations~\cite{Akhmedov:1998qx,Asaka:2005an}.
More precisely, we have considered the minimal extension of the SM  by two sterile states with  couplings leading to an Inverse Seesaw  mass structure. Being insufficient to accommodate neutrino data, instead of adding further sterile states, we have completed the  scenario  with a Linear Seesaw mass term (see for instance ~\cite{Gavela:2009cd}), violating total lepton number also by two units. 
We have conducted a thorough analytical and numerical analysis investigating both neutrino mass hierarchies, normal (NH) and inverted (IH), for the neutrino mass spectrum.  To this end we have implemented and solved a system of Boltzmann equations 
 and have also derived an analytical expression for the baryon asymmetry in the weak washout regime, supporting our understanding of the behaviour of the numerical solutions. 
Our studies reveal that this scenario can incorporate a successful leptogenesis through oscillations between the two mostly sterile states while accommodating the observed neutrino data. 
We have also investigated if our scenario can be probed by SHiP~\cite{Alekhin:2015byh} or FCC-ee~\cite{Blondel:2014bra}.

In the second part of this work, we have considered  the possibility of having the pure and minimal Inverse Seesaw mechanism with four or five extra neutrinos  - that is the ISS(2,2) (or ISS(2,3)) scenario with two RH neutrinos plus two (three) steriles states~\cite{Abada:2014vea} - at work for an efficient leptogenesis through  oscillations. Notice  that the ISS(2,3) scenario can provide in principle a DM candidate~\cite{Abada:2014zra}.
We find that the required mass splitting between the pseudo-Dirac pairs is too large to achieve a successful leptogenesis in the weak washout regime while accommodating neutrino data.

This work is organised as follows: 
in Section~\ref{sec_analytic} we first describe the idea by means of a toy model framework of one generation (one flavour) and two sterile states, with two sources of lepton number violation corresponding  to a combined scenario of the ISS and the LSS. In the second part of this section, we  extend the toy model to the full three flavour case  and obtain estimates for the scales of our set of parameters. 
Section~\ref{parametrization} is devoted, in its first part,  to describing our numerical study of the viable parameter space, taking into account the various constraints on the sterile states. Here we make use of an analytical expression for the BAU in the weak washout regime, which is derived in Appendix~\ref{app_leptogenesis}. The second part of Section~\ref{parametrization} confronts this expression with the full numerical results for some instructive benchmark points. 
The results derived in Section~\ref{parametrization} are discussed in Section~\ref{Sec:Results.discussion}. Section~\ref{Sec:strong washout} is devoted to the scenario of strong washout regime for the Yukawa couplings.
Before concluding, we  dedicate Section \ref{ISS} to  the pure minimal ISS. 
 Finally, we provide  in   Appendix B the relevant numerical input parameters for all the solutions discussed in this work. 

%% file: analytic0720.tex
\section{Leptogenesis and lepton number violation \label{sec_analytic}}

In Ref.~\cite{Akhmedov:1998qx}, followed up and refined e.g.\ in Refs.~\cite{Asaka:2005pn, Shaposhnikov:2008pf, Asaka:2010kk, Asaka:2011wq, Canetti:2012kh}, a compelling mechanism  accommodating neutrino data, the dark matter abundance and providing a successful mechanism for leptogenesis at the electroweak scale was proposed. In its simplest setup, this mechanism requires two additional heavy and nearly mass-degenerate neutrinos $N_R^{1,2}$ (with masses in the MeV - GeV range~\cite{Canetti:2012kh}) with sufficiently small Yukawa couplings to the SM, ensuring that these states have not yet reached thermal equilibrium at the electroweak phase transition. Starting from a zero initial abundance, these heavy states are produced thermally as the Universe approaches the electroweak phase transition. Oscillations between these two states produce a CP-asymmetry which induces particle-antiparticle asymmetries in the individual lepton flavours, which are produced in the decay of these states. These asymmetries in the active sector act as a background potential for the sterile flavours (similar to the MSW effect for neutrino oscillations in matter~\cite{langacker:1986jv}), resulting in particle-antiparticle asymmetries for the sterile states. The two heavy states have opposite $CP$ and form a pseudo-Dirac pair, thus for negligible Majorana masses no total lepton asymmetry is induced,
 as the total lepton asymmetry in the active sector balances the one in the sterile sector. At the electroweak phase transition, $T \sim T_W$, the SM sphaleron processes freeze out, converting the asymmetry in the active sector (and only the active sector) to a net baryon asymmetry. To summarise, the Sakharov conditions~\cite{Sakharov:1967dj} necessary for a successful baryogenesis are fulfilled at $T \gtrsim T_\text{W}$ because (i) the heavy states have not yet reached thermal equilibrium, (ii) the oscillations of the heavy states violate CP and (iii) sphaleron processes  violate baryon number. We review this mechanism in more detail in Appendix~\ref{app_leptogenesis}, deriving analytical expressions for the produced individual asymmetries.

A crucial ingredient for this mechanism is the small mass splitting between the two sterile states, with a relative degeneracy in the pair at the per mille level, $\Delta m/M \lesssim 10^{-3}$~\cite{Canetti:2012kh}. This small mass splitting is obtained naturally if there is a symmetry which imposes fully degenerate masses for the heavy states, the small mass splitting is then linked to the small breaking parameters of this symmetry. A simple choice is an additional global $U(1)$ symmetry with opposite charges assigned to the new fields. As a result, the two states $N_R^{1,2}$ form a Dirac spinor $\Psi= N_R^{1}+ (N_R^{2})^c$, and the $U(1)$ global symmetry mimics the lepton number one~\cite{Shaposhnikov:2006nn}. The small breaking of lepton number is moreover a promising source to explain the light neutrino masses, studied in detail in, for instance,  Refs.~\cite{Antusch:2015mia, Gavela:2009cd, Kersten:2007vk}.
In the following, we will thus focus on models with approximately conserved lepton number and investigate their viability for leptogenesis through neutrino oscillations. 

\subsection{An instructive toy model \label{sec_toymodel}}
Let us recapitulate the simplest implementation of this idea, by adding to the SM two sterile fermions with opposite lepton number, cf.\ also \cite{Gavela:2009cd}. In order to obtain clear analytical results we first consider a toy model with only one active neutrino. In this case the lepton conserving part of the mass matrix is, in the basis $(\nu_L, {N_R^{1}}^c, {N_R^{2}}^c)$,
\bee
M_0 &=& \left(\begin{array}{ccc} 0 & \frac{1}{\sqrt{2}}Y v & 0\\
\frac{1}{\sqrt{2}} Y v & 0 & \Lambda \\
0 & \Lambda & 0
\end{array}\right),
\eee
with $v$ denoting the vacuum expectation value of the Higgs boson, $v = 246$~GeV  after the EW phase transition, $Y$ denoting the Yukawa coupling of the sterile state with lepton number (+1) to the SM lepton and Higgs doublets and $\Lambda$ denoting a new mass parameter which will set the scale for the masses of the additional heavy states.
The mass spectrum resulting from this mass matrix is
\begin{equation}
m_\nu =0 \,,\quad
M_{1,2} = \sqrt{|\Lambda|^2 +\frac{1}{2}|Y v|^2}\ .
\label{eq_M}
\end{equation}

Let us consider now all possible patterns for breaking the global lepton number in $M_0$. A term in the $(1,1)$ entry breaks gauge invariance, and can only be generated in non minimal models, for example by adding an isospin triplet of Higgs fields. Since we are not interested in such a case, there are 3 possible patterns to perturb $M_0$:
\begin{align}
\Delta M_{ISS}=\begin{pmatrix} 0 & 0 & 0\\
0 & 0 & 0\\
0 & 0 & \xi \,\, \Lambda
\end{pmatrix}, \,
\Delta M_{LSS}=\begin{pmatrix} 0 & 0 & \frac{\epsilon}{\sqrt{2}} \, Y' v\\
0 & 0 & 0\\
\frac{\epsilon}{\sqrt{2}} \, Y' v & 0 & 0 
\end{pmatrix}, \,\,
\Delta M_{lp}=\begin{pmatrix} 0 & 0 & 0\\
0 & \xi' \, \Lambda & 0\\
0 & 0 & 0 
\end{pmatrix}.
\label{eq_DMs}
\end{align}
Here $\epsilon$, $\xi$ and $\xi'$ are small dimensionless parameters accounting for the breaking of lepton number and $Y' \sim Y$ is a new Yukawa coupling. Without loss of generality we can choose $|Y'| = |Y|$,  keeping $\epsilon$ as a free parameter.
The first possibility generates the usual Inverse Seesaw pattern~\cite{Wyler:1982dd, Mohapatra:1986bd}, the second one corresponds to the so called Linear Seesaw~\cite{Malinsky:2005bi}, while the third one does not generate neutrino masses at tree level but does it at loop level~\cite{Dev:2012sg,LopezPavon:2012zg}.
However loop corrections are only relevant in the regime of a large lepton number violation, $\xi' \gtrsim 1$, and since we focus on models with an approximate lepton number conservation we will concentrate on the first two possibilities
.\footnote{This structure can be obtained dynamically by extending the particle content of the SM, e.g.\ it is possible to generate a small $\Delta L=2$ mass $\sim  \xi \, \Lambda$ as in the general formulation of the Inverse Seesaw~\cite{Mohapatra:1986bd}, where the smallness of  $\xi$ was attributed to the supersymmetry breaking effects in a (superstring inspired) $E_6$ scenario. In the context of a non-supersymmetric $SO(10)$ model, which contains remnants of a larger $E_6$ symmetry, $ \xi \, \Lambda$ is generated at two-loop while $\xi' \, \Lambda$ is generated at higher loops, justifying its smallness compared to $\sim  \xi \, \Lambda$~\cite{Ma:2009gu}.} 
Here $M$  contains only a single physical complex phase (after absorbing  three complex phases by rotating the three fields of the toy model), which we will assign to $Y'$ in the following,  taking  $\Lambda $, $\xi$, $\epsilon$ and $Y$ to be real and positive.

Perturbatively diagonalising $M_0 + \Delta M$ yields expressions (at leading order in the small lepton number violating parameters $\epsilon,\xi \ll 1$) for the two quantities relevant for leptogenesis and neutrino mass generation: the mass-scale of the active neutrinos $m_\nu$ and the mass splitting between the two (heavy) states, $\Delta m^2$. For the Inverse Seesaw scenario, we find
\begin{align}
 m_\nu &= \frac{\xi  \, (Y v)^2 \Lambda}{2 \Lambda^2 + (Y v)^2 } + {\cal O}({\xi^2}) \simeq \frac{ \xi  \, (Y v)^2}{2 \Lambda } \,, \label{eq_mnuiss}\\
 \Delta m^2 &=  \frac{2 \, \xi \, \Lambda^3}{\sqrt{\Lambda^2 + \frac{1}{2}(Y v)^2 }} + {\cal O}({\xi^2})  \simeq 2 \xi \Lambda^2 \,, \label{eq_M12iss}
\end{align}
whereas the Linear Seesaw yields
\begin{align}
  m_\nu &=  \frac{2 \epsilon Y^2 v^2 \Lambda}{2 \Lambda^2 + (Y v)^2 } + {\cal O}({\epsilon^2}) \simeq  \frac{\epsilon \, (Y v)^2}{\Lambda } \,,\label{eq_mnulss}\\
 \Delta m^2 &=  \frac{4 \epsilon (Y v)^2 \Lambda}{\sqrt{2 \Lambda^2 + (Y v)^2 }}  + {\cal O}({\epsilon^2})  \simeq 2 \epsilon (Y v)^2 \,. \label{eq_M12lss}
\end{align}
Here after expanding in $\xi$ or in $\epsilon$, we have exploited that the heavy neutrinos cannot be fully thermalised for a viable leptogenesis scenario, implying an upper bound on their Yukawa couplings,
$ Y \, v \ll \Lambda$.

From these expressions we can draw several important conclusions. Firstly, comparing Eqs.~\eqref{eq_mnuiss} and \eqref{eq_mnulss} we note that both the Inverse and the Linear Seesaw realisations require the same degree of lepton number violation in order to reproduce the observed light neutrino masses,
\begin{equation}
\epsilon, \xi \simeq (m_\nu M_{1,2})/m_D^2  \,,
\label{eq_epsmu}
\end{equation}
with $m_D = Y v/\sqrt{2} = |Y'| v/\sqrt{2}$. 

\noindent Secondly,  looking at Eqs.~\eqref{eq_M12iss} and \eqref{eq_M12lss}, we note that above the EW phase transition, where $\langle v \rangle = 0$, the mass splitting induced by the Linear Seesaw vanishes, contrary to the one induced by the Inverse Seesaw. This is a relevant detail since successful leptogenesis can occur only above the EW phase transition, where the sphalerons can efficiently convert a lepton asymmetry into a baryon asymmetry.  Thermal effects during the oscillation process might alleviate this difficulty~\cite{Canetti:2012kh}. For simplicity we will in the following focus on the situation where the mass splitting is related to the Majorana mass term $\xi \Lambda$.
In the pure Inverse Seesaw model this however implies
\begin{equation}\label{eq:dmISS}
 (\Delta m^2)_{ISS}^{1/2} \simeq \left( \frac{2 \, m_\nu M_{1,2}}{m_D^2} \right)^{1/2} M_{1,2}\ .
\end{equation}
For example for $Y  < \sqrt{2} \times 10^{-7}$, $m_\nu = 0.05$~eV and $M_{1,2} = 1$~GeV, this yields ${\Delta m/M_{1,2} \gtrsim 0.4}$ - a value far too large for a successful leptogenesis\footnote{This upper bound on $Y$ forces the heavy states to be out-of-equilibrium~\cite{Akhmedov:1998qx} and washout processes to be negligible. 
The numbers quoted here a priori only apply to the toy model discussed in this section, and not to realistic, more elaborate versions of the Inverse Seesaw mechanism. We will return to this point in Section~\ref{ISS}.}.

In conclusion, the minimal setup to accommodate acceptable light neutrino masses $m_\nu$ and a sufficiently small mass splitting $\Delta m^2$  is obtained  by simultaneously considering both  $\Delta M_{LSS}$ and  $\Delta M_{ISS}$, with $\epsilon > \xi$: the leading order contribution to the light neutrino masses stems from $\epsilon$, with the scale of $\epsilon$ determined by Eq.~\eqref{eq_epsmu}. Above the EW phase transition, the leading order contribution to the mass splitting is in turn set by Eq.~\eqref{eq_M12iss} and can be sufficiently small for $\xi \ll \epsilon$:
\begin{equation}
 m_\nu \simeq 2 \epsilon \frac{m_D^2}{M_{1,2}} \,, \quad \Delta m^2 \simeq 2 \xi M_{1,2}^2 \,.
 \label{eq_mueps}
\end{equation}

This analysis suggests that the minimal viable realisation of our ansatz is given by the mass matrix $M = M_0 +  \Delta M_{ISS} + \Delta M_{LSS}$.
Notice that the ordering of the second and third column/row  of Eqs.~\eqref{eq_M} and \eqref{eq_DMs} arises from the assignment of lepton number 1 and -1, respectively. Choosing $\epsilon > 1$ and $|Y| \simeq |Y'|$ correspondingly smaller, implies switching this assignment. Hence very large values of $\epsilon \gg 1$ also correspond to a small violation of lepton number, and there is an approximate symmetry under $\epsilon \rightarrow 1/\epsilon$ which becomes exact when $\xi, \xi' \rightarrow 0$. Accounting for solutions with $\epsilon \gg 1$ is equivalent to considering the mass matrix $M=M_0 +\Delta M_{LSS}+ \Delta M_{lp}$, which represents a minimal setup as well. The main difference between the two possibilities is that the Majorana mass term $\Delta M_{ISS}$ breaks the lepton number by $\Delta L=2$, i.e.\ by the same amount of the violation in the Yukawa sector given by $\Delta M_{LSS}$, while $\Delta M_{lp}$ carries $\Delta L=-2$. For simplicity we will focus on the case $\epsilon \ll 1$ in the remainder of this section, but our numerical study in Section~\ref{parameter_scan} will  cover the entire range for $\epsilon$. At leading order, the corresponding expressions for the perturbative expansion in $\epsilon' = 1/\epsilon \ll 1$ can be obtained by replacing $\epsilon \rightarrow 1/\epsilon$  and $Y \rightarrow \tilde{Y}\equiv \epsilon Y$ in the expressions below.

A further important parameter, which is  particularly relevant for leptogenesis, is the mixing between the two heavy neutrino mass eigenstates. To estimate this, we consider the effective potential for the heavy states arising from the interactions with the SM lepton and Higgs doublets in the surrounding hot thermal plasma~\cite{Asaka:2011wq},
\begin{equation}
 V_N = \frac{1}{8} T (Y^\text{eff})^\dagger Y^\text{eff}\,,
 \label{eq_VN}
\end{equation}
where $T$ denotes the temperature and $Y_\text{eff}$ are the Yukawa couplings of the heavy mass eigenstates $m_j,\ j=2,3$,
\begin{equation}
Y^\text{eff}_{\alpha j} = Y_{\alpha I} U_{I j} \,.
\label{eq_Yeff}
\end{equation}
Here $U$ is the matrix which diagonalises $M = M_0 + \Delta M$, $U^T M U = \text{diag}(m_\nu, \, M_1, \, M_2)$, $\alpha$ runs over the active flavours (only one flavour in this toy model), $I$ denotes the sterile flavours and $Y_{\alpha 1} = Y_\alpha = Y$ and $Y_{\alpha 2} = \epsilon Y'_\alpha = \epsilon Y'$. Hence for the toy model of this section with $\beta \equiv \text{Arg}(Y')$, 
$Y^\text{eff} = Y (   - 1 +   \epsilon \, e^{i \beta}, \, 1 + \epsilon \, e^{i \beta} )/\sqrt{2} + {\cal O}(\xi/\Lambda)$, implying $|Y^\text{eff}| \simeq Y$. The eigenvectors of the potential $V_N$  above the EW phase transition (i.e.\ for $v = 0$) are given by
\begin{equation}
 v_1 \simeq (1 + e^{i \beta} \epsilon,\, 1 - e^{i \beta} \epsilon)\,, \quad v_2 \simeq (- 1 + e^{i \beta} \epsilon, \,  1 + e^{i \beta} \epsilon ) \ ,
 \label{eq_sterile_mixing}
\end{equation}
up to corrections of order $\xi$, $\epsilon^2$. This indicates that in the parameter region of interest, which corresponds to $\xi \ll \epsilon$, maximal mixing between the heavy mass eigenstates with a mixing angle of $\theta_{PD} \simeq 45^\circ$ and hence particularly efficient oscillations are obtained for $\epsilon \rightarrow 0$.

In this regime, which we will refer to as ``perturbative'', since viable neutrino masses and a small enough splitting between the heavy states are obtained through a small violation of the lepton number, the condition for a successful leptogenesis can be casted as:
\begin{equation}
\label{eq:perturbative}
 |Y'| = Y = 10^{-7} \left(\frac{M_{1,2}}{1 \text{ Gev}}\right) \left( \frac{0.1}{\epsilon} \right)^{1/2} {\lesssim} \sqrt{2} \times 10^{-7},  \, \quad \xi < \frac{1}{2} \left( \frac{100 \text{ keV}}{M_{1,2}}\right)^2 \,,
\end{equation}
or, equivalently, for the flipped assignment of lepton charges, corresponding to $Y \rightarrow \tilde{Y} = \epsilon Y$:
\begin{equation}
 {|\tilde Y'| = \tilde Y = 10^{-7} \left(\frac{M_{1,2}}{1 \text{ Gev}}\right) \left( \frac{\epsilon}{10} \right)^{1/2}{\lesssim} \sqrt{2} \times 10^{-7}, \quad \xi < \frac{1}{2} \left( \frac{100 \text{ keV}}{M_{1,2}}\right)^2 \,.}
\end{equation}

In the more realistic model with three active flavours, the situation is more complicated, as cancellations in the matrix equations can arise. In particular, $\epsilon$ may be of order one and still yield viable solutions, though this in some sense goes against the spirit of our ansatz, linking the small mass splitting to an approximate symmetry. {In the following, we will refer to this latter type of viable parameter points, approximately identified by the condition $0.1 \lesssim \epsilon \lesssim 10 $, as ``generic'', as opposed to the ``{perturbative}'' solutions identified above.} This section served to clarify the  parameter region of interest which requires no matrix cancellations. We will proceed in the next section with a rigorous perturbative expansion of the full three-flavour model in the perturbative region, before turning to a numerical study in Sections~\ref{parametrization}, \ref{Sec:Results.discussion} and \ref{Sec:strong washout}. In Section~\ref{ISS} we will revisit the pure Inverse Seesaw scenario, and investigate if the conclusions above can be circumvented by considering the three active flavours.

\subsection{Perturbative expansion of the full model}

In the previous section, we illustrated a symmetry inspired ansatz for the neutrino mass matrix by means of a 3 $\times$ 3 toy model. In this section, we extend this analysis to a full model taking into account the three active flavours, confirming that the estimates for the scales which were obtained in the toy model (one flavour) remain  also valid in the full model.
Consider hence the following mass matrix,
\begin{equation}
 M =\ \Lambda \, \begin{pmatrix}
      0 & 0 & 0 & \frac{1}{\sqrt{2}}Y_1  v/ \Lambda & \frac{1}{\sqrt{2}} \epsilon Y'_1  v/\Lambda \\
      0 & 0 & 0 & \frac{1}{\sqrt{2}} Y_2  v/\Lambda & \frac{1}{\sqrt{2}} \epsilon Y'_2  v/\Lambda \\
      0 & 0 & 0 & \frac{1}{\sqrt{2}} Y_3  v/\Lambda& \frac{1}{\sqrt{2}} \epsilon Y'_3  v/\Lambda\\
      \frac{1}{\sqrt{2}} Y_1  v/\Lambda &  \frac{1}{\sqrt{2}} Y_2  v/\Lambda &  \frac{1}{\sqrt{2}} Y_3  v/\Lambda & 0 & 1 \\
    \frac{1}{\sqrt{2}}  \epsilon Y'_1  v/\Lambda &  \frac{1}{\sqrt{2}} \epsilon Y'_2  v/\Lambda  &   \frac{1}{\sqrt{2}} \epsilon Y'_3  v/\Lambda  & 1  & \xi
     \end{pmatrix}  \,. \label{eq_Mpertexp}
\end{equation}
In the parameter region of interest, as identified in Section~\ref{sec_toymodel}, all entries of this matrix except for the (4,5) and (5,4) entries are small,
\begin{equation}
 |\frac{1}{\sqrt{2}} Y_\alpha v/\Lambda|, \, |\frac{1}{\sqrt{2}}  \epsilon Y'_\alpha  v/\Lambda |, \, |\xi| \ll 1, \qquad \alpha=\{1,2,3\}\,,
\end{equation}
thus justifying a perturbative approach. In this setup we have two additional physical complex phases, whose assignment will be discussed in the next section. Expanding the eigenvalues of $M^\dagger M$ to fourth order in any combination of the perturbative parameters (including mixed terms), we can identify the leading order contributions to the decisive quantities,  $m_\nu$ and $\Delta m^2$. For the masses of the heavy states this yields
\begin{equation}
 M_{1,2}^2 \simeq |\Lambda|^2 \pm \frac{1}{2} |\xi| |\Lambda|^2 + \frac{1}{2}|\vec{Y}|^2 v^2 + \frac{1}{2} |\xi|^2 |\Lambda^2| + \frac{1}{2}|\epsilon|^2 |\vec{Y'}|^2 v^2 \ ,
\end{equation}
up to third order terms in $\{ \frac{1}{\sqrt{2}} |Y_\alpha v/\Lambda|, \,\frac{1}{\sqrt{2}} | \epsilon Y'_\alpha v/\Lambda |, \, |\xi| \}$. Here $|\vec{Y}|^2 \equiv \sum_{\alpha = 1}^3 |Y_\alpha|^2$. 
As in the toy model of Section~\ref{sec_toymodel}, the overall scale is hence determined by $|\Lambda|$ and the  leading order contribution to the mass splitting is $\Delta m^2 \simeq |\xi \Lambda^2|$.  

Proceeding to the light neutrino masses, we notice that one state remains exactly massless while the other two obtain small masses. This scale is given, up to fourth order in   $\{ \frac{1}{\sqrt{2}} |Y_\alpha v/\Lambda|, \, \frac{1}{\sqrt{2}}| \epsilon Y'_\alpha  v/\Lambda |, \, |\xi| \}$, by
\begin{equation}
 m_\text{tot}^2 \equiv \sum_{i = 1}^3 m_i^2  \simeq \frac{1}{2} |\epsilon|^2  \frac{v^4}{|\Lambda|^2} \left(\sum_{\alpha = 1}^3 |Y_\alpha|^2 |Y'_\alpha|^2  + \sum_{\alpha = 1}^3 \sum_{\beta = 1}^3 |Y_\alpha|^2 |Y'_\beta|^2     \right) \,,
\end{equation}
again in agreement with the expectation from the one flavour toy model studied in Section~\ref{sec_toymodel}.

%% file: parametrization0720.tex
\section{Computation of the baryon abundance \label{parametrization}}
In this section we investigate the impact of  requiring  a successful leptogenesis on our scenario. To achieve this task we will compute the baryon abundance for a large set of model realisations, complying with the experimental constraints on the active neutrinos, as well as with limits from possible signatures of the extra sterile fermions in laboratory searches. In order to perform an efficient exploration of the parameter space, we adopt the parametrisation  of the neutrino mass matrix introduced in~\cite{Gavela:2009cd}, reviewed in detail in the next subsection.
An accurate determination of the baryon density would require the solution of a system of coupled Boltzmann equations, like the ones introduced in~\cite{Asaka:2005pn,Asaka:2011wq,Canetti:2012kh}, in the entire parameter space. Unfortunately this task is computationally demanding. For this reason we first focus our analysis on a subset of the parameter space, corresponding to very suppressed values of the Yukawa couplings of the new neutrinos (see below for details). We will refer to this scenario as ``weak washout'' regime. Here all the heavy neutrinos are far below thermal equilibrium during the entire leptogenesis process; as a consequence there is no depletion of the produced baryon asymmetry from washout processes. In this regime the system of Boltzmann equations can be perturbatively solved (see details on the derivation in Appendix~\ref{app_leptogenesis}), resulting in an analytical expression for the baryon abundance $Y_B$ well approximating the full numerical result.

We have further implemented the numerical solution of Boltzmann equations to validate and complement our analytical study,  extending our analysis beyond the reach of the analytical estimates, cf. Section~\ref{Sec:strong washout} for an analysis of the ``strong washout'' regime. This regime is characterised by higher values of the entries of the Yukawa matrix of the heavy neutrinos such that they reach thermal equilibrium at temperatures between the initial production of the lepton asymmetry (i.e. the temperature $T_L$ defined in Appendix~\ref{app_leptogenesis}) and $T_{\rm W}$. This entails a depletion of the lepton asymmetry.

For the sake of clearness we will present in the following the main results, while the details of both the analytical and numerical computations will be reviewed in the appendix.

\subsection{Parametrisation of the mass matrix}\label{sec:parametrization:mass}

\noindent
We consider the neutrino mass matrix introduced in Eq.~\eqref{eq_Mpertexp}. As discussed in Section~\ref{sec_toymodel},
the lepton number violation is represented by the dimensionless parameters $\epsilon$ and $\xi$. The entries of the mass matrix associated to these parameters violate the lepton number by the same amount, namely $\Delta L=2$. Although in a ``natural'' scenario these two parameters would be expected to be of the same order of magnitude (and both small), we will stick to a more generic case taking them to be free and independent among each other. In particular we will here also allow for large lepton number violation $\epsilon, \xi \sim 1$, going beyond the perturbative expansion of Section~\ref{sec_toymodel}.

In order to identify a minimal set of parameters for a numerical scan, we have adopted 
the parametrisation introduced in~\cite{Gavela:2009cd}. The Yukawa matrices are expressed as a function of two free parameters $y$ and $y'$, of an additional parameter $\rho$ given by:
\begin{equation}
\rho=\frac{\sqrt{1+r}-\sqrt{r}}{\sqrt{1+r}+\sqrt{r}}\, , \quad r=\frac{|\Delta m_{\rm solar}^2|}{|\Delta m_{\rm atm}^2|}\ , 
\end{equation}
and of the elements of the PMNS matrix as:
\begin{align}
\label{eq:yukawa_parameters}
& Y_{\alpha}=\frac{y}{\sqrt{2}} \left[U^{*}_{\alpha 3} \sqrt{1+\rho} + U^{*}_{\alpha 2} \sqrt{1-\rho}\right] \ ,\nonumber\\
& Y_{\alpha }^{\prime}= \tilde{Y}_{\alpha }^{\prime}+\frac{k}{2}Y_{\alpha }\ ,\nonumber\\
& \tilde{Y}_{\alpha}^{\prime}=\frac{y^{\prime}}{\sqrt{2}}\left[U^{*}_{\alpha 3} \sqrt{1+\rho} - U^{*}_{\alpha 2} \sqrt{1-\rho}\right]\ , \nonumber\\
& k=\frac{\xi}{\epsilon}\ .
\end{align}
The three physical phases
in the mass matrix~\eqref{eq_Mpertexp} are conveniently assigned as follows: the Dirac phase $\delta_{CP}$ and the unique Majorana phase $\alpha$ of the PMNS matrix\footnote{The second Majorana phase in the PMNS matrix can be rotated away since in this case one neutrino is massless.} appear in $Y$ and $Y'$ through Eq.~\eqref{eq:yukawa_parameters}, the third `high-energy' phase is assigned to $\Lambda$. The parameters $\epsilon$, $\xi$, $v$, $y$ and $y'$ can then be taken to be real and positive and the $\Delta L=2$ Majorana mass term is taken equal to $\xi \left|\Lambda\right|$.
Using this parametrisation the mass eigenstates coincide with the expressions in the limit of pure Linear Seesaw (the Majorana mass parameter $\xi$ is encoded in the definition of the Yukawa matrices) and are thus given by~\cite{Gavela:2009cd}: 
\begin{align}
\label{eq:mass_eigen}
 |m_1|=0 \,, \qquad
 |m_2|= \frac{\epsilon y y^{'} \left(1-\rho\right) v^2}{2\left|\Lambda\right|}\,, \qquad
 |m_3|= \frac{\epsilon y y^{'} \left(1+\rho\right)v^2}{2\left|\Lambda\right|} \ ,
\end{align}
while in the limit $|\Lambda| \gg |Y| v, \epsilon |Y^{\prime}| v$, which is the one relevant for leptogenesis, the masses of the two mostly sterile states are given by:
\begin{equation}
M_{1,2}=|\Lambda| \left(1 \mp \xi \right)\ .
\end{equation}
Notice that this parametrisation generates neutrino masses only according to a normal hierarchy. An inverted hierarchical spectrum can be obtained by modifying the definition of $\rho$ as:
\begin{equation}
\rho=\frac{\sqrt{1+r}-1}{\sqrt{1+r}+1}\ ,
\end{equation}
and by replacing $U_{\alpha 3} \rightarrow U_{\alpha 2}$ and  $U_{\alpha 2} \rightarrow U_{\alpha 1}$ in Eq.~(\ref{eq:yukawa_parameters}).

%% file: numerics0720.tex
\subsection{Parameter scan \label{parameter_scan}}

As seen in the previous section, the neutrino mass spectrum  depends on 6 parameters, $y,y^{\prime},\Lambda, \epsilon, \xi, k$. These parameters are actually not independent among each other. We can reduce the number of free parameters by imposing the correct values for the neutrino masses (as can be inferred by the atmospheric and solar mass squared differences) through the expressions~(\ref{eq:mass_eigen}). We can, for example, determine $\epsilon$ by imposing a normal hierarchy for the light neutrino masses\footnote{For simplicity we are reporting here just the case of a normal hierarchy regarding the light neutrino mass spectrum. An analogous procedure has been employed in the case of an  inverted hierarchy. \smallskip} leading to:
\begin{equation}
\label{eq:epsilon_neutrino}
\epsilon=\frac{2 \,  m_3 \,  \left|\Lambda\right|}{y y'\left(1+\rho\right) v^2}\ .
\end{equation}
The last line in Eq.~(\ref{eq:yukawa_parameters}) implies that only one of the two parameters $\xi$ and $k$ is a free parameter, which we choose to be $k$.
We thus generate a set of models by scanning over $y,y',\Lambda,k$ within the following ranges:
\begin{align}
& 100\,\mbox{MeV} < {|}\Lambda{|} < 40\,\mbox{GeV}\ ,\nonumber\\
& 10^{-10} < y,y^{\prime} <1 \ ,\nonumber\\ 
& 10^{-10} < k <100 \ ,
\end{align}
where, in the spirit of the model, we chose $y$ and $y^{\prime}$ to be of the same order of magnitude for each generated realisation (see Section~\ref{sec_toymodel}).
All the generated points are required, besides complying with the correct neutrino mass and mixing pattern~\cite{Gonzalez-Garcia:2014bfa}, to comply with all laboratory constraints, including LFV processes, searches of heavy neutrinos in rare decay processes and direct production at colliders, as well as a cosmological bound, consisting on the requirement for the lifetime of heavy neutrinos of being below the typical time scale of onset of BBN. 
The bounds used in our numerical study are based on Ref.~\cite{Alekhin:2015byh}.
Finally we calculate the baryon asymmetry generated in the oscillations of the sterile neutrinos (cf.\ Appendix~\ref{app_leptogenesis})\footnote{We follow the notation of Refs.~\cite{Asaka:2005pn,Asaka:2011wq,Canetti:2012kh} for our expression for the baryon asymmetry.\smallskip}:
\begin{equation}
\label{eq:baryo_analytical}
Y_{\Delta B}=\frac{n_{\Delta B}}{s}= \frac{945}{2528} \frac{\, 2^{2/3}}{  \,\,  3^{1/3} \, \pi^{5/2}  \,  \Gamma(5/6)} \frac{1}{g_s}\sin^3 \phi \, \frac{M_0}{T_{\rm W}} \frac{M_0^{4/3}}{ \left(\Delta m^2\right)^{2/3}} \, Tr\left[ F^\dagger \delta F\right] \ ,
\end{equation}
where $F=Y^\text{eff} $ with $Y^\text{eff} $  defined in Eq.~(\ref{eq_Yeff}), 
 $\Delta m^2=M_2^2-M_1^2$ is the mass squared splitting of the heavy neutrinos, $T_{\rm W}$ is the temperature of the EW phase transition - set to 140 GeV, $g_s$ counts the degrees of freedom in the thermal bath at $T = T_\text{W}$, $M_0 \approx 7 \times 10^{17}\,\mbox{GeV}$, $\sin\phi \sim 0.012$ and $\delta = \text{diag}(\delta_\alpha)$ is the CP asymmetry in the oscillations defined as:
\begin{equation}\label{eq:deltaCP}
\delta_{\alpha}=\sum_{i >j} \text{Im}\left[F_{\alpha i} \left(F^{\dagger} F\right)_{ij} F^{\dagger}_{j\alpha}\right]\ .
\end{equation}
As before the index $\alpha$ corresponds to a flavour index, while the indices $i,j$ run over the sterile mass eigenstates. The derivation of this expression, firstly introduced in~\cite{Akhmedov:1998qx,Asaka:2005pn}\footnote{The expression~(\ref{eq:baryo_analytical}) differs by an $O(1)$ factor with respect to the one given in these references. The origin of this difference will be clarified in Appendix~\ref{app_leptogenesis}.}, is carefully revisited in the appendix. In the next section, this analytical expression will be confronted with the numerical solution of suitable Boltzmann equations, also detailed in the appendix.
Eq.~\eqref{eq:baryo_analytical} is valid under the assumption that the baryon asymmetry is produced with maximal efficiency, which is achieved if the heavy sterile neutrinos never reach thermal equilibrium during the generation process and, consequently, washout effects are always negligible. This requirement can be expressed, as rule of thumb, through the condition $\left|Y^\text{eff} _{\alpha i}\right| < \sqrt{2}\times 10^{-7}$~\cite{Akhmedov:1998qx} (the condition applies to all the elements of the matrix $Y^\text{eff} $).

We consider models as viable if Eq.~(\ref{eq:baryo_analytical}) yields a value for $Y_{\Delta B}$ such that $3 \times 10^{-11} \leq Y_{\Delta B} \leq 2.5 \times 10^{-10}$. The choice of this broad range, compared to the rather precise experimental determination~\cite{Ade:2015xua}, $Y_{\Delta B}= \left(8.6 \pm 0.01 \right) \times 10^{-11}$, is motivated by the need  to account for deviations with respect to the determination of $Y_{\Delta B}$ from the numerical solution of the Boltzmann equations. We expect, in particular, that the analytical expression~(\ref{eq:baryo_analytical}) overestimates the baryon asymmetry for values of $Y^\text{eff} $ close to the equilibrium value $\sqrt{2}\times 10^{-7}$, since in this case the late time equilibration of the heavy neutrinos will reduce the total baryon asymmetry. The choice of the range of allowed values of $Y_{\Delta B}$ allows also to account for uncertainties in the determination of the production/destruction rates of the heavy neutrinos (see appendix A).

\subsection{Comparison with numerical results}
\label{sec:numerical_ly}

\begin{figure}[htb]
\begin{center}
\subfloat{\includegraphics[width=6.0 cm]{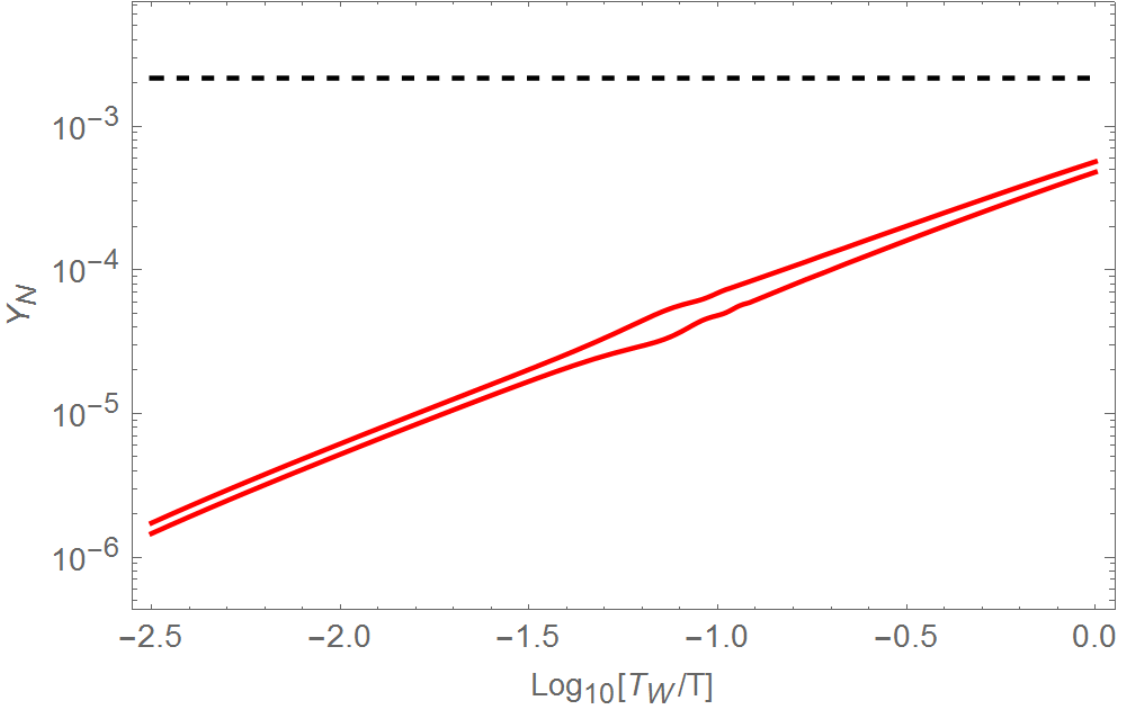}} \hfil
\subfloat{\includegraphics[width=6.0 cm]{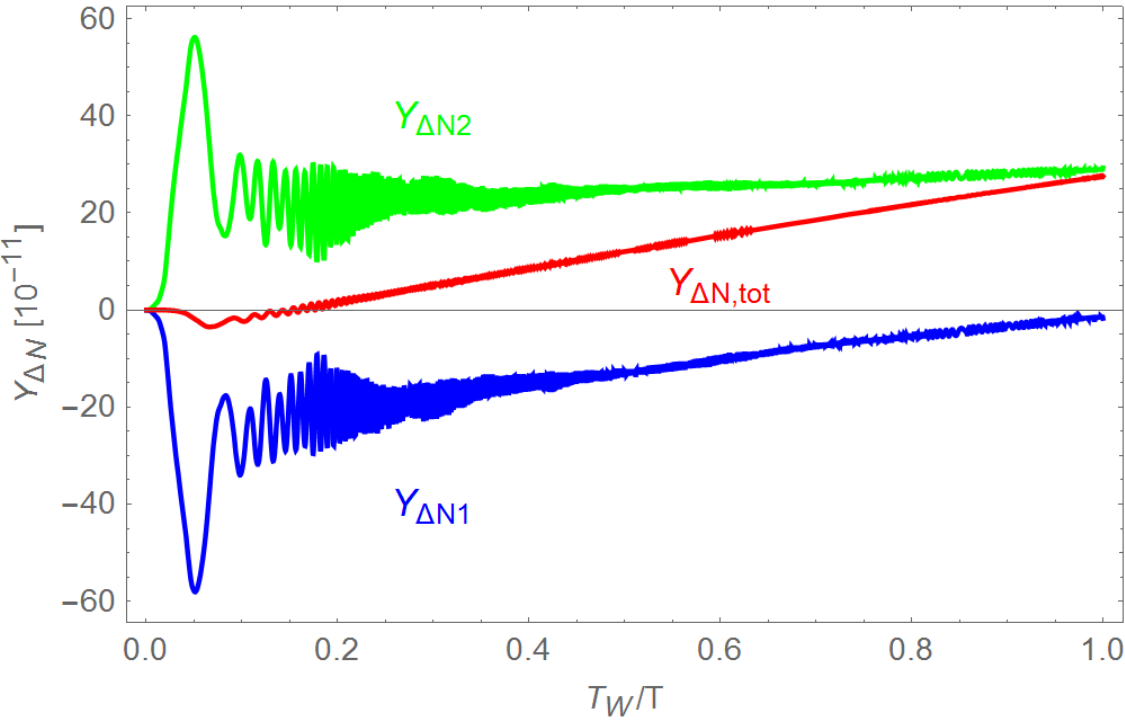}}\\
\subfloat{\includegraphics[width=6.5 cm]{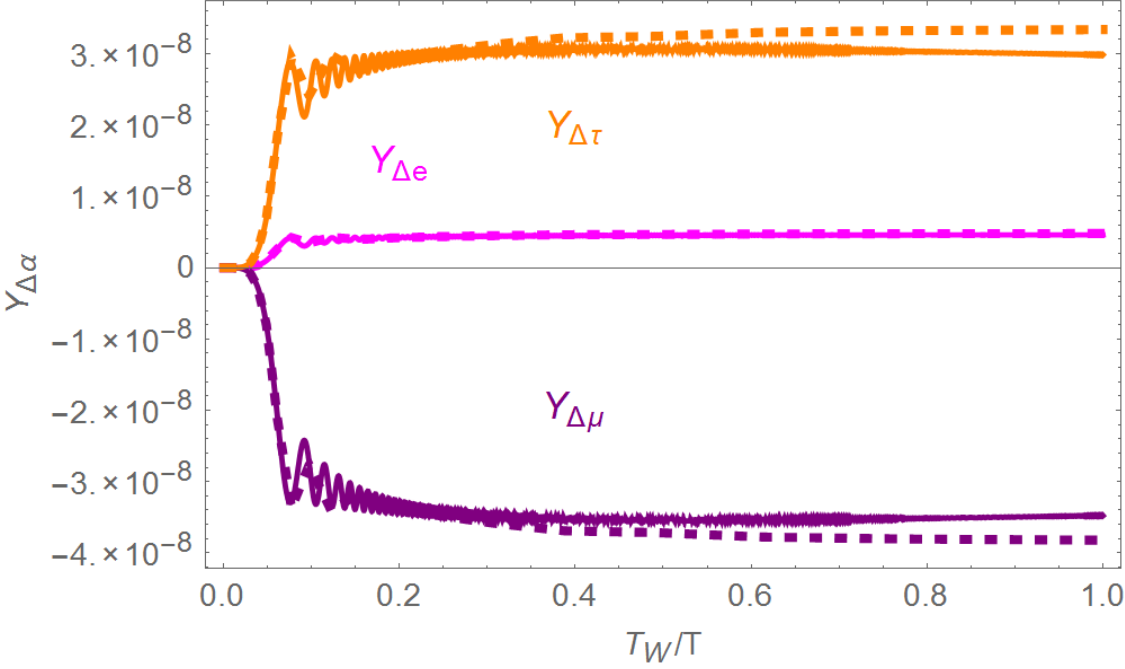}}\hfil
\subfloat{\includegraphics[width=6.5 cm]{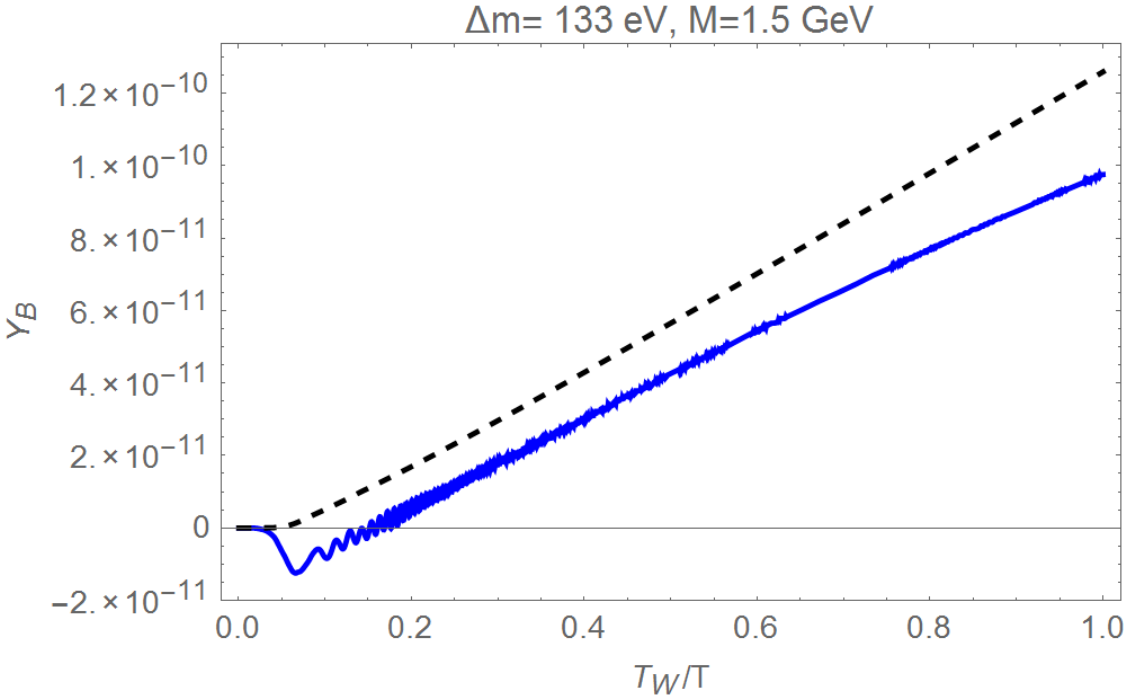}}
\caption{\footnotesize{Evolution of some relevant quantities as function of the temperature, obtained from our numerical treatment for a benchmark model with parameters $\epsilon$ and $\xi$ lying in the ``perturbative'' regime. Top left panel: Evolution of the abundance of the heavy neutrinos (red solid lines) compared with the equilibrium value (dashed black line) $Y_{{N0}}$, defined in the appendix. Top right panel: Evolution of the total (red line) and individual (blue and green lines) asymmetries in the two heavy neutrinos. 
Bottom left panel: Evolution of the asymmetries in the active flavours according the numerical solution of the Boltzmann equations (solid lines) and the analytical estimate (dashed lines). 
Bottom right panel: Evolution of the baryon abundance (blue line) compared with its analytical determination (dashed black line).}}
\label{fig:bench_natural}
\end{center}
\end{figure}

\begin{figure}[htb]
\begin{center}
\subfloat{\includegraphics[width=6.15 cm]{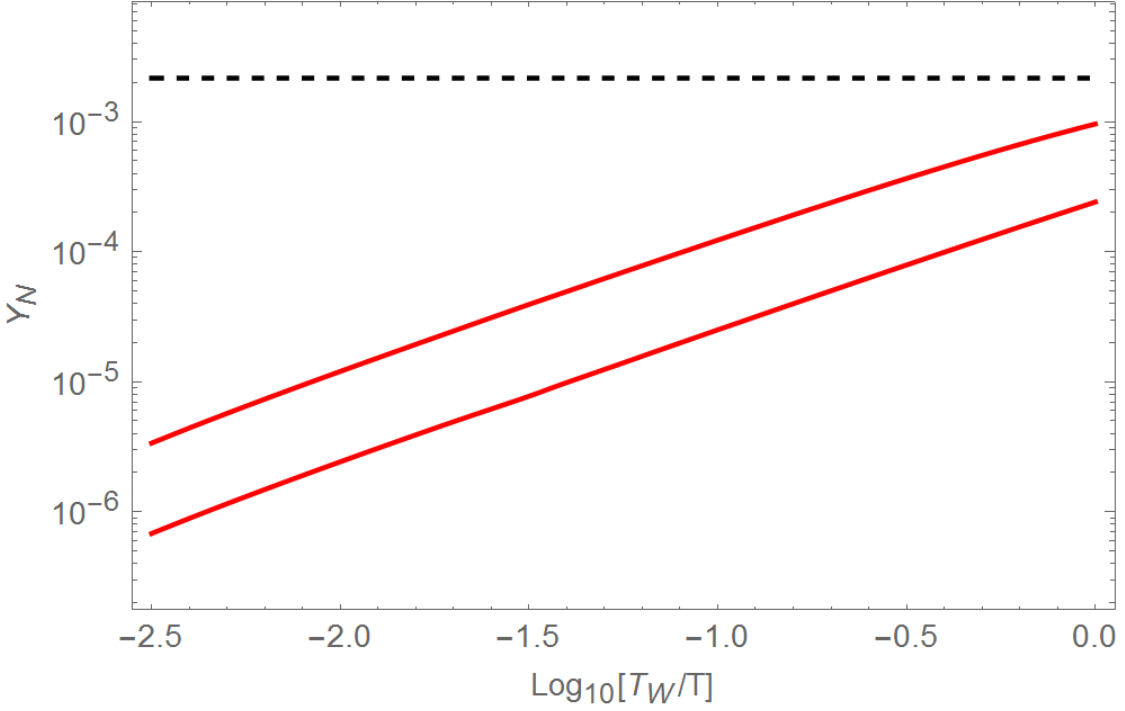}} \hfil 
\subfloat{\includegraphics[width=6.0 cm]{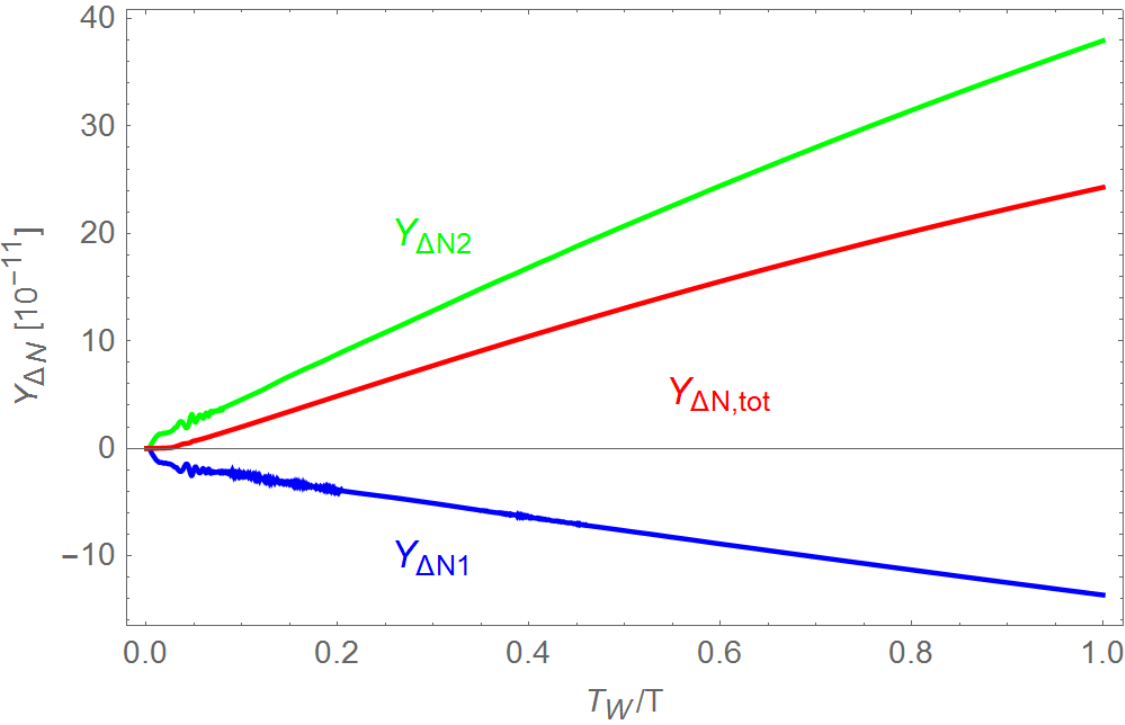}}\\
\subfloat{\includegraphics[width=6.5 cm]{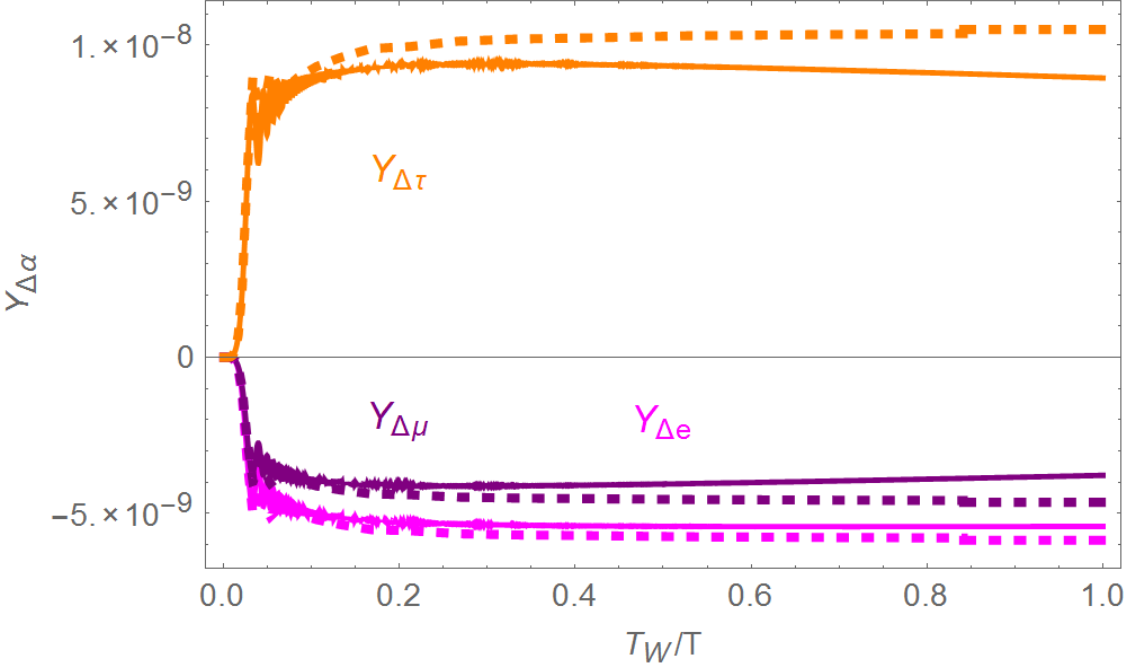}} \hfil
\subfloat{\includegraphics[width=6.5 cm]{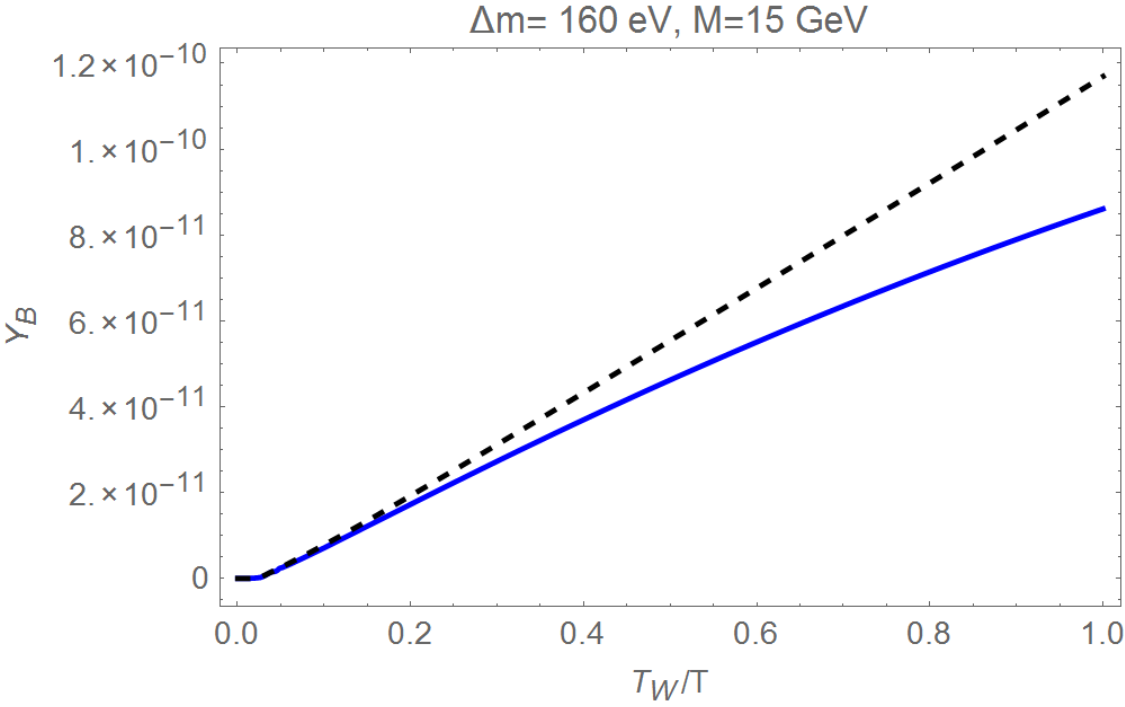}}
\caption{\footnotesize{As in Fig.~\ref{fig:bench_natural} but for a model with large $\epsilon$ belonging to the ``generic'' regime.}}
\label{fig:bench_numsm}
\end{center}
\end{figure}

\begin{figure}[htb]
\begin{center}
\subfloat{\includegraphics[width=6.2 cm]{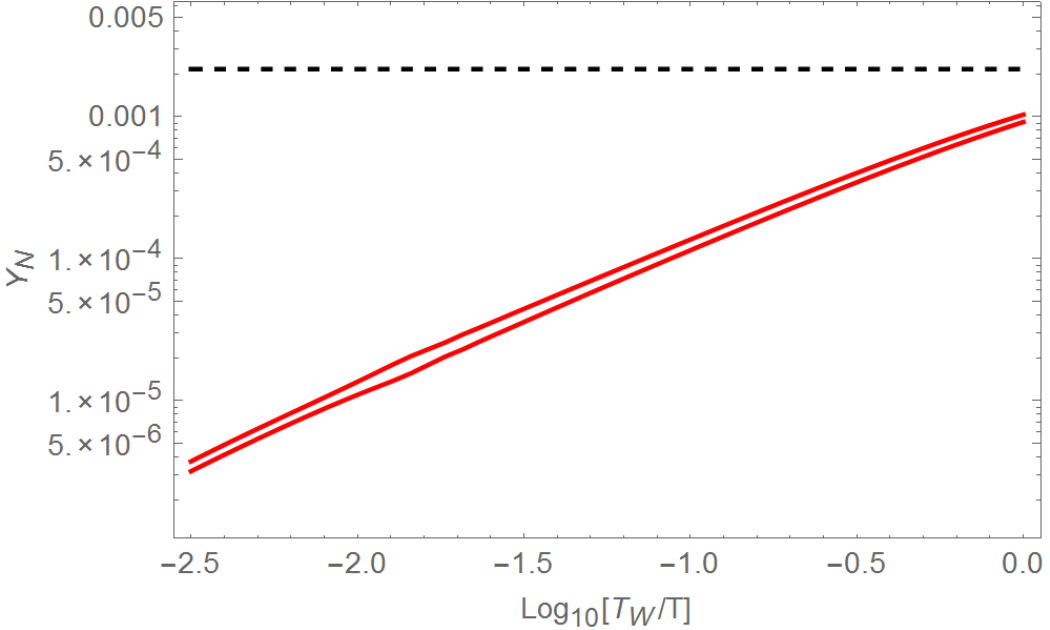}} \hfil
\subfloat{\includegraphics[width=6.0 cm]{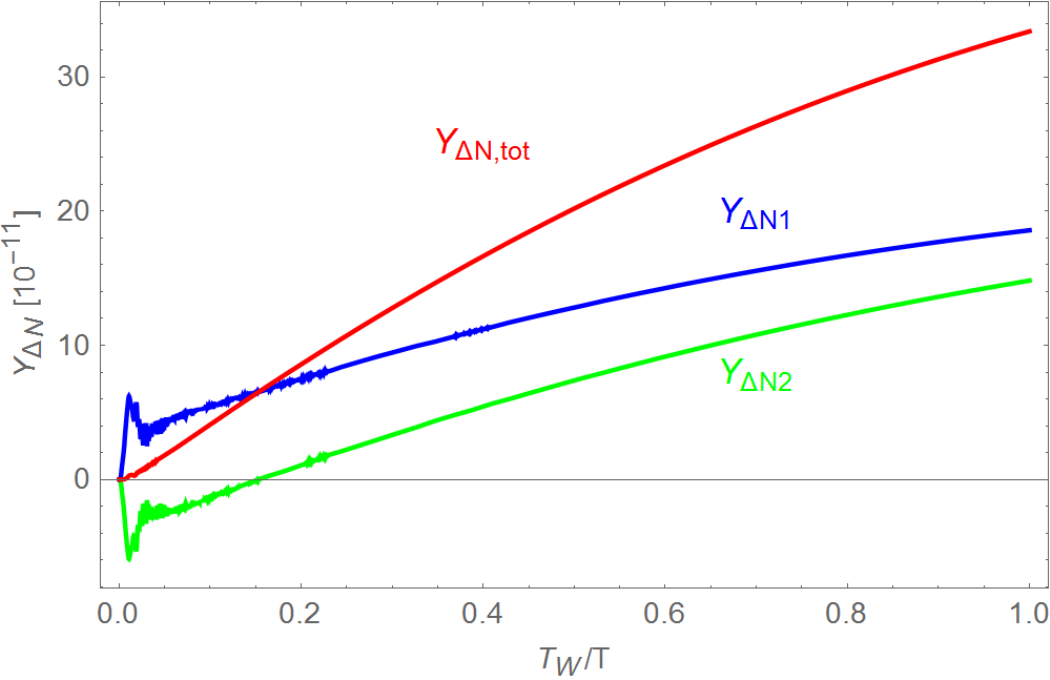}}\\
\subfloat{\includegraphics[width=6.5 cm]{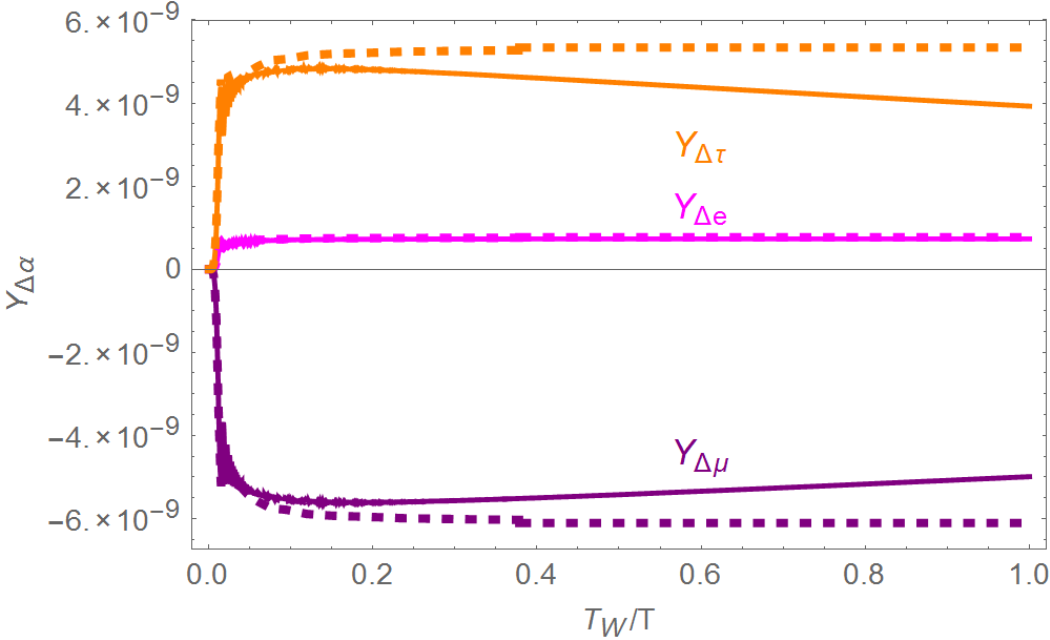}} \hfil
\subfloat{\includegraphics[width=6.5 cm]{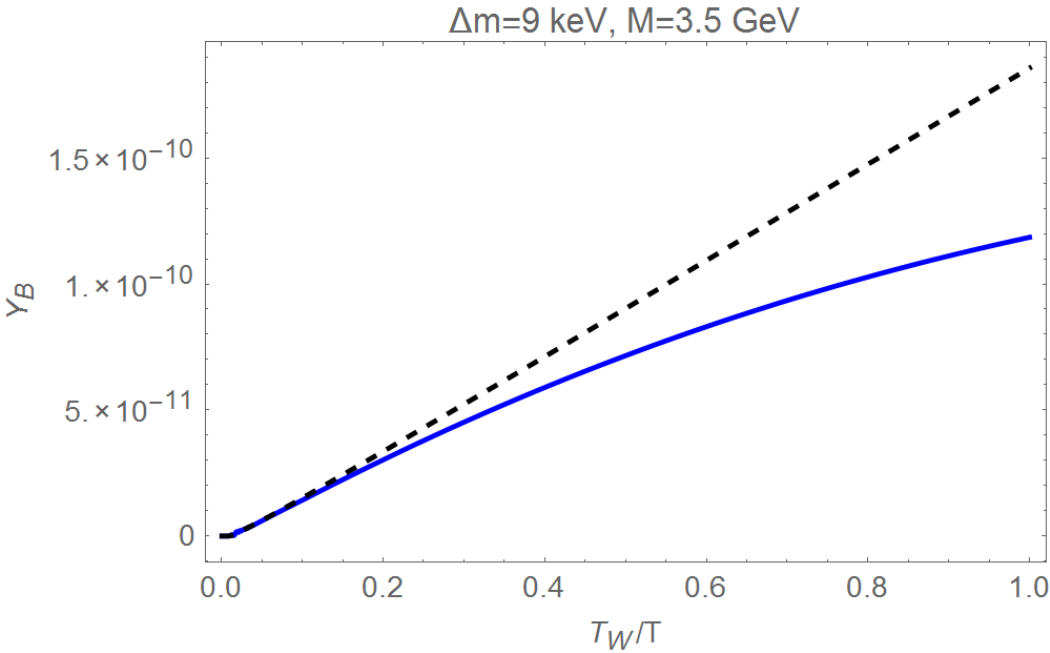}}
\caption{\footnotesize{As in Fig.~\ref{fig:bench_natural} but for a model in the ``perturbative'' regime with entries of $Y^\text{eff} $ close to the equilibration value. }}
\label{fig:bench_thermal}
\end{center}
\end{figure}

In this section we compare the analytical expression~(\ref{eq:baryo_analytical}) for the baryon asymmetry in the weak washout regime with the numerical solution of the Boltzmann equations describing this process, cf.~Eq.~\eqref{eq:full_system} in Appendix~\ref{app_leptogenesis}, for a set of benchmark points (see also Ref.~\cite{Asaka:2011wq}).
In most of the cases we have found a good agreement, with deviations ranging between 5 and 15 $\%$. Larger deviations arise if the  entries of $Y^\text{eff} $ are very close to the out-of-equilibrium condition. 
An explicit comparison between the numerical and analytical determination of the baryon density is shown in Figs.~\ref{fig:bench_natural},~\ref{fig:bench_numsm} and~\ref{fig:bench_thermal}. The three corresponding benchmark points represent, respectively, a model satisfying the ``perturbative'' regime, featuring $\epsilon \sim 0.01$, a model in the ``generic'' regime, with $\epsilon \sim 1$, and, finally, a model with the entries of $Y^\text{eff} $ very close to the out-of-equilibrium condition. The  mass scale $M=(M_1+M_2)/2$ of the heavy neutrinos and their mass splitting $\Delta m$ are reported in the fourth panel of each figure. Moreover, the entries of the matrices $Y^{\rm eff}$ are listed in Appendix~\ref{app_benchmarks}.

\begin{figure}[htb]
\begin{center}
\includegraphics[width=7.0 cm]{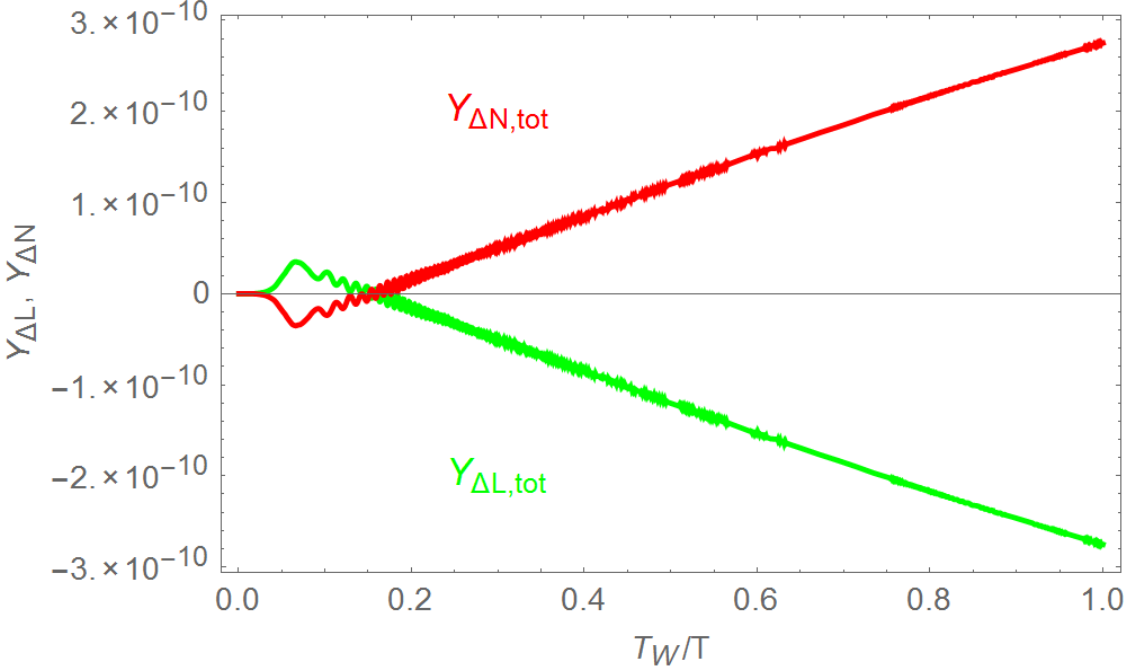}
\caption{\footnotesize{Total asymmetries in the sterile (red curve) and active (green curve) sectors as function of the temperature, for the first benchmark model (cf.\ Fig.~\ref{fig:bench_natural}). As expected from the conservation of the total lepton number, these are equal and opposite.}}
\label{fig:numerical_cross_check}
\end{center}
\end{figure}

For each benchmark point we show a set of four plots describing the evolution of the abundance of the two heavy neutrinos, the individual and total asymmetries in the sterile sector, the individual asymmetries in the active sector and, finally, the baryon asymmetry $Y_B$. 
The baryon asymmetries are also compared with their analytical estimates, represented as dashed lines, whose derivation is described in detail in the appendix.
These plots illustrate the main features of the leptogenesis mechanism at work here: the abundance of sterile neutrinos (top left panel) grows according to Eq.~\eqref{eq:first_step}, but does not reach the equilibrium value before sphaleron processes convert the lepton asymmetry into a baryon asymmetry at $T \sim T_\text{W}$, thus suppressing washout processes. The oscillations of these sterile states source an asymmetry in the individual active and sterile flavours (bottom left and top right panel). These are described by Eq.~\eqref{eq:secondstep} and \eqref{eq:third_step}, respectively, after inserting Eqs.~\eqref{Eqrho}, \eqref{eq:YN0} and \eqref{eq:RN}.\footnote{Notice that the comparison between the numerical determination of the individual asymmetries in the active sector and the analytical expression~\eqref{eq:secondstep} should be regarded with care.  Besides deviations appearing at low temperatures in Figs.~\ref{fig:bench_numsm}-\ref{fig:bench_thermal} due to the fact that the heavy neutrinos approach thermal equilibrium (see main text), the two quantities should not exactly coincide. Indeed in the analytical derivation a net baryon asymmetry appears, as a higher order effect, only in Eq.~\eqref{eq:third_step} (see also~\cite{Asaka:2005pn}). This effect, on the contrary, is already automatically encoded in the numerical determination of $Y_{\Delta {\alpha}}$.} We note that the asymmetries in the individual flavours, in particular in the active sector,  are typically much larger than the total asymmetry in that sector. We have confirmed that the analytical expressions presented in the appendix are well adapted to describe both the individual asymmetries, as well as the total asymmetries. 
As expected from global lepton number conservation, the total asymmetries in the active and sterile sector are equal but with opposite sign, as demonstrated in Fig.~\ref{fig:numerical_cross_check} for the benchmark point of Fig.~\ref{fig:bench_natural}. The sphaleron processes however only act on the active flavours, yielding a total baryon asymmetry described by Eq.~\eqref{eq:baryo_analytical} and depicted in the fourth panel of Figs.~\ref{fig:bench_natural}, \ref{fig:bench_numsm} and~\ref{fig:bench_thermal}.

We will now discuss in more detail the main features of each of the considered benchmarks. 
The model represented in Fig.~\ref{fig:bench_natural} is characterised by $\epsilon \sim 0.01$. As discussed in a more systematic way in the next subsection, this setup corresponds to a very strong, although not complete, superimposition of the two heavy neutrino states, see also Eq.~\eqref{eq_sterile_mixing}. This is the source of the nearly equal abundances of the two heavy states in the top left panel. In such a scenario essentially equal (large) and opposite in sign asymmetries are stored in the sterile states. A non-vanishing asymmetry arises at later times as a small difference between the individual asymmetries. This behaviour can be understood as follows. At very early times (corresponding to high temperatures) the heavy neutrino pair essentially behaves as a single Dirac neutrino, thus carrying an approximately vanishing lepton asymmetry. A net asymmetry is created only after thermal/matter effects cause the oscillations to enter into the resonant regime. The net asymmetry increases at lower temperatures due to the not exact overlap between the neutrino states. As can be seen in  the bottom right panel, the analytical estimate does not provide a correct description of the early time behaviour of the numerical solution but provides nonetheless a good approximation of the total net asymmetry such that there is a relative difference of the order of 10 $\%$ between the numerical and analytical determination of $Y_B$.

Rather different is the case of the benchmark presented in Fig.~\ref{fig:bench_numsm}, which features $\epsilon \sim 1$. As suggested by the top left plot of the figure, there is little overlap between the two heavy neutrinos. Contrary to the previous scenario, the two states acquire individual and uncorrelated net asymmetries which grow essentially monotonically in time. As evident from the plot there is a very good agreement with the analytical estimate at early times (high temperatures) while small deviations arise at later times since one of the two neutrinos gets very close to thermal equilibrium (see dashed black line in top left panel of Fig.~\ref{fig:bench_numsm}), causing a slight depletion of the asymmetries in the $\mu$ and $\tau$ flavours, which in this benchmark come with $Y^\text{eff}$ values close to the equilibrium value.

The late time depletion of the baryon asymmetry is more evident in Fig.~\ref{fig:bench_thermal}, where all the entries of the matrix $Y^{\rm eff}$ are close to the equilibrium value. This translates into a relative difference of approximately 40 $\%$ between the analytical and numerical solutions. With a value of $\epsilon \sim 0.01$ for this point, we notice  in the top left panel (as in Fig.~\ref{fig:bench_natural}) that the two heavy neutrinos appear strongly overlapped. Contrary to the first benchmark model, we observe here a good agreement between the analytical and numerical solutions at early times. This is due to the fact that the oscillations enter rather early in the resonant regime and thus only a small mismatch between the analytical and numerical solutions arises. More generally we have found that in the regime of $Y^{\rm eff}$ close to the equilibrium value, the analytical determination of the baryon density can overestimate the correct value by up to a factor 3. This  motivates the choice of a broad range of allowed values for the baryon density in our scan of the parameter space.

Finally  we remark that, as already argued in~\cite{Drewes:2012ma}, the asymmetries in the individual active flavours are in general much higher in magnitude, with respect to the total net lepton asymmetry converted by the sphalerons. We notice in particular that larger (in magnitude) asymmetries are stored in the $\mu$ and $\tau$ flavours. This is because the considered benchmark model features a normal hierarchical neutrino spectrum. In such a case the matrix $Y^\text{eff} $ features as well a hierarchical structure with larger values of the entries associated to the $\mu$ and $\tau$ flavours, which thus achieve larger amounts of asymmetry.

%% file: numerics_results0805.tex
\section{Discussion of the weak washout regime}\label{Sec:Results.discussion}

In this section we discuss the key results obtained from the dedicated parameter scan in the weak washout regime described in Section~\ref{parameter_scan}.

\begin{figure}[htb]
\begin{center}
\begin{tikzpicture}
 \node at (-4.5,0) {\includegraphics[width=6.5 cm]{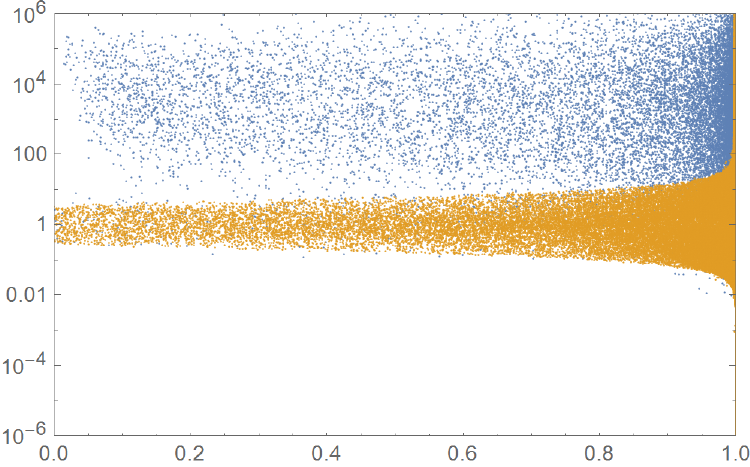}};
\node[rotate = 90] at (-8,0) {\footnotesize{$\epsilon$}};
\node at (-4.5, -2.4) {\footnotesize {$\sin^2(2 \theta_{PD}) $}};
\node at (-0.5, 0.3) {\textcolor{orange}{\footnotesize \bf $\xi < 0.1$}};
\node at (-0.5, -0.3) {\textcolor{mathblue}{\footnotesize \bf $\xi > 0.1$}};
 \node at (4,0) {\includegraphics[width=6.5 cm]{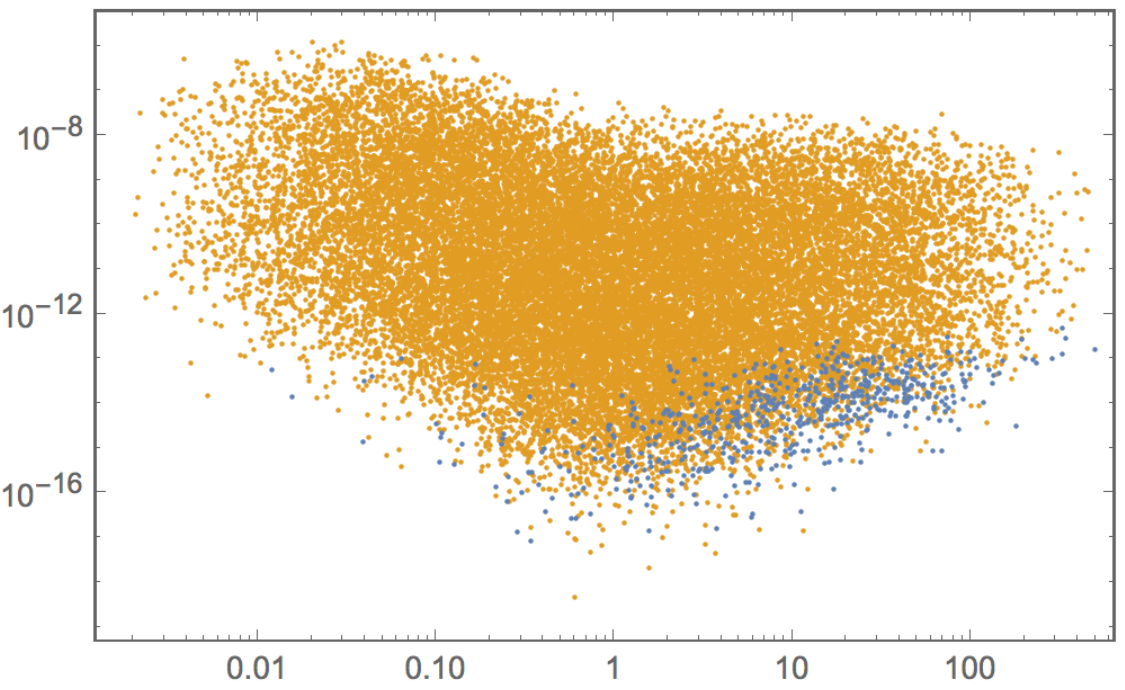}};
\node[rotate = 90] at (0.4,0) {\footnotesize{$Y_B$}};
\node at (4, -2.4) {\footnotesize{ $\epsilon$}};
\end{tikzpicture}
\caption{\footnotesize{Viable parameter points for $\xi > 0.1$ (in blue) and $\xi < 0.1$ (in orange) after imposing constraints from neutrino masses and mixings as well as laboratory searches for sterile fermions. Left panel:  In the plane of the lepton number violating parameter $\epsilon$ and the mixing angle $\theta_{PD}$ between the two heavy mass eigenstates. Right panel:  In the plane of $\epsilon$ and the generated baryon asymmetry $Y_B$, after having imposed the out-of-equilibrium condition $|Y^\text{eff}_{\alpha j}| < \sqrt{2} \times 10^{-7}$.}}
\label{fig:befYB}
\end{center}
\end{figure}
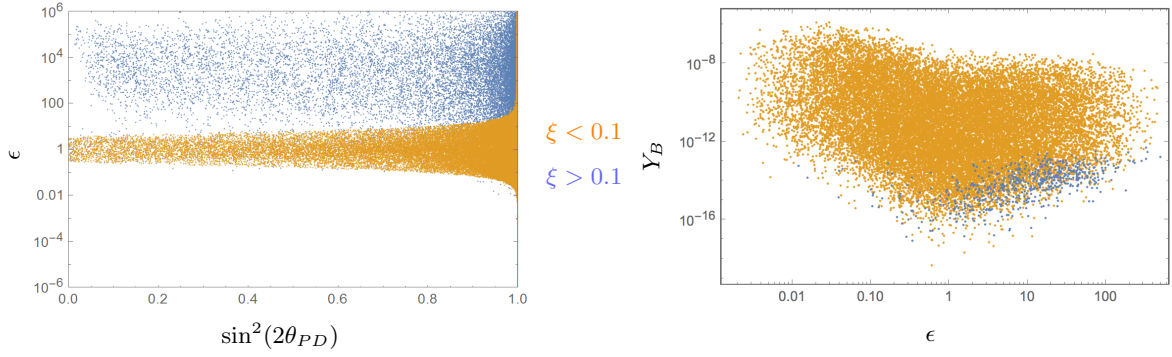

In Fig.~\ref{fig:befYB} we depict some instructive properties  of the solutions found in the scan over the parameter space before imposing the constraint on the baryon asymmetry. In the left panel, we show these solutions in the plane $(\epsilon, \sin^2(2 \theta_{PD}))$. Here $\theta_{PD}$ is the mixing angle between the two heavy mass eigenstates resulting from the potential in Eq.~\eqref{eq_VN}. A large mixing angle enhances the oscillations among the heavy states and hence the produced baryon asymmetry. As discussed in Section~\ref{sec_toymodel}, for small values of the lepton number violating parameter in the Majorana mass term, $\xi < 0.1$ (yellow points), the resulting distribution is approximately symmetric under the transformation $\epsilon \rightarrow 1/\epsilon$, corresponding to switching the lepton number assignments of the two additional states. In fact this will be the region we will focus on in the following, since large values of $\xi$ (blue points, $\xi > 0.1$) imply a mass splitting between the two heavy states too large to accomplish a successful leptogenesis, cf.\ Eq.~\eqref{eq_mueps}. Moreover, among the points with $\xi < 0.1$, we can distinguish two types of solutions. For $\epsilon < 0.1$ or $1/\epsilon < 0.1$ the mixing between the two heavy states is found to be close to maximal, $\sin^2(2 \theta_{PD}) \simeq 1$. This corresponds to the solutions found in the perturbative expansion of the toy model discussed in Section~\ref{sec_toymodel}, dubbed ``perturbative'' solutions. On the other hand, for $0.1 < \epsilon < 10$ any value of the mixing angle $\theta_{PD}$ can be obtained. This is what we referred to as ``generic'' solutions.

The right panel of Fig.~\ref{fig:befYB} shows the dependence of the resulting baryon asymmetry on the two lepton number violating parameters $\epsilon$ and $\xi$, after imposing the out-of-equilibrium condition $|Y^\text{eff}| < \sqrt{2} \times 10^{-7}$.  As anticipated from the previous figures, values of $\epsilon$ much larger or much smaller than one lead to a large mixing of the heavy states, rendering leptogenesis through the oscillations of these states very effective. Moreover, large values of $\xi$ imply a large mass splitting between the heavy states, rendering leptogenesis less effective (blue points versus yellow points). In fact above the EW phase transition, in the regime relevant for leptogenesis, the correlation between the mass splitting $\Delta m^2$ and $\xi$ is very well described by Eq.~\eqref{eq_mueps} in the regime $\xi \leq 1$.

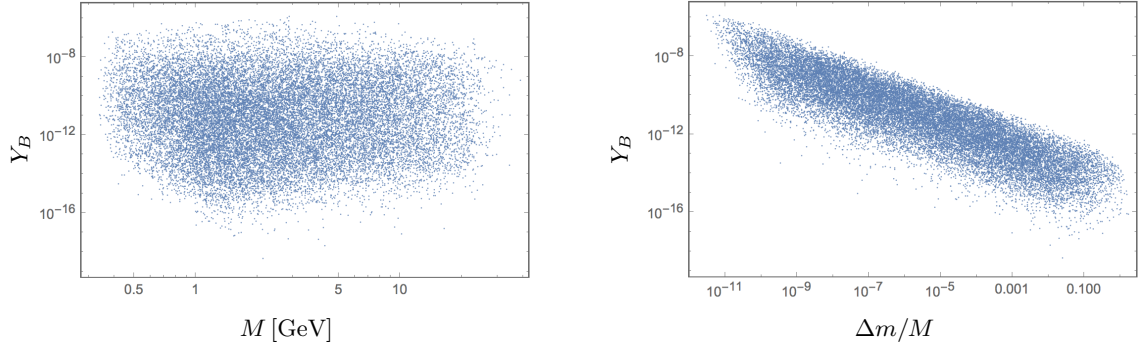
\begin{figure}[htb]
\begin{center}
\begin{tikzpicture}
 \node at (-4,0) {\includegraphics[width=6.5 cm]{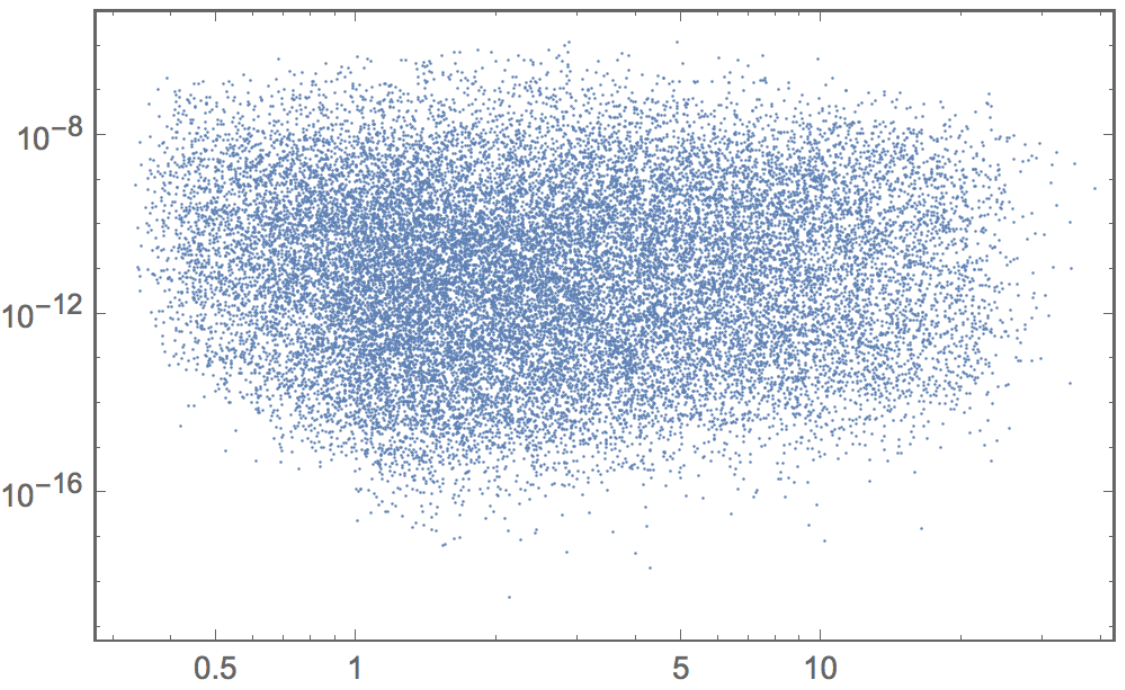}};
\node[rotate = 90] at (-7.5,0) {\footnotesize{$Y_B$}};
\node at (-4, -2.4) {\footnotesize{$M \, [ \text{GeV} ]$}};
 \node at (4,0) {\includegraphics[width=6.5 cm]{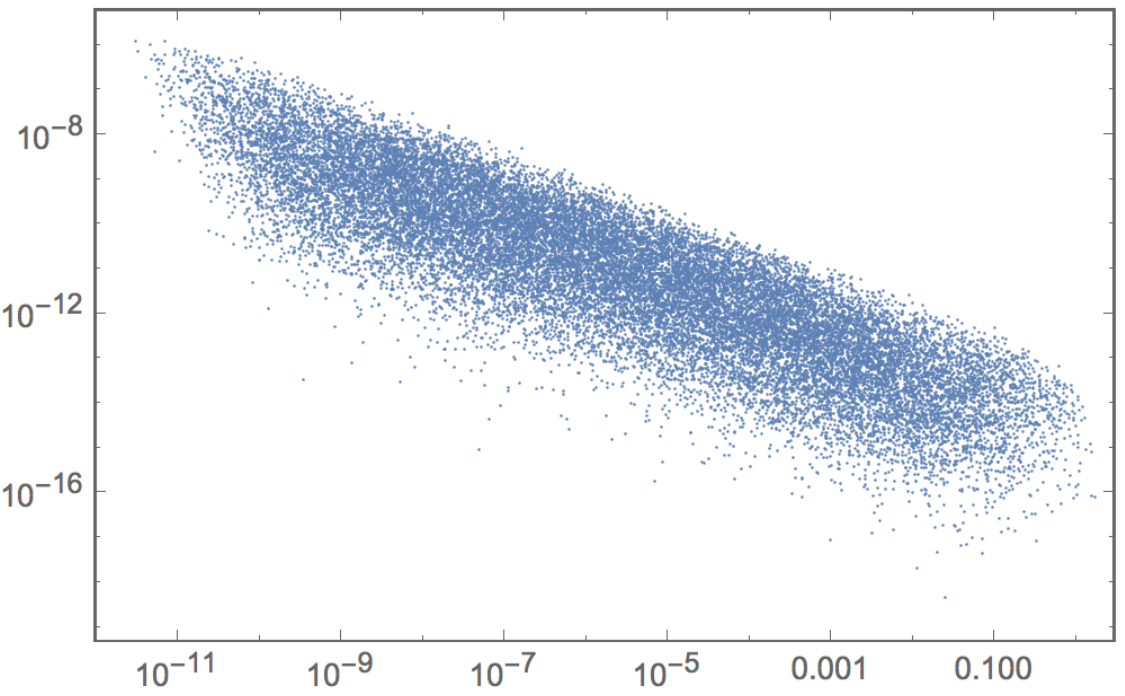}};
\node[rotate = 90] at (0.4,0) {\footnotesize{$Y_B$}};
\node at (4, -2.4) {\footnotesize{$\Delta m/M$}};
\end{tikzpicture}
\caption{\footnotesize{Viable parameter points after imposing the constraints from neutrino masses and mixings, from laboratory searches for sterile fermions and the out-of-equilibrium condition $|Y^\text{eff}_{\alpha j}| < \sqrt{2} \times 10^{-7}$, but before imposing the constraint on the baryon asymmetry. We show the baryon asymmetry $Y_B$ as function of the scale of the heavy neutrinos $M=(M_1+M_2)/2$ (left panel) and of the relative mass splitting $\Delta m/M$ between the two heavy states (right panel).}}
\label{fig:MdM}
\end{center}
\end{figure}

In Fig.~\ref{fig:MdM} our results have been re-expressed as function of the dimensionful parameter $M=(M_1+M_2)/2$, i.e. the mass scale of the heavy neutrinos, 
and of the relative mass splitting between the two heavy states, $\Delta m/M$, to give an impression of the viable parameter space. 
Here we show only solutions which obey the out-off equilibrium condition, $|Y^\text{eff}| < \sqrt{2} \times 10^{-7}$. We find solutions within the assumed viable range of values of the baryon asymmetry, i.e. $ 3 \times 10^{-11} < Y_B < 2.5 \times 10^{-10} $, for basically the entire range of heavy neutrino mass scales considered, $0.3~\text{GeV} \lesssim M \lesssim 35~\text{GeV}$ and relative mass splitting within  $10^{-11} \lesssim \Delta m/M \lesssim 10^{-3}$, 
with a lower bound on the mass splitting $\Delta m \gtrsim 10^{-2} \, \mbox{eV}$. \footnote{This lower limit on the mass splitting is not actually originated by the requirement of viable leptogenesis but has been imposed, as an additional constraint, in the parameter scan. It follows from the requirement $T_L > T_W$, where $T_L$ defined in Eq.~(\ref{eq:TL}) is the temperature at which the production of the lepton asymmetry is peaked. Although lower mass splittings are not excluded for a viable  leptogenesis~\cite{Canetti:2012kh}, since thermal effects can modify the value of the mass splitting inferred by the diagonalization of the mass matrix, we have imposed this lower bound in order to compare the numerical and the analytical determinations of the baryon abundance. As clarified in Appendix~\ref{app_leptogenesis}, the latter relies on the assumption that the temperature dependent mass splitting originated during oscillations is subdominant with respect to the one sourced by $\xi$.} 
\begin{figure}[htb]
\begin{center}
\subfloat{
\begin{tikzpicture}
 \node at (0,0) {\includegraphics[width=8.2 cm]{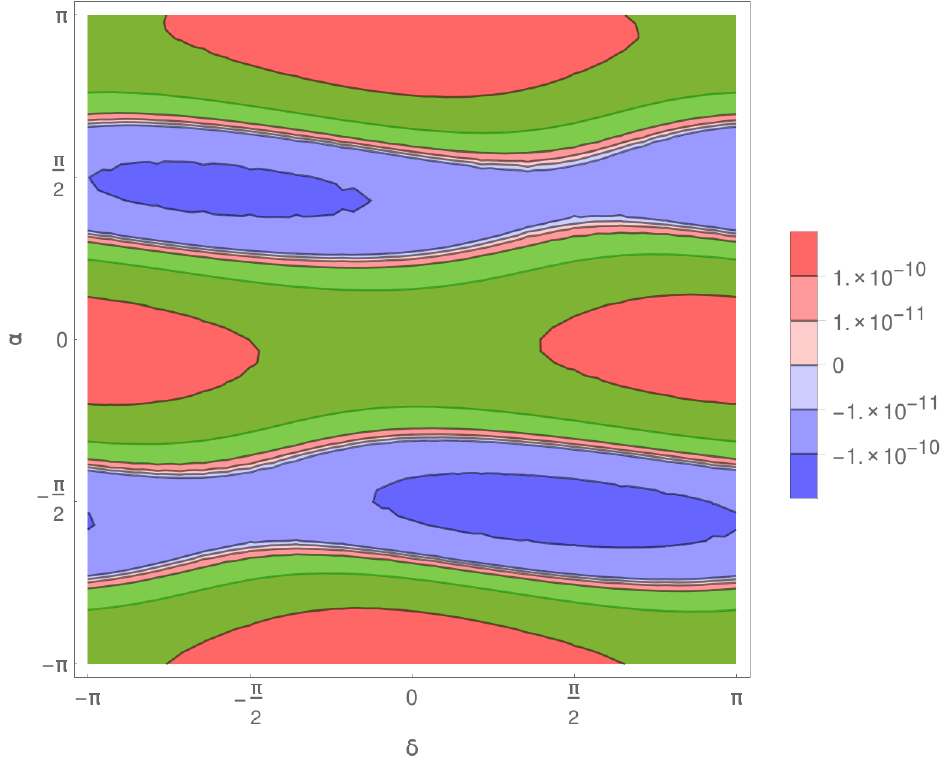}};
 \node[rotate = 90, fill = white] at (-4,0.25) {\footnotesize $\alpha$};
\node[fill = white] at (-0.2,-3.1) {\footnotesize $\delta_{CP}$};
\node at (3.0,1.8) {\footnotesize $Y_B$};
 \node at (7.5,0) {\includegraphics[width=6.4 cm]{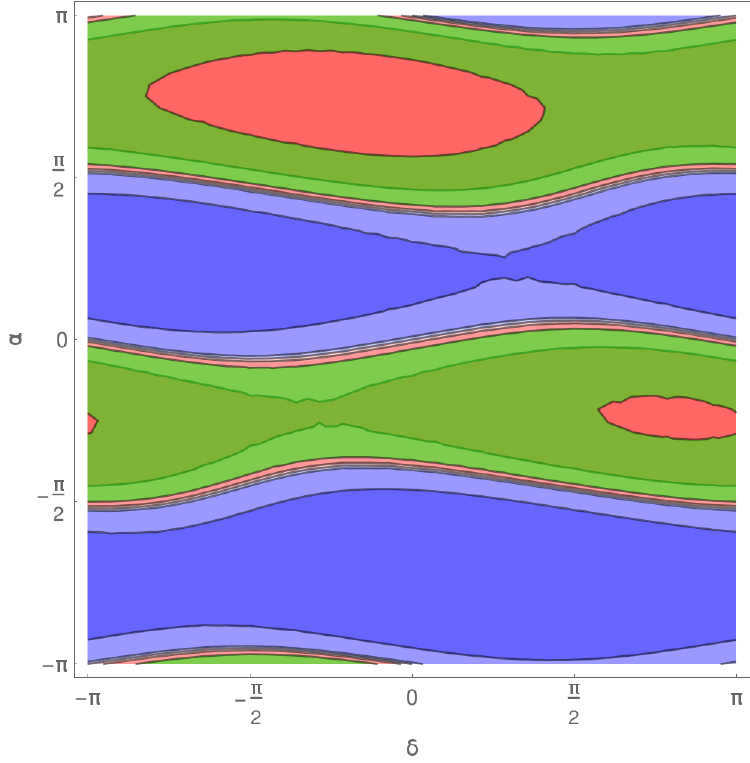}};
\node[rotate = 90, fill = white] at (4.4,0.25) {\footnotesize $\alpha$};
\node[fill = white] at (7.6,-3.1) {\footnotesize $\delta_{CP}$};
\end{tikzpicture}
} 
\caption{Contour plot of the baryon asymmetry $Y_B$ in terms of the Dirac phase $\delta_{CP}$ and the Majorana phase $\alpha$, for a fixed $(|Y_{\alpha i}|, \epsilon, \xi, |\Lambda|)$ parameter point, with a fixed non-zero phase assigned to $\Lambda$ (Arg($\Lambda) = 0.44 \,  \pi$, left panel) and a real value for $\Lambda$ (right panel). Negative (positive) asymmetry is marked in blue (red), overlayed by the green region marking a baryon asymmetry in agreement with observation, $ 3 \times 10^{-11} < Y_B < 2.5 \times 10^{-10} $.}
\label{fig:phases}
\end{center}
\end{figure}
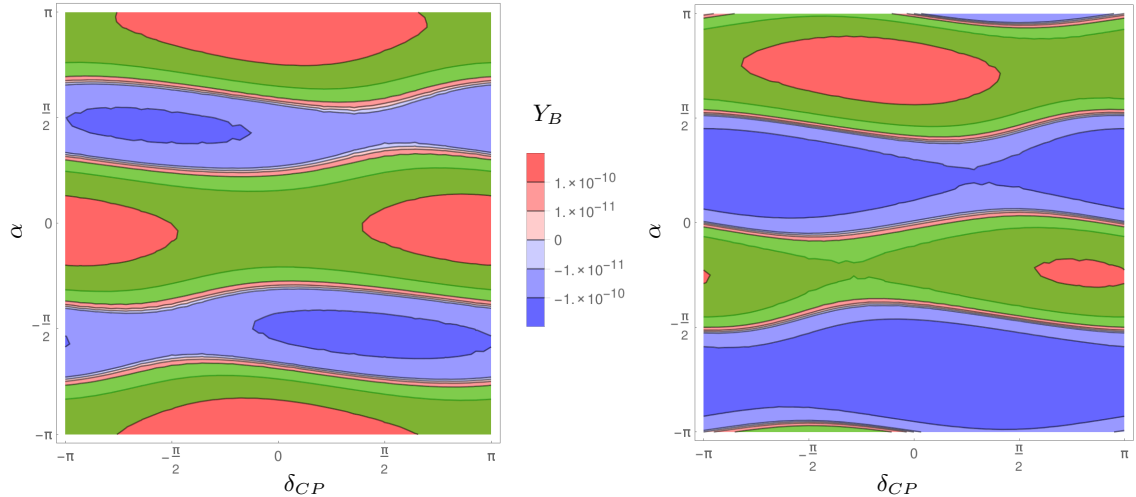

In Fig.~\ref{fig:phases} we depict the impact of the Dirac phase $\delta_{CP}$ and the Majorana phase $\alpha$ (defined in Section~\ref{sec:parametrization:mass}) on the determination of the baryon abundance $Y_B$.  To this purpose we have considered a fixed choice of the model parameters, namely  $\xi \simeq 6.7 \times 10^{-7}$, $\epsilon \simeq 0.075$, and $M_1 \simeq M_2 \simeq 4.4~\text{GeV}$, yielding  values for $Y_B$ in the allowed range and fulfilling all bounds on the active neutrinos (normal hierarchy). 
This  picture qualitatively  holds throughout the parameter space, the position of the allowed bands however varies significantly, since the third ``high-energy'' phase related to the parameter $\Lambda$ affects the value of $\delta_\alpha$ in Eq.~(\ref{eq:deltaCP}), rendering all values of the ``low-energy'' phases $\alpha$ and $\delta_{CP}$  possible when considering the entire parameter space. This third phase is also responsible for the non-zero value of the asymmetry even if $\delta_{CP}$ and $\alpha$ are zero, cf.\ left panel of Fig.~\ref{fig:phases}. { Vice versa, the correct value for the baryon asymmetry can also be obtained if this high-energy phase is zero, i.e.\ only through the phases of the PMNS matrix, as depicted in the right panel of Fig.~\ref{fig:phases}. }

\begin{figure}[htb]
\begin{center}
\subfloat{
\begin{tikzpicture}
\node at (0,0) { \includegraphics[width=6.5 cm]{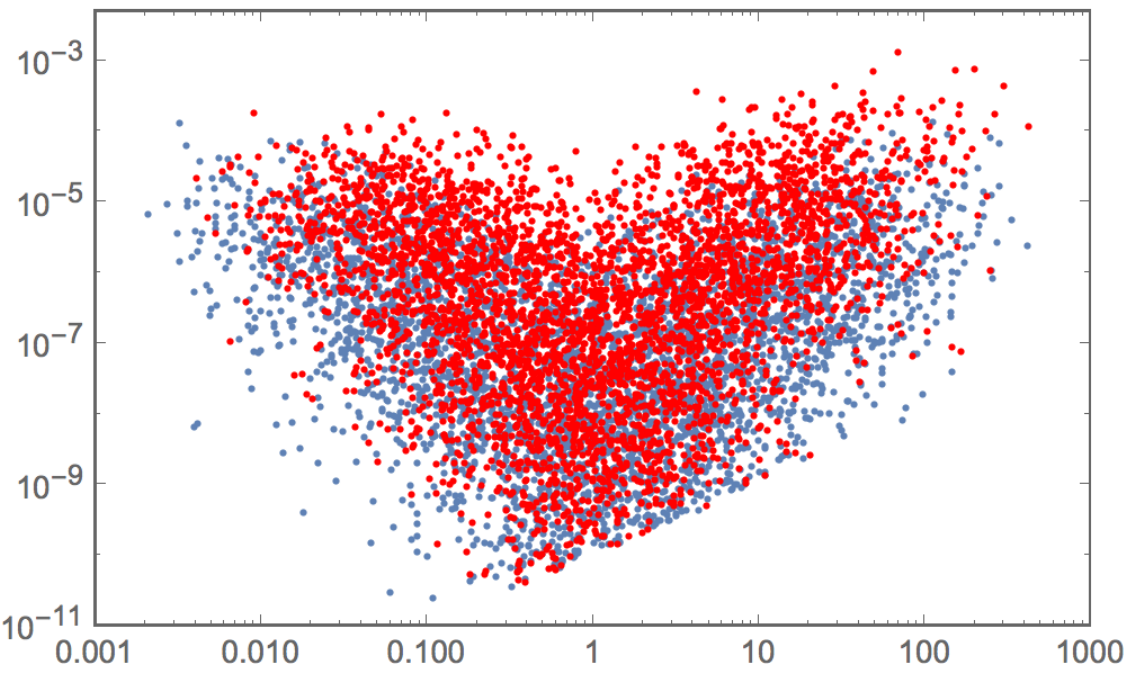}};
\node[rotate = 90] at (-3.5,0) {\footnotesize $\xi$};
\node at (0,-2.2) {\footnotesize $\epsilon$};
\node at (3.8, 0.3) {\textcolor{red}{NH}};
\node at (3.8, -0.3) {\textcolor{mathblue}{IH}};
\end{tikzpicture}
} 
\subfloat{
\begin{tikzpicture}
\node at (0,0) { \includegraphics[width=6.3 cm]{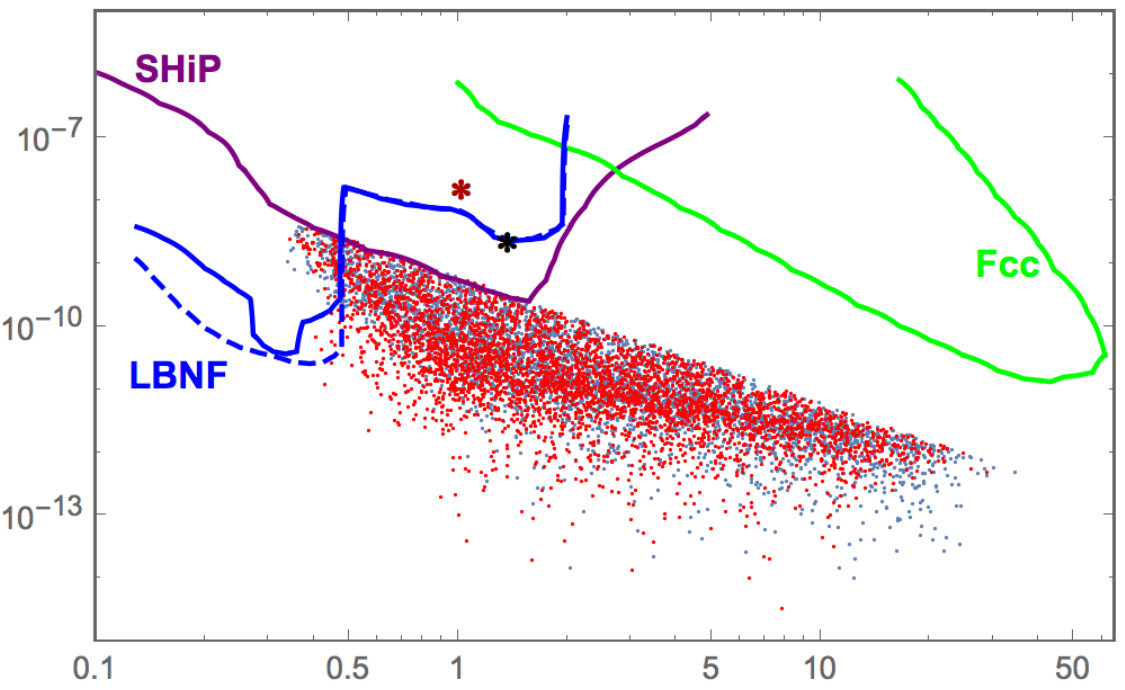}};
\node[rotate = 90] at (-3.5,0) {\footnotesize $|U_{\mu 4}|^2$};
\node at (0,-2.2) {\footnotesize $M_1$ [GeV]};
\end{tikzpicture}
}
\caption{\footnotesize{Left panel: Set of model points giving a viable baryon abundance in the weak washout regime, in the plane $(\epsilon, \xi)$. The red and blue points refer, respectively, to solutions with normal and inverted hierarchy for the active neutrino mass spectrum. Right panel: Models featuring a viable baryon abundance in the plane $\left(|U_{\rm \mu 4},M_1|\right)$ where $U_{\rm \mu 4}$ is the mixing between the lightest of the two exotic neutrinos with the $\mu$ flavour while $M_1$ is its mass. The color code is the same as in the left panel. The points are the result of a scan over the parameter space of weak washout regime. The asterisks refer to the benchmark solutions in the strong washout regime, characterised by the Yukawa couplings in Eqs.~(\ref{eq:benchSW1}) (red) and (\ref{eq:benchSW2}) (black).}}
\label{fig:aftYB}
\end{center}
\end{figure}
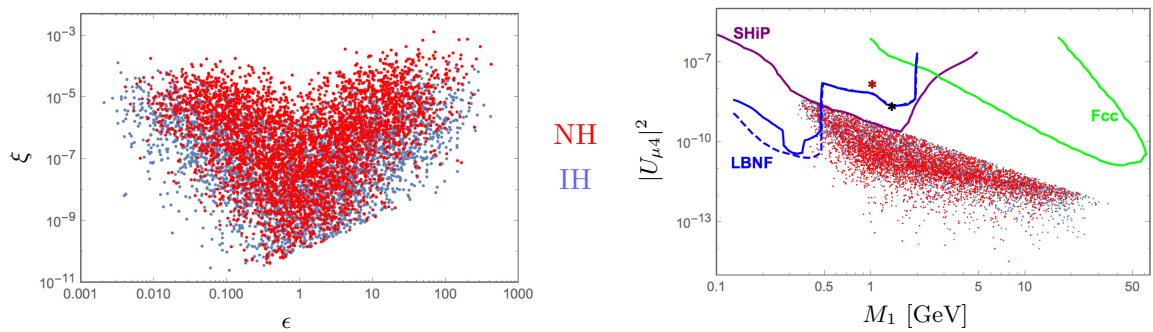

The results of our analysis are finally summarised in Fig.~\ref{fig:aftYB}, which shows the distribution of the parameter points featuring a viable baryon asymmetry, in addition to the constraints above.
In the left panel of this figure we display the distribution of the viable parameter points in the plane of the lepton number violating parameters $\epsilon$ and $\xi$. The shape of this region can be well understood in terms of the toy model presented in Section~\ref{sec_toymodel}. Again we note the  approximate symmetry under $\epsilon \rightarrow 1/\epsilon$.
This figure demonstrates that  the parameter $\xi$ appearing in the Majorana mass term must be very small, in order to ensure a sufficiently small mass splitting between the two heavy neutrinos. Indeed all viable points are found to be within the range $\xi < 10^{-3}$. On the other hand, we find viable solutions for a large range of values of $\epsilon$, and moreover $\xi \lesssim 5\times  10^{-2}\ \epsilon$ in the entire parameter range.
This follows from the fact that the value of $\epsilon$ is inversely proportional to the size of the Yukawa couplings, cf. Eq~(\ref{eq:epsilon_neutrino}), and the requirement $\left| Y^\text{eff}\right| < 10^{-7}$ translates to the bound $10^{-3} \lesssim \epsilon \lesssim 10^3$. On the other hand the requirement of a sufficiently small mass splitting puts an upper bound on the possible values of $\xi$, resulting in the aforementioned bound on the ratio of the two parameters. Lower values of $\epsilon$, and consequently larger $\xi/\epsilon$ ratios, are nonetheless allowed for values of $Y^\text{eff} > 10^{-7}$, cf.\ Section~\ref{Sec:strong washout}. Similar parameter ranges have been found in a purely numerical study in Ref.~\cite{Canetti:2012kh} and we find excellent agreement as far as the parameter space considered overlaps.

In the right panel of Fig.~\ref{fig:aftYB}, we show the mixing between the active and the sterile sector, parametrised by the mixing matrix element $|U_{\mu 4}|$, as a function  of $M_1$, the mass of the lighter of the two heavy states. Similar results hold, of course, for all other $U_{\alpha I}$, and in fact the total active-sterile mixing $\sum_{\alpha I} |U_{\alpha I}|^2$ is bounded from below by the seesaw relation (see e.g.~\cite{Shaposhnikov:2006nn,Asaka:2011pb,Canetti:2012kh}).
The active-sterile mixing $U_{\alpha I}$ is a particularly interesting quantity, since it is in principle experimentally accessible through experiments such as SHiP~\cite{Anelli:2015pba,Alekhin:2015byh}, FCC-ee~\cite{Blondel:2014bra} and LBNF/DUNE~\cite{Adams:2013qkq}. Unfortunately, the viable parameter points for solutions in the weak washout regime are found to be below the expected sensitivity of these experiments, with the exception of a very small region of particularly light sterile states, $M_1 \lesssim 500$~MeV, which can be reached by LBNF/DUNE. We remark however that our study has been limited, up to now, to a subset of the parameter space, due to the limitation of the analytical expression of Eq.~(\ref{eq:baryo_analytical}). In the next section we will extend (at least partially) our analysis to regions characterised by higher values of $Y^\text{eff}$ and, consequently, higher values of the mixing between the heavy and the active neutrinos, which can be possibly probed in future facilities. We anticipate in Fig.~\ref{fig:aftYB} two solutions associated to  a viable leptogenesis in the strong washout scenario, whose active-sterile mixing is represented by the two asterisks. It is evident that these model realisations can be probed by both SHiP and LBNF/DUNE.

%% file: strongwashout_0720.tex
\section{Solutions in the strong washout regime}
\label{Sec:strong washout}

\begin{figure}[htb]
\begin{center}
\subfloat{\includegraphics[width=6.2 cm]{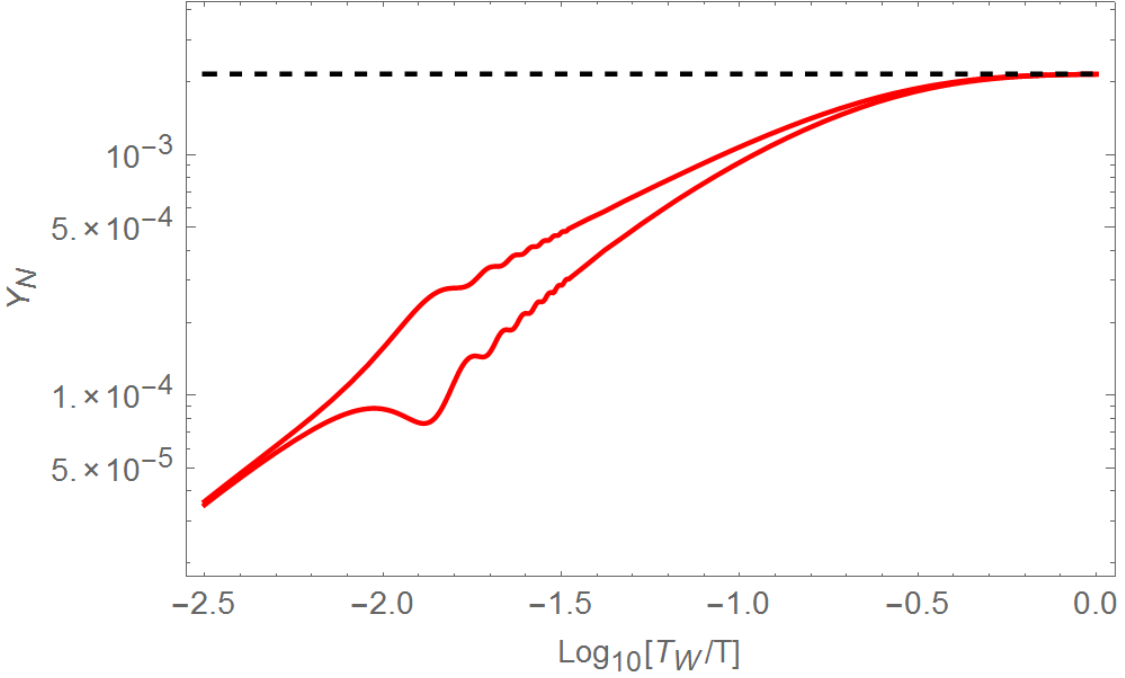}} \hfil
\subfloat{\includegraphics[width=6.0 cm]{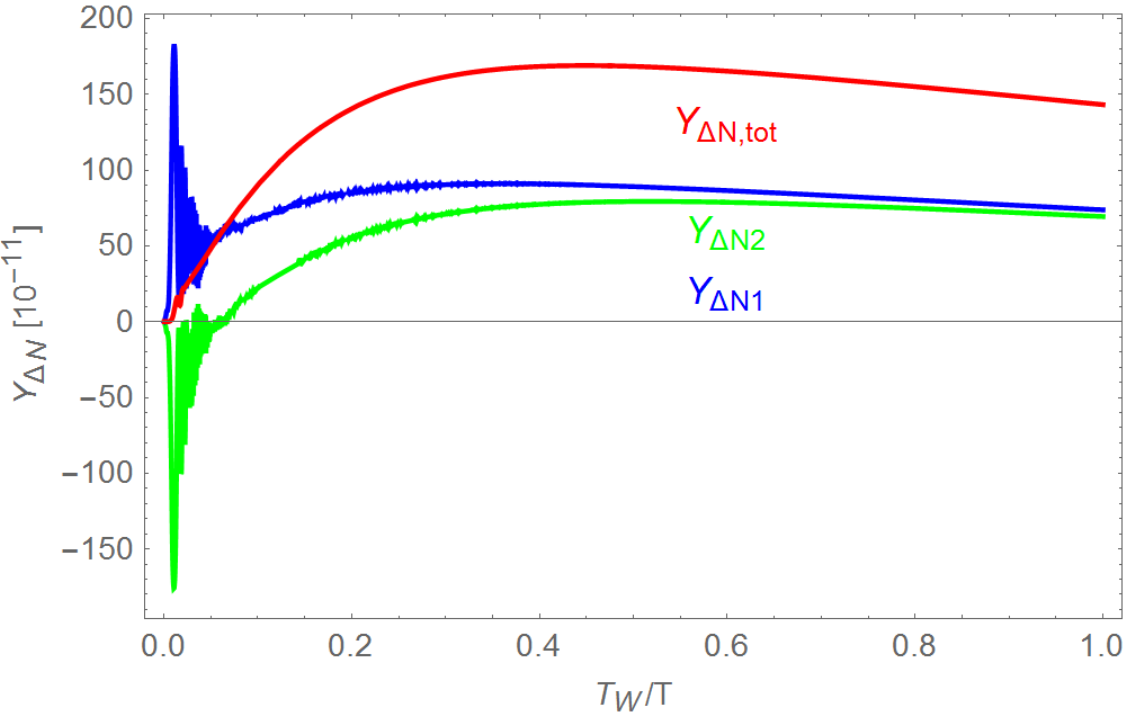}}\\
\subfloat{\includegraphics[width=6.5 cm]{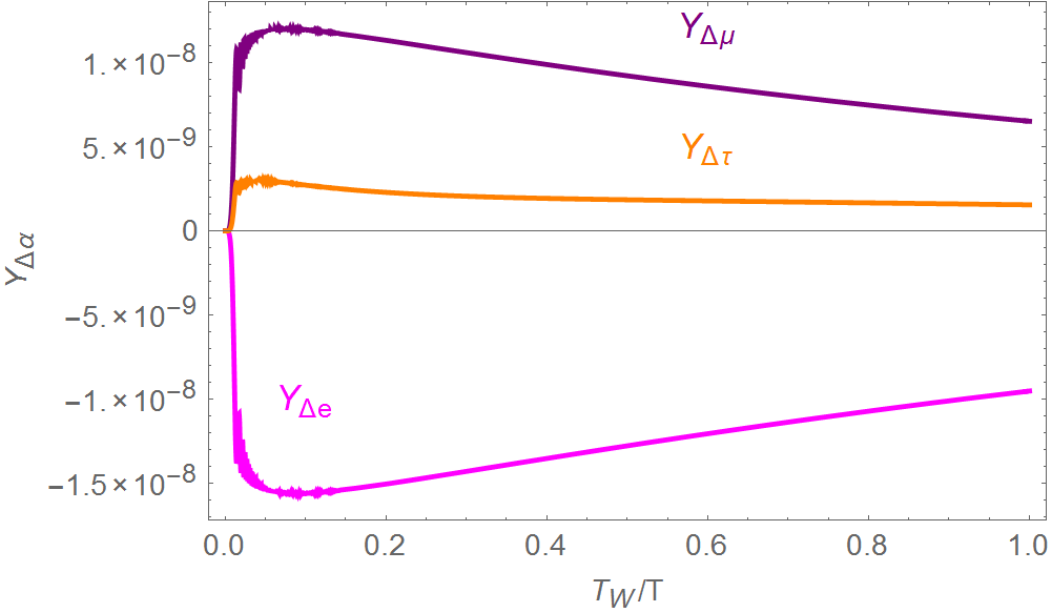}} \hfil
\subfloat{\includegraphics[width=6.5 cm]{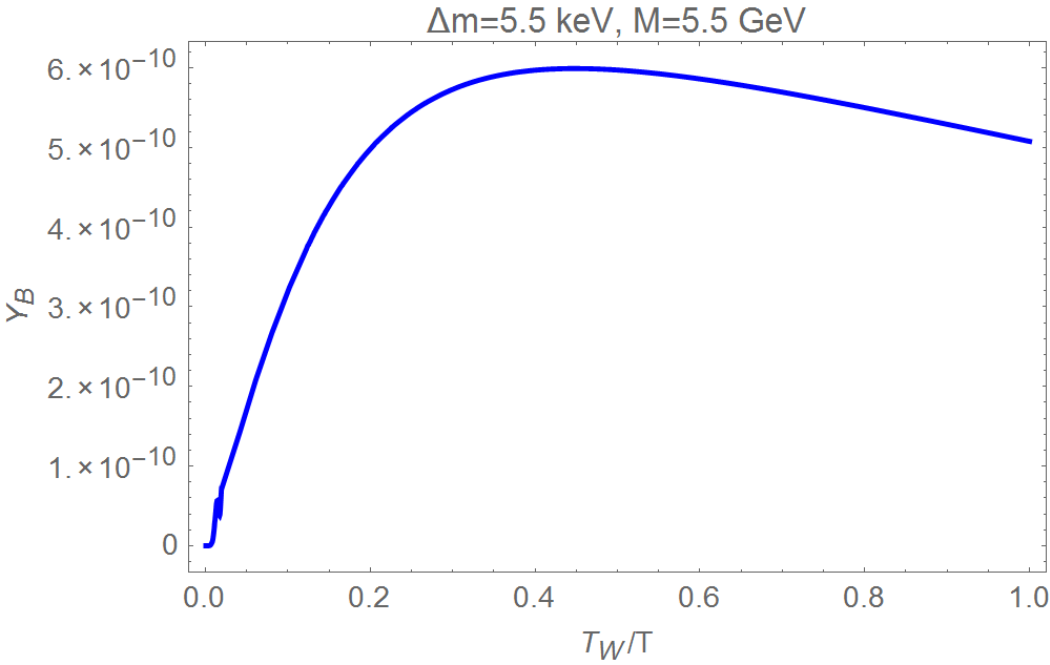}}
\caption{\footnotesize{Evolution of the baryon asymmetry (bottom right panel) as well as the individual lepton asymmetries in the active (bottom left panel) and sterile (top right panel) sectors. Top left panel: evolution of the abundances of the heavy neutrinos. For the parameter point chosen the entries of the matrix $Y^\text{eff}$ exceed the equilibration value by ${\cal{O}}(1)$ amounts. The asymmetries are depleted at late times but a sizeable residual baryon asymmetry of the order of the observed value is nonetheless present. }}
\label{fig:benchhy1}
\end{center}
\end{figure}

In this section we will investigate the possibility of achieving a successful leptogenesis in the case where the matrix $Y^\text{eff}$ has entries above the equilibration value $\sqrt{2} \times 10^{-7}$. In this situation the analytical solution of Eq.~(\ref{eq:baryo_analytical}) is not valid since we can no longer neglect the depletion of the baryon abundance when the heavy sterile neutrinos are in thermal equilibrium. At the same time higher entries of the $Y^\text{eff}$ correspond to a more efficient production of the sterile neutrinos which translates into an accordingly more efficient generation of a lepton asymmetry. The correct amount of the baryon abundance  might hence be in principle obtained even in a strong washout regime, provided that a sufficiently high initial lepton asymmetry is created. 

\begin{figure}[htb]
\begin{center}
\subfloat{\includegraphics[width=6.2 cm]{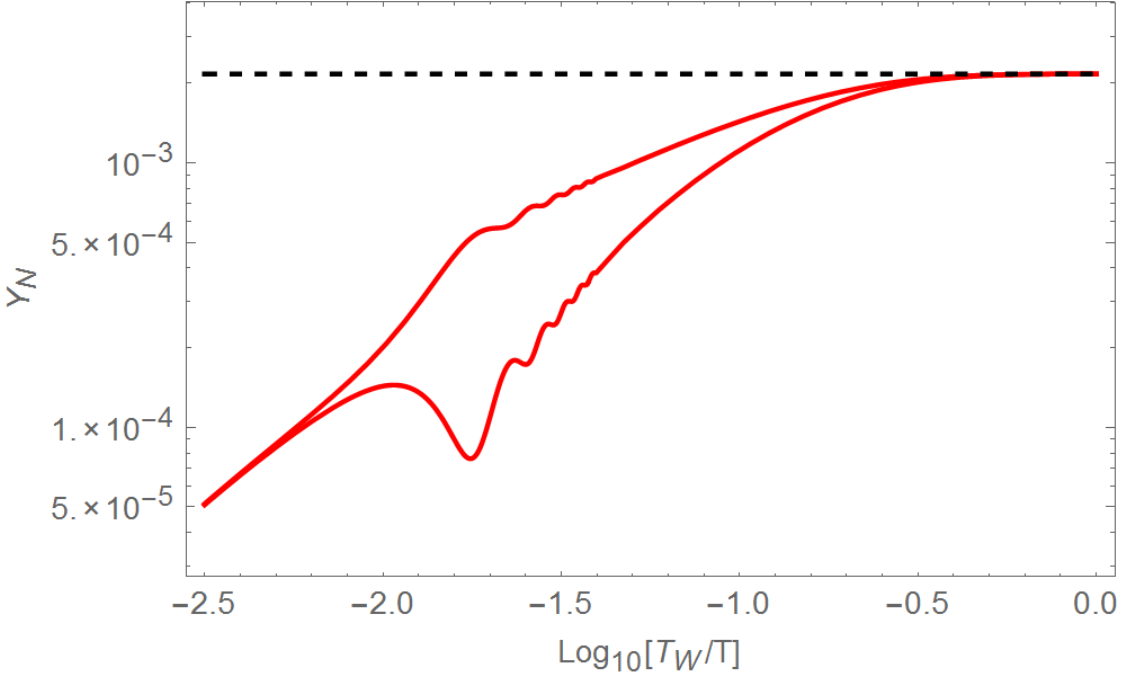}} \hfil
\subfloat{\includegraphics[width=6.0 cm]{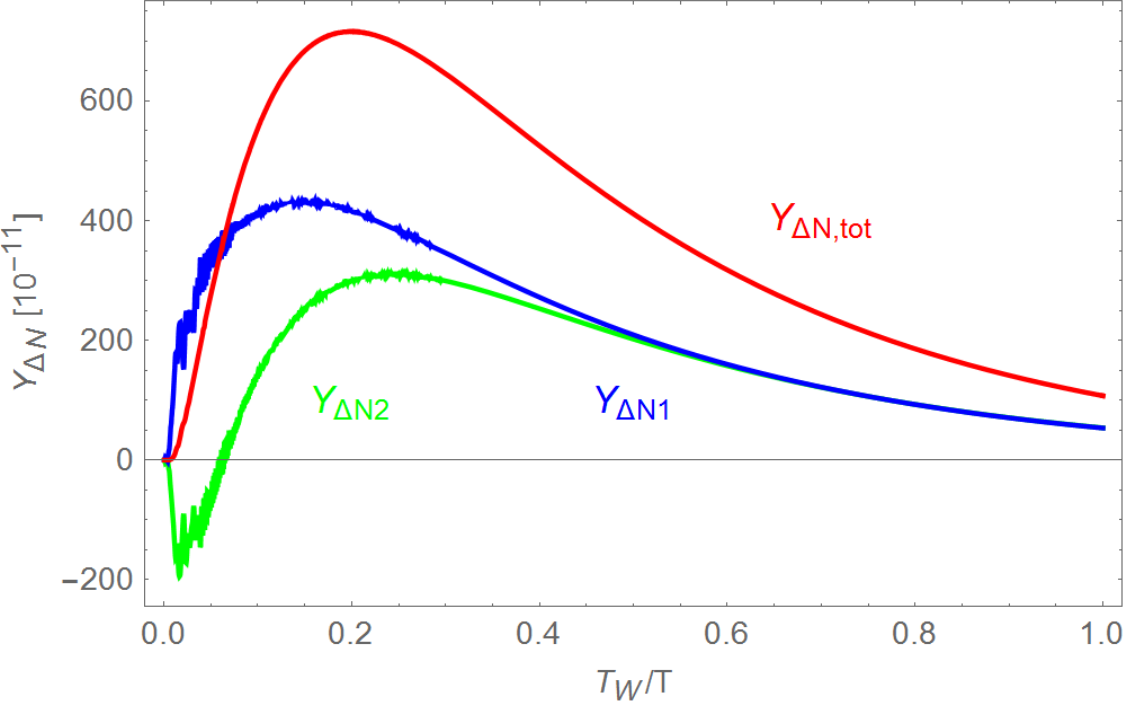}}\\
\subfloat{\includegraphics[width=6.5 cm]{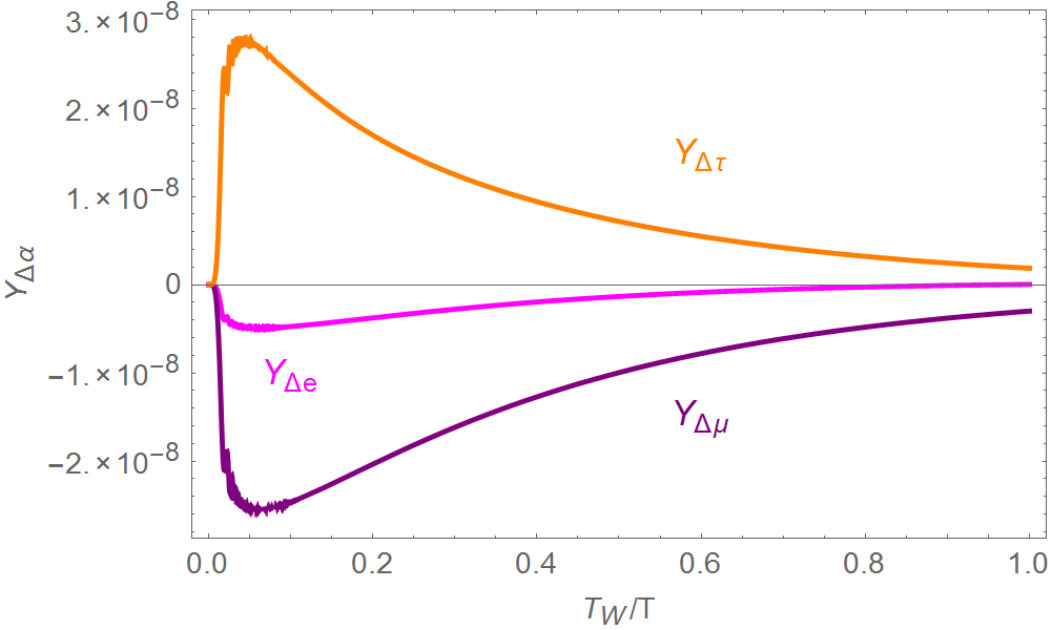}} \hfil
\subfloat{\includegraphics[width=6.5 cm]{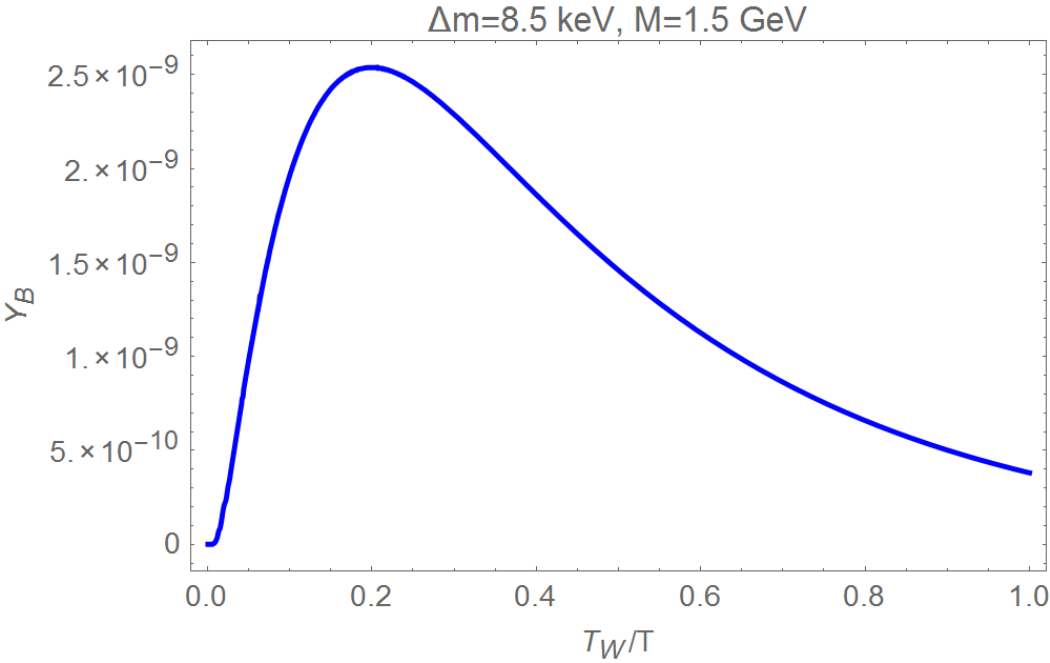}}
\caption{\footnotesize{As in Fig.~\ref{fig:benchhy1} but for a model with higher values of the entries of $Y_{\rm eff}$, but still below $10^{-6}$ (see Appendix~\ref{app_benchmarks} for details).}}
\label{fig:benchhy2}
\end{center}
\end{figure}

\begin{figure}[htb]
\begin{center}
\subfloat{\includegraphics[width=6.0 cm]{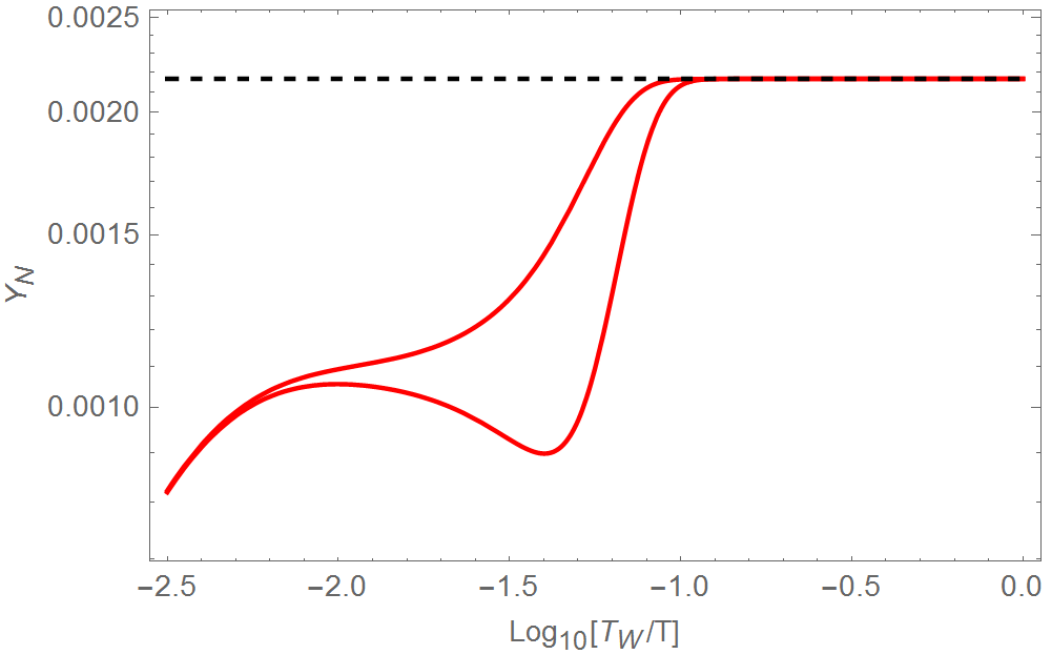}} \hfil
\subfloat{\includegraphics[width=6.0 cm]{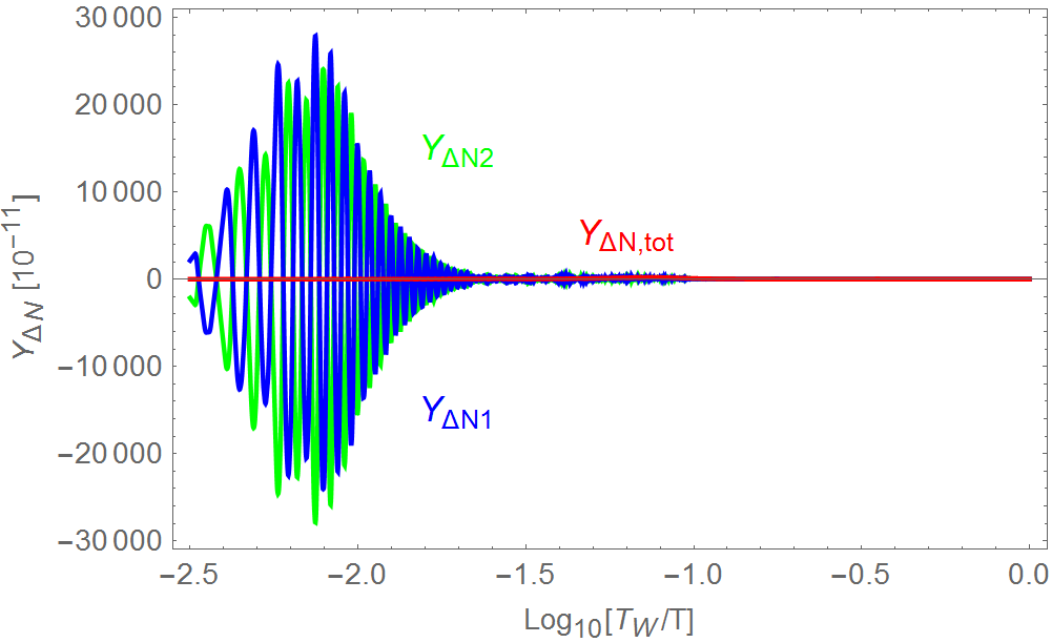}}\\
\subfloat{\includegraphics[width=6.5 cm]{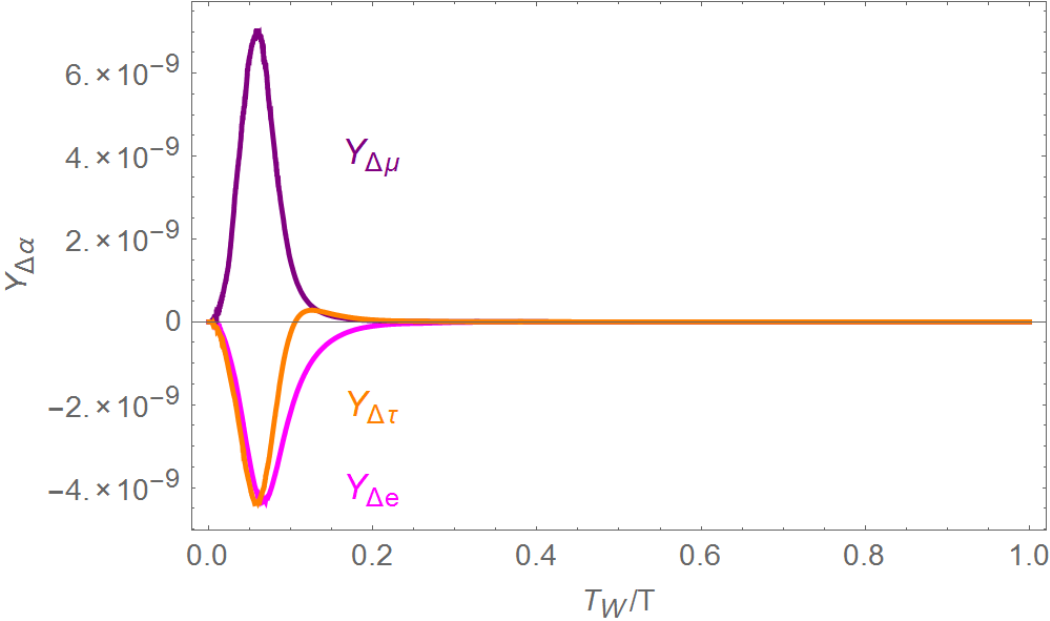}} \hfil
\subfloat{\includegraphics[width=6.5 cm]{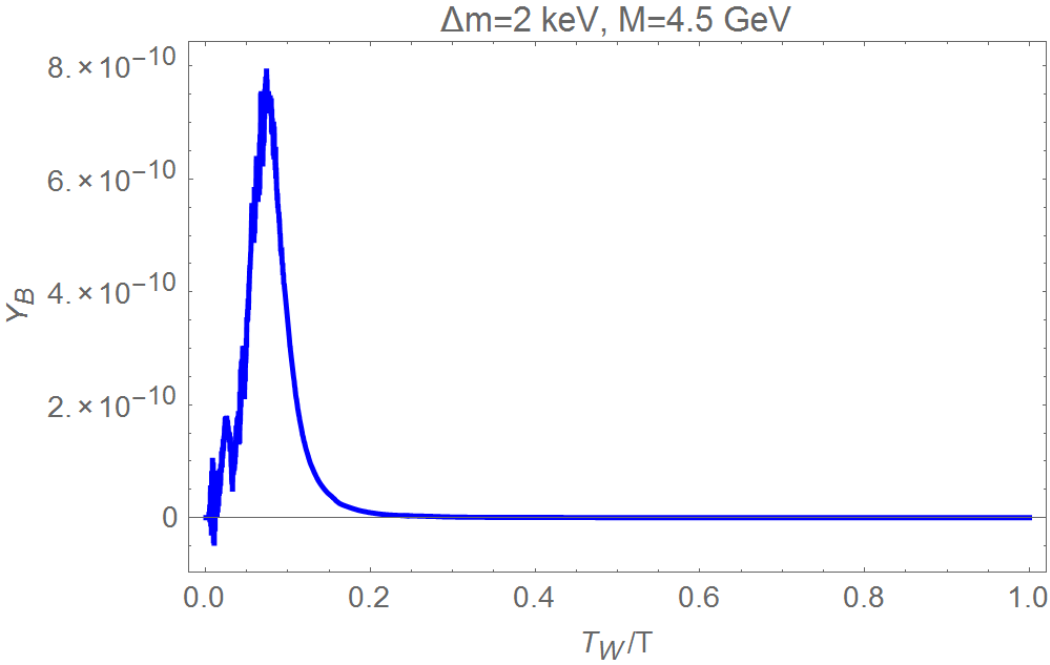}}
\caption{\footnotesize{As in Fig.~\ref{fig:benchhy1} but with entries of $Y^\text{eff}$ exceeding the equilibration value by more than one order of magnitude. In this case initially created asymmetries are completely depleted at later times and the final baryon abundance is negligible.}}
\label{fig:benchhy3}
\end{center}
\end{figure}

A full numerical exploration of the parameter space is computationally very demanding. We will thus limit our analysis to some relevant benchmark points which will be used to infer the main trends of the numerical solutions of the Boltzmann equation. We have, to this purpose, traced the evolution of the baryon abundance for three benchmarks characterised by increasing values of the entries of $Y^{\rm eff}$, ranging from $|Y^{\rm eff}_{\alpha i}| \sim \sqrt{2} \times 10^{-7}$ to $|Y^{\rm eff}_{\alpha i }| \approx 3 \times 10^{-6}$ (cf.\ Appendix~\ref{app_benchmarks}). Our results are reported in Figs.~\ref{fig:benchhy1}-\ref{fig:benchhy3}. Each figure reports the  same relevant quantities  as those chosen  in Section~\ref{sec:numerical_ly}. Before discussing the individual benchmarks we notice that all the plots indicate a very strong overlap between the neutrino states, except in very pronounced resonance regions. As can be inferred by Eq.~(\ref{eq:mass_eigen}), higher Yukawa couplings correspond to lower values of $\epsilon$ (higher values in the flipped regime $\epsilon > 1$). In the strong washout regime we thus expect the heavy neutrinos to typically form pseudo-Dirac pairs.

The first benchmark point, reported in Fig.~\ref{fig:benchhy1}, has values of $|Y^\text{eff}_{\alpha i}|$ slightly above the equilibration condition. In this case the sterile neutrinos reach thermal equilibrium only at a rather late time. The depletion of the lepton asymmetry is limited and a value of $Y_B$ above the observed value is obtained, demonstrating the feasibility of leptogenesis in this regime.

The second benchmark point, cf.\ Fig.~\ref{fig:benchhy2}, features higher entries of the Yukawa matrices but still exceeding  the equilibrium conditions by less than one order of magnitude, i.e. $\sqrt{2} \times 10^{-7} < |Y_{\alpha i}^\text{eff}| < 10^{-6}$. As already anticipated, the expected stronger washout effects are compensated by the higher initially produced lepton asymmetry leading to $Y_B(T_{\rm W}) \sim 4 \times 10^{-10}$, again sizeably above the observed value.

Finally we consider a benchmark point with $|Y^\text{eff}_{\alpha i}| \gtrsim 10^{-6}$, cf.\ Fig.~\ref{fig:benchhy3}. In this last case depletion effects are largely dominant and the final value of the baryon abundance is several orders of magnitude below the correct one, indicating that here the washout is too strong to allow for successful leptogenesis. { Notice that an upper bound on the Yukawa coupling in the strong washout regime  has been derived in Ref.~\cite{Shaposhnikov:2006nn}, $|Y^\text{eff}| < 1.2 \times 10^{-6} $. }

\begin{figure}[htb]
\begin{center}
\subfloat{\includegraphics[width=6.0 cm]{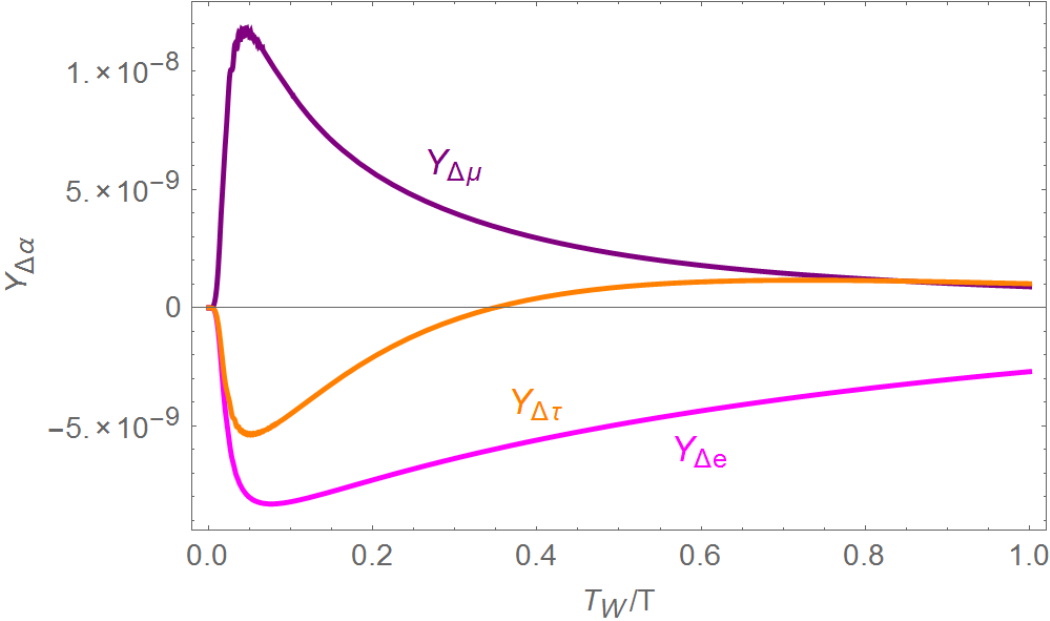}} \hfil
\subfloat{\includegraphics[width=6.0 cm]{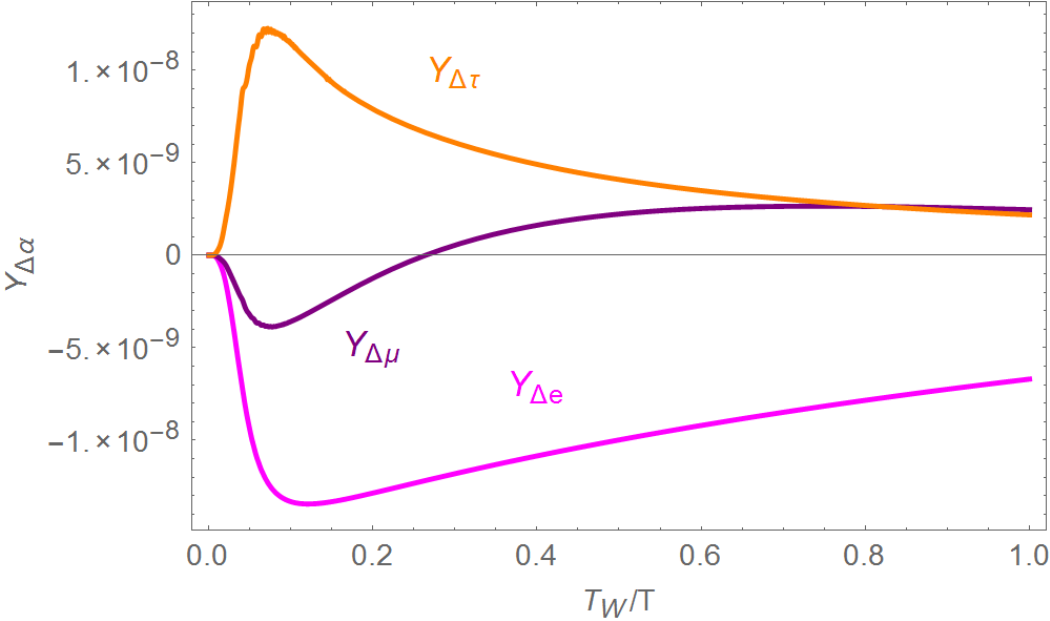}}\\
\subfloat{\includegraphics[width=6.5 cm]{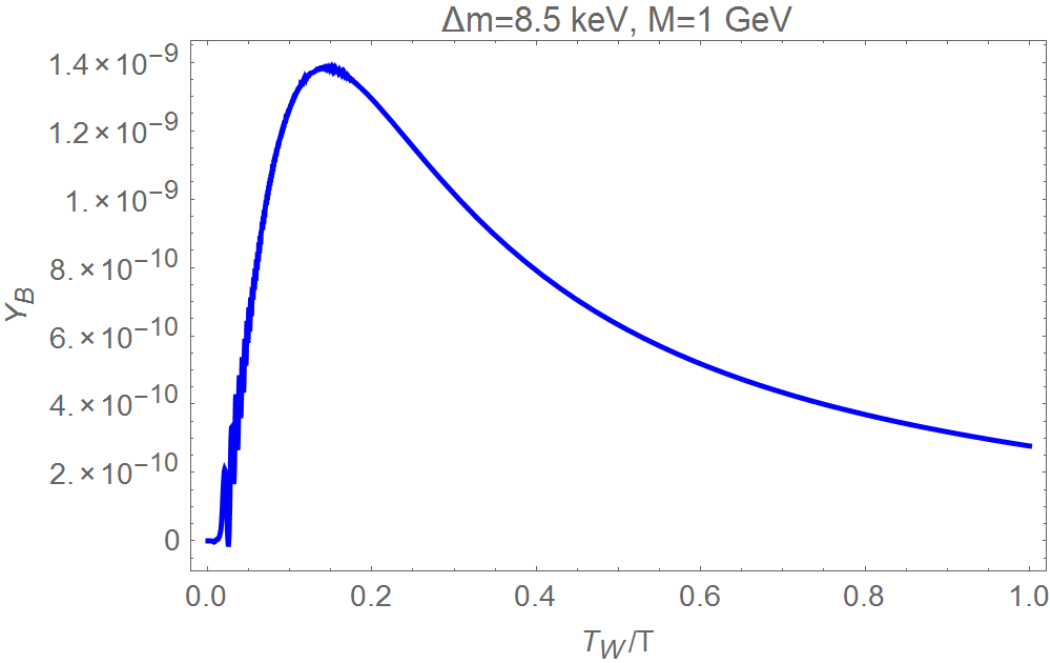}} \hfil
\subfloat{\includegraphics[width=6.5 cm]{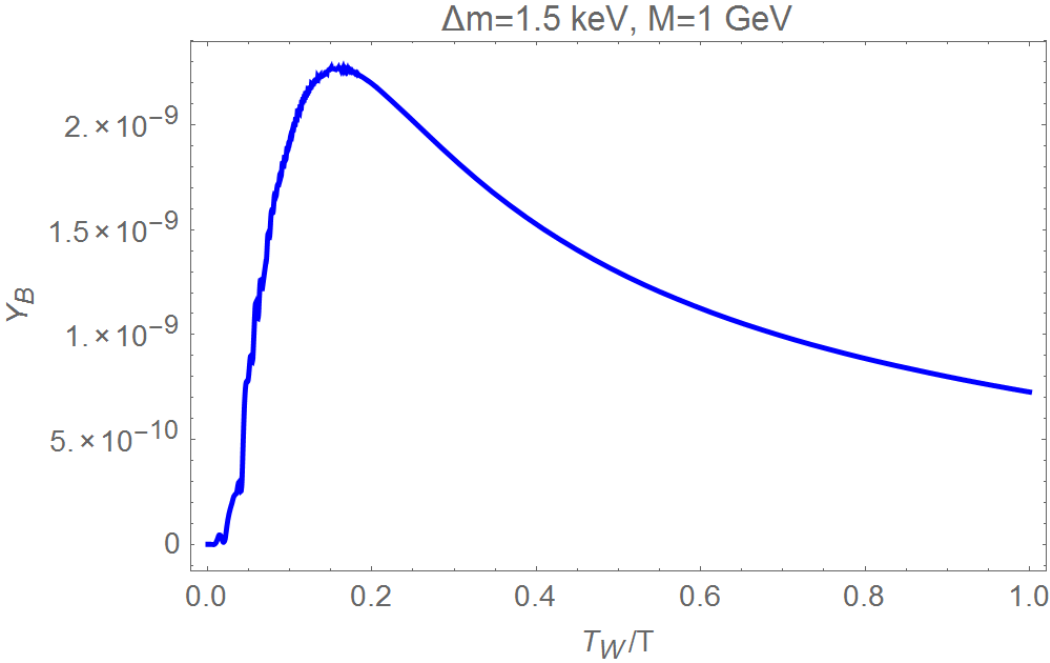}}
\caption{\footnotesize{Evolution of the asymmetries in the leptonic flavours (top panels) and of the total baryon abundance (lower panels) for two benchmarks featuring a hierarchical structure in the matrix $Y^\text{eff}$: The entries $Y^{\rm eff}_{ei}$ are  below the equilibration value whereas the other entries are significantly above.
The value of $Y_B$ for the two benchmarks is $2.5 \times 10^{-10}$ and $7.5 \times 10^{-10}$, respectively.}}
\label{fig:bench_flavored}
\end{center}
\end{figure}

The benchmark points presented so far were characterised by matrices $Y^\text{eff}$ with entries of similar size (cf.\ Appendix~\ref{app_benchmarks}). This implies that the lepton asymmetry is generated with similar efficiency for all the three neutrino flavours. On the other hand, a viable neutrino spectrum can be obtained, in our scenario, also in the case of ``hierarchical'' structure, i.e.\ when there is a sizeable separation, possibly greater than one order of magnitude, between the entries of $Y^\text{eff}$ corresponding to different active flavours. In this case, it is possible to have realisations with $Y^\text{eff}$  entries below and above the equilibration condition. 

Two relevant examples are shown in Fig.~\ref{fig:bench_flavored}. They show two benchmark scenarios with  $Y^\text{eff}_{ei} \leq \sqrt{2} \times 10^{-7}$ and $Y^\text{eff}_{\mu (\tau) i } \geq \sqrt{2} \times 10^{-7}$. As can be seen in the top panels of the figure, the asymmetry in the electron flavour features a much weaker depletion than the other two flavours and tends, at late times, to become the dominant contribution  for the total baryon asymmetry. As evident from the bottom panels of the figure, in both cases the final value of $Y_B$ exceeds the experimental value, demonstrating the possibility of having an efficient leptogenesis in this kind of setup.  Interestingly the two benchmark points have values of $|U_{\mu i}|^2$ of, respectively, $10^{-8}$ and $2 \times 10^{-9}$, lying within the expected sensitivity of SHiP (cf. Fig.~\ref{fig:aftYB}). The shown plots refer to normally ordered spectra of active light neutrinos. Our result partially resembles the scenario of flavoured leptogenesis discussed in~\cite{Drewes:2012ma}. However contrary to the case discussed in this reference (where three right-handed neutrinos are involved in the generation of the BAU), in presence of only two heavy neutrinos, complying with all low-energy neutrino data implies that the hierarchy between the entries of $Y^{\rm eff}$ can hardly exceed one order of magnitude and the flavour effects are less efficient, still requiring approximate degeneracy between the heavy neutrinos. Although models with hierarchical structure for the Yukawa matrix $Y^{\rm eff}$ are present both for normal and  inverted hierarchy for the spectrum of active (light) neutrinos, we find that this kind of setup favours the normal hierarchy spectrum.

As evident from the analysis presented in this section, the viable parameter space is enlarged with respect to the one shown in Fig.~\ref{fig:aftYB}, towards larger values of the Yukawa couplings, at least $O\left(10^{-6}\right)$.
A conclusive statement on the extension of this parameter space requires a (computationally very demanding) numerical analysis. We have nevertheless shown that  our scenario can provide successful leptogenesis in the strong washout-out regime, with  values of the mixing between light active and heavy neutrinos within the sensitivity region of future facilities like SHiP, as has been found also in the three neutrino extension of the SM, cf.\ \cite{Canetti:2014dka}. Promising parameter points lie both in the region of hierarchical and non-hierarchical Yukawa couplings, e.g.\ $|U_{\mu i}|^2 \sim 10^{-8} - 10^{-9}$ for the points reported in Figs.~\ref{fig:benchhy2} and \ref{fig:bench_flavored}.

%% file: iss0805.tex
\section{A special case: the inverse seesaw \label{ISS}}

A special case of the ansatz introduced in Section~\ref{sec_analytic} arises for $\epsilon \rightarrow 0$, referred to as the Inverse seesaw.
As discussed in Ref.~\cite{Abada:2014vea}, minimal realisations of this scenario in agreement with neutrino oscillation data, laboratory and unitarity constraints, as well as  constraints from lepton flavour violating observables, require four or five additional heavy states (referred to as ISS(2,2) and ISS(2,3), respectively).  
 With respect to Eq.~\eqref{eq_Mpertexp}, the fourth row/column is extended to contain two ``right-handed neutrino'' fields and the fifth row/column is extended to two or three ``sterile'' fields, respectively. Schematically, the mass matrix can be written as
 \begin{equation}
  M = \begin{pmatrix}
   0  & \frac{1}{\sqrt{2}} Y v & 0 \\
  \frac{1}{\sqrt{2}} Y^T v & 0 & Z \Lambda \\
   0 & Z^T \Lambda & \xi \Lambda
  \end{pmatrix}.
 \end{equation}
Here in the ISS$(I,J)$ setup $Y$ and $\xi$ are understood as $3\times I$ and $J\times J$ matrices. $Z$ is a $I\times J$ matrix with entries of order unity. 
The entries of the $Z$ and $Y$ matrices are taken complex and the matrix $\xi$ can be taken real and diagonal~\cite{Abada:2014vea}.
 
The new states form two heavy pseudo-Dirac pairs of mass ${\cal O}(\Lambda)$ with  squared mass-splittings of order $\mathcal{O}(\xi \Lambda^2)$,  the ISS(2,3) case features in addition a sterile state at the scale $\mathcal{O}(\xi \Lambda)$. In the mass ranges of eV or keV, the latter is an interesting candidate to address anomalies in the neutrino oscillation data or to explain dark matter~\cite{Abada:2014zra}, respectively. The two heavier pseudo-Dirac pairs are promising candidates for generating a baryon asymmetry, as discussed in Section~\ref{sec_toymodel}. Given that we are now dealing with a $7\times 7$ or $8\times 8$ mass matrix, there are clearly many possibilities for cancellations in the equations and we can no longer trust the simple estimates of Section~\ref{sec_toymodel}, which as we recall, lead us to disfavour the pure ISS due to a too large mass splitting of the heavy states. Indeed, a detailed scan of the ISS(2,3) parameter space performed in \cite{Abada:2014zra} found solutions for the light sterile state in the sub-eV to 100 keV range, pointing to mass-splittings of order $\Delta m^2 \sim (10~\text{keV})^2 - (10~\text{MeV})^2$. In addition, suitable Yukawa couplings
 below the critical value of $\sqrt{2} \times 10^{-7}$ are indeed achievable for light neutrino masses and mixings in agreement with current observations. This renders this scenario very promising for a minimal low-energy setup to simultaneously explain neutrino masses, dark matter and leptogenesis. In the following we revisit this setup, clarifying that, despite the large number of parameters, a successful leptogenesis in the weak washout regime cannot be achieved.
 
We have shown in Section~\ref{sec_toymodel} that for the ISS toy model the requirements $Y<\sqrt{2}\times 10^{-7}$, $m_\nu = 0.05$~eV and $\Lambda = 1$~GeV, imply $\Delta m^2 \gtrsim (0.4~\text{GeV})^2$, a value too large for leptogenesis.
Let us now generalise this result using the full matrix equations and considering first an ISS(3,3) setup, for which all the relevant submatrices are invertible $3\times 3$ matrices.
 It is possible to define the $3\times3$ PMNS mixing matrix $N$ as~\cite{Antusch:2009pm,Ibarra:2010xw}
\bee
N= (1+\eta)V,
\eee
where $V$ is a unitary matrix and $\eta$ parametrises the deviation from unitarity,
\bee
\eta\simeq -\frac{1}{4} \frac{v^2}{\Lambda^2} Y^* {Z^{-1}}^\dagger {Z^{-1}} Y^T,
\eee 
which is hermitian.
Retaining only the first order terms in the non-unitarity parameters, which are expected to be small, one has
\bee
NN^\dagger = (1+\eta)VV^\dagger (1+\eta)\simeq 1+2\eta.
\eee
In terms of the unitary $9\times 9$ leptonic mixing matrix $U$ we have (no sum over $\alpha$) 
\bee
{\sum_{i = 4}^9 \,} |U_{\alpha i}|^2= 1-\sum_{k=1}^3 |U_{\alpha k}|^2= 1- \left(NN^\dagger\right)_{\alpha \alpha}\simeq -2\eta_{\alpha \alpha},
\eee
implying for the active-sterile mixing
\bee
{\sum_{i=4}^9} \sum_\alpha |U_{\alpha i}|^2\simeq  \frac{v^2}{2 \Lambda^2} \Tr{Y^*\ {Z^\dagger}^{-1}\ {Z^{-1}}\ Y^T}=\frac{v^2}{2 \Lambda^2} \Tr{\left|Y\ {Z^T}^{-1}\right| \left| {Z^{-1}}\ Y^T\right|},
\label{eq:Ubound}
\eee
since $\Tr{A^\dagger A} = \Tr{|A^T| |A|}$.
The  effective Yukawas are related to the mixing matrix and to the mass eigenstates by (no sum over i)
\bee\label{eq:yuk_mass}
Y^\text{eff}_{\alpha i}= \sqrt{2}
U_{\alpha i}^* \frac{M_i}{v}\simeq \sqrt{2}
U_{\alpha i}^* \frac{\Lambda}{v}.
\eee
In the ISS scenario,  the neutrino mass matrix is given by
\bee\label{eq:mnuISS}
m_\nu \simeq - \frac{v^2}{2 \Lambda} Y\ {Z^{-1}}^T\ \xi \ {Z^{-1}}\ Y^T,
\eee
and in a basis in which $\xi$ is positive and diagonal
\bee\label{eq:xibound}
\max{\left[\xi_{jj}\right]} \frac{v^2}{2 \Lambda} \Tr{\left|Y\ {Z^{-1}}^T\right| \left|{Z^{-1}}\ Y^T\right|} &\geq&    \frac{v^2}{2 \Lambda}  \Tr{\left|Y\ {Z^{-1}}^T\ \xi\ {Z^{-1}}\ Y^T\right|}\non 
&=&\Tr{\left|m_\nu\right|}\ge 0.05 \text{ eV}\,.
\eee
From Eqs.~(\ref{eq:Ubound}, \ref{eq:yuk_mass}, \ref{eq:xibound}), the Yukawa couplings between the active flavours  and the heavy mass eigenstates are then bounded from below by
\begin{equation}
 {\sum_{i=4}^9}  \sum_{\alpha=1}^3 |Y^\text{eff}_{\alpha i}|^2 \geq 2 \,
 \frac{0.05~\text{eV}}{\text{max}|\xi_{jj}|} \frac{\Lambda}{v^2}\,.
 \label{eq_Ybound}
\end{equation}
Finally, imposing the lower bound in Eq.~(\ref{eq_Ybound}) to lie below the out-of equilibrium condition, $ \sum_{\alpha} |Y^\text{eff}_{\alpha i}|^2 < 2 \times 10^{-14}$ for all the heavy states $i$, implies $\max\left[ \xi_{jj}\right] \gtrsim 0.07 $ for $\Lambda = 1\text{ GeV}$, corresponding to a mass splitting $\mathcal{O}(\xi \Lambda^2) \gtrsim \left(0.25 \text{ GeV}\right)^2$, in good agreement with the estimation obtained in the toy model using Eq.~(\ref{eq:dmISS}). In conclusion, the ISS(3,3) model yields a mass splitting which is significantly too large for viable leptogenesis in the weak washout regime, which requires $\Delta m^2 \lesssim \text{MeV}^2$~\cite{Drewes:2012ma}. Moreover, the scale $\text{max}|\xi_{jj}| \Lambda$ which sets the scale of the  potential  DM candidate in the ISS(2,3) model  is found to be unpleasantly large: X-ray observations exclude sterile neutrinos heavier than about 100~keV contributing significantly to the DM abundance~\cite{Boyarsky:2005us}.

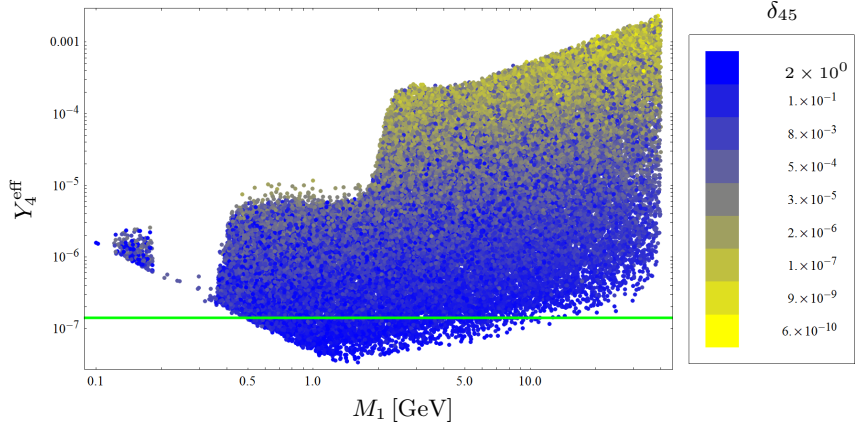
\begin{figure}[htb]
\centering
\begin{tikzpicture}
 \node at (0,0) { \includegraphics[width = 0.7 \textwidth]{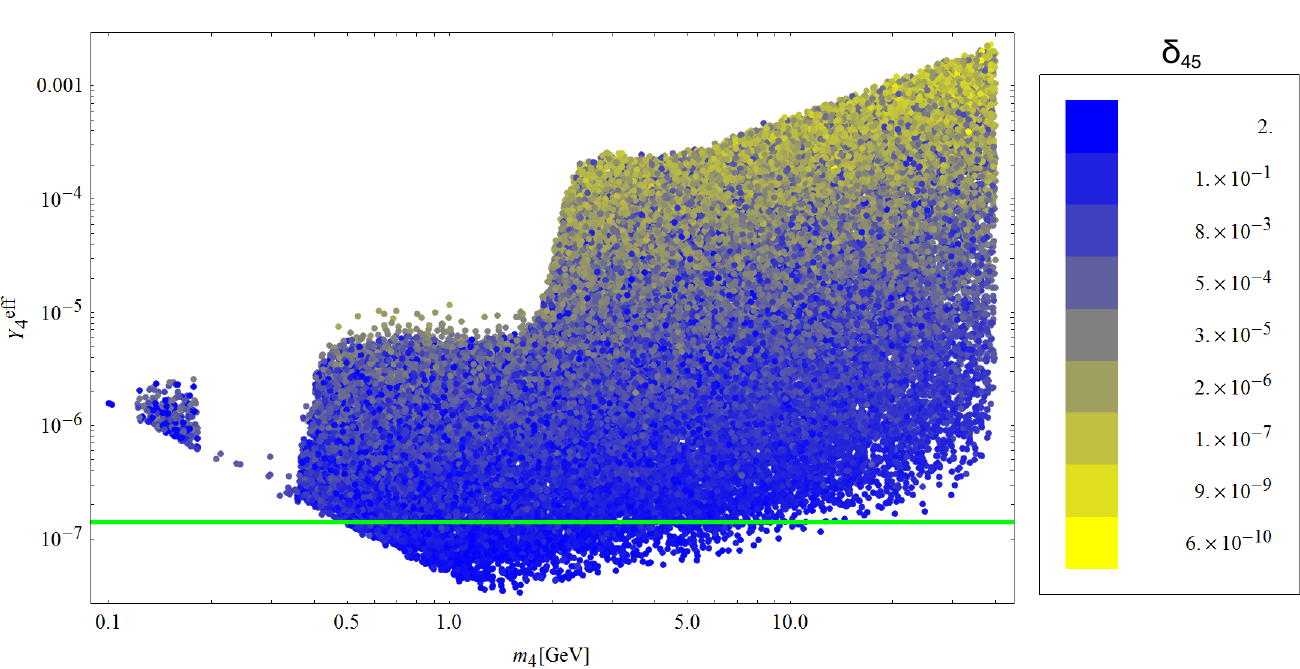}};
\node[fill = white] at (-0.5, -2.8) {\footnotesize $M_1  \, [\text{GeV}]$};
\node[fill = white, rotate = 90] at (-5.5, 0) {\footnotesize $Y_4^\text{eff}$};
\draw[fill = white, draw = white] (4.9, 1.5) rectangle (5.2, 2.1);
\node at (4.95,1.7) {\tiny $2 \times 10^0$};
\node[fill = white] at (4.5,2.5) {\footnotesize{$\delta_{45}$}};
\end{tikzpicture}
 \caption{Effective Yukawa coupling $Y^\text{eff}_4$ and mass  $M_1$ for the lightest sterile state in the ISS(2,2). The colour coding refers to the relative mass degeneracy $\delta_{45}$ with a high (low)  degeneracy marked in yellow (blue). }
 \label{Fig:ISS22}
\end{figure}

Since the lower bound in Eq.\ (\ref{eq_Ybound}) relies on the assumption $M_i \simeq \Lambda$ for every $i>3$, one may expect that the above conclusions are invalidated if a large mass difference among the different pseudo-Dirac pairs is present. In order to probe the feasibility of this configuration we performed a numerical scan of the the simpler phenomenologically viable realisation, the ISS(2,2). We generated the entries of the complex submatrices $Z\Lambda$ and $\xi \Lambda$ in the ranges $100 \text{ MeV} \le \left|Z_{ij}\right|\Lambda\le 40 \text{ GeV}$ and $1 \text{ eV} \le \left|\xi_{ij}\right|\Lambda\le 10 \text{ GeV}$, taking the different entries in each submatrix to be of the same order of magnitude.
The Dirac submatrix was generated using a modified version of the Casas-Ibarra parametrisation~\cite{Casas:2001sr} adapted for the ISS(2,2)
\bee
\frac{Y v}{\sqrt{2}} = N_\text{PMNS}^*\ \sqrt{\hat{m}_{\nu}}\ R\ \sqrt{\xi^{-1}}\ Z^T\Lambda\ ,
\eee
where the ``orthogonal'' matrix $R$ is defined as
\bee
R\left(\theta\right) = \left( \begin{array}{cc} 1 & 0 \\ \cos \theta & \sin \theta \\ -\sin \theta & \cos \theta \end{array} \right),
\eee
and the complex angle $\theta$ is randomly varied in the range $0\le \left| \theta\right|\le 2\pi$. Each solution is required to accommodate neutrino oscillation data, laboratory bounds on direct searches of sterile fermions and BBN bounds.  The values of the effective Yukawas for the lightest sterile state, $Y_{4}^\text{eff}=\sum_{\alpha}\left|Y_{\alpha 4}^\text{eff}\right|$ as a function of the lightest sterile mass $M_1$ are reported in Fig.~\ref{Fig:ISS22}. The horizontal green line represents the out of equilibrium value $Y_{4}^\text{eff} = \sqrt{2}\times 10^{-7}$ while the colour code is related to the mass degeneracy in the lighter pseudo-Dirac pair
\bee
\delta_{45} \equiv 2\frac{M_2-M_1}{M_1+M_2} \,.
\label{eq:delta45}
\eee

\begin{figure}[htb]
\centering
 \includegraphics[width = 0.55\textwidth]{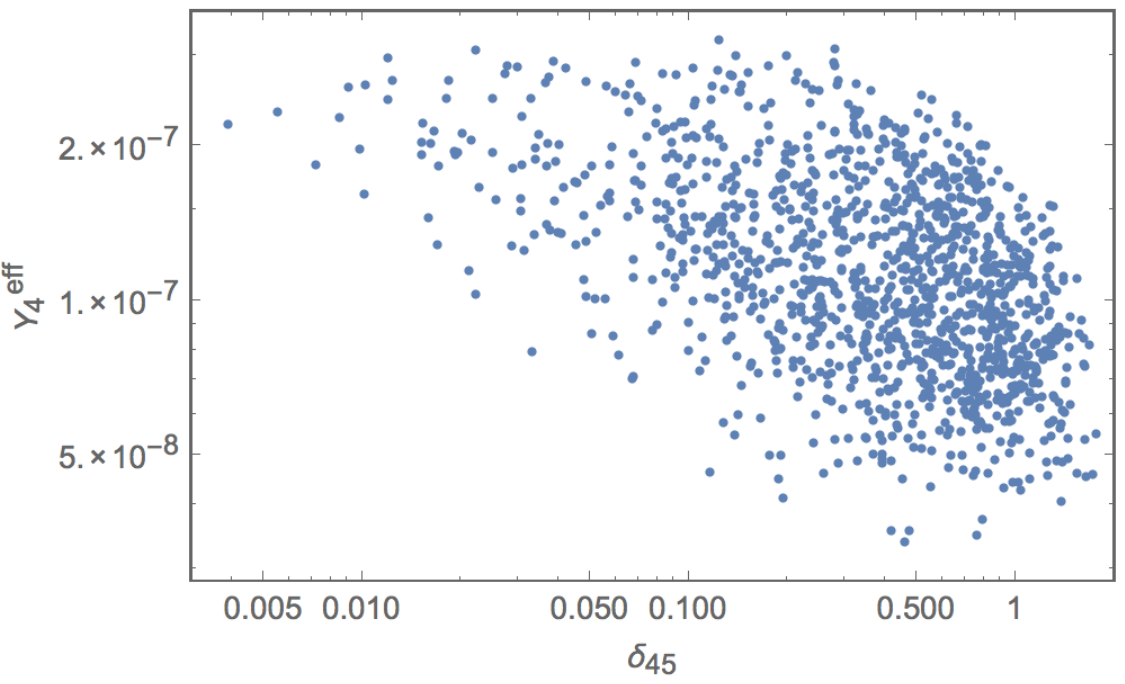}
 \caption{Effective Yukawa coupling $Y^\text{eff}_4$ and  relative mass degeneracy $\delta_{45}$ for the points in Fig.~\ref{Fig:ISS22} which lie below the equilibration value, $\left| Y^\text{eff}_{\alpha i}\right| < \sqrt{2}\times 10^{-7}$. In the weak washout regime a higher mass degeneracy is hard to obtain.}
 \label{Fig:degISS22}
\end{figure}

As is evident from this figure, smaller values of the Yukawa couplings are related to larger mass splittings in the pseudo-Dirac pair. Moreover the lower values for the Yukawa couplings are strongly limited by the BBN constraints in the region $M_1 \lesssim 1$ GeV, and by the seesaw relation Eq.~(\ref{eq:mnuISS}) in the region $M_1 \gtrsim 1$ GeV, leaving only a small out-of-equilibrium region in the mass range $500 \text{ MeV} \lesssim M_1 \lesssim 10 \text{ GeV}$. Focusing now on the weak washout regime of Fig.~\ref{Fig:ISS22}, i.e.\  requiring $\left| Y^\text{eff}_{\alpha i}\right| < \sqrt{2}\times 10^{-7}$ for all $\alpha=e,\mu,\tau$, $i=4,\dots,7$, we depict in  Fig.~\ref{Fig:degISS22} the quantities governing the efficiency of leptogenesis:\footnote{In the ISS(2,2) scenario, generally both heavy pseudo-Dirac pairs contribute to leptogenesis. We report here only the relevant parameters for the lighter pair, similar conclusions hold analogously for the heavier pair.} the mass degeneracy $\delta_{45}$ and  effective Yukawa coupling $Y^\text{eff}_4$.
As it is evident the condition $\delta_{45}<10^{-3}$ is not reached, and the parameter space of the model prefers the region $0.1 \lesssim \delta_{45} \lesssim 1$.
 Besides of being ineffective for leptogenesis, these realisations are clearly outside the ``natural'' region of the inverse seesaw, in which $\xi \ll 1$. {We conclude that in the weak washout regime, and in particular in the regime of approximate lepton number conservation, $\xi \ll 1$, solutions which are able to accommodate both neutrino oscillation data and a viable leptogenesis are hard, if not impossible, to find.}

{Although we have conducted a detailed numerical study only for the ISS(2,2) scenario, we expect similar results to hold for the ISS(2,3) setup. The additional state in the ISS(2,3) with a mass of ${\cal O}(\xi \Lambda)$ comes with correspondingly suppressed Yukawa couplings, implying a negligible effect on leptogenesis, which relies only on the heavier states. Moreover, since the weak washout regime prefers large values of $\xi$, cf.\ Fig.~\ref{Fig:ISS22}, the particularly attractive parameter range of the ISS(2,3), which can simultaneously account for DM, is disfavoured since the potential DM candidate would be too heavy to comply with the aforementioned X-ray constraints.}

We remark that this however does not exclude the ISS as a viable setup for a low-scale leptogenesis, since solutions in the strong washout regime may be allowed. This region of the parameter space is however outside the range of validity of our analytical solution Eq.~(\ref{eq:baryo_analytical}). We leave the numerical exploration of the relevant parameter space of the model for a future study.

%% file: conclusion0720.tex
\section{Conclusion}

In this work we have proposed a minimal extension of the Standard Model by adding two sterile fermions with opposite lepton number, forming a setup with an approximate lepton number conservation. The new fields form a pseudo-Dirac pair and are coupled to the active leptons via mixing terms.  The small mass splitting within this pair, as well as the smallness of the active neutrino masses, are due to two sources of lepton number violation with $\Delta L=2$,  corresponding to an Inverse Seesaw framework extended by a Linear Seesaw mass term. The main goal was to  study the feasibility of simultaneously having a very low scale seesaw mechanism - typically at $1 - 10$ GeV - at work for generating neutrino masses and mixings as well as an efficient leptogenesis through oscillations at the electroweak scale within this ``natural'' and minimal framework. Here the naturally arising pseudo-Dirac state ensures a highly efficient leptogenesis due  to its small mass splitting. We have also considered the (pure) Inverse Seesaw mechanism in which several pseudo-Dirac states arise naturally.

We have conducted a comprehensive analytical and numerical analysis investigating both neutrino mass hierarchies, normal (NH) and inverted (IH), for the neutrino mass spectrum and exploring the different washout regimes for the baryon asymmetry of the Universe.
To this end we have implemented and solved a system of Boltzmann equations 
 and have additionally derived an analytical expression for the baryon asymmetry,  providing a better understanding of the behaviour of the solutions.

Our studies reveal that our scenario (SM extended by two right-handed neutrinos
with two sources of lepton number violation 
by 2 units) 
is efficient to generate a successful leptogenesis through oscillations between the two mostly sterile states while complying with all available data. Our analytical expression is valid in the weak washout regime and agrees with the results obtained by numerically solving the system of Boltzmann equations. In the regime of strong washout, which is numerically very demanding, we have nevertheless proven that  our scenario can provide successful leptogenesis, with values of the active-sterile mixing that can be probed by future facilities such as SHiP.

We have conducted the same study for the pure Inverse Seesaw setup, in which case 
we find that the mass splitting between the states in the pseudo-Dirac pairs is too large to achieve a successful leptogenesis in the weak washout regime while accommodating the neutrino data. 
This analysis is however not conclusive to discard the ISS scenario since it relies on the (severe) condition that all the Yukawa couplings are below the equilibration value.
A complete analysis of the whole parameter space in this case is numerically very demanding, and will be the purpose of a future study.

%% file: app_leptogenesis_0720_v2.tex
\section{Analytical determination of the baryon asymmetry \label{app_leptogenesis}}

In this appendix we review the main details regarding the analytical and numerical determinations of the baryon asymmetry. The starting point is a system of coupled Boltzmann equations for the density matrices $\rho_{AB}$ with $A,B = \{N, \bar N, L , \bar L\}$ associated, respectively, to sterile neutrinos, active leptons and their anti-particles. This kind of system has been originally introduced in~\cite{Asaka:2005pn}, and has been reduced to a  system of ordinary differential equations.
A more refined version of this system, also adopted in this work, retaining the full dependence on the momentum $k$ of the density matrices, has been successively proposed in~\cite{Asaka:2011wq}. In this case one has to solve a system of integro-differential equations (cf. Eqs.~(12) and (14) of~\cite{Asaka:2011wq}) of the form:
\begin{align}
 \frac{d\rho_N}{dt} = & -i\left[H_N(k_N),\rho_N\right]
 -\frac{3}{2}\gamma_N^d(k_N) \left\{ F^\dagger\,F,\rho_N\right\}-\frac{1}{2}\gamma_N^d \left\{F^\dagger \left(A^{-1}-I\right)F,\rho_N\right\} \nonumber\\
& +3 \gamma_N^d(k_N) \rho_{\rm eq}(k_N) F^\dagger\,F+2 \gamma^d_N(k_N) \rho_{\rm eq}(k_N) \left[F^\dagger \left(A-I\right)F\right] \,,
\label{eq_N}
\end{align}
\begin{align}
 \frac{d\mu_{\alpha}}{dt}= & -\frac{3}{2}\gamma^d_\nu(T) FF^\dagger \tanh\mu_\alpha
 -\frac{\gamma_\nu^d (T)}{4}\left(1+\tanh\mu_\alpha\right) \int_0^\infty \frac{d k_N k_N}{T^2} F\left(\rho_{\bar N}^T -\rho_{\rm eq}\right)F^\dagger \nonumber\\
& +\frac{\gamma_\nu^d (T)}{4}\left(1-\tanh\mu_\alpha\right) \int_0^\infty \frac{d k_N k_N}{T^2} F^{*}\left(\rho_{N}^T -\rho_{\rm eq}\right)F^T \nonumber\\
& + \frac{\gamma_\nu^d (T)}{2 \cosh\mu_\alpha} \int_0^\infty \frac{dk_L}{T^2}\int_0^{k_L} dk_N \frac{\rho_{\rm eq}(k_L)}{\rho_{\rm eq}(k_N)}\left[F\rho_N F^\dagger-F^{*}\rho_{\bar N} F^T\right]\nonumber\\
&+\frac{\gamma_\nu^d (T)}{2 \cosh\mu_\alpha} \int_0^\infty \frac{dk_L}{T^2}\int_{k_L}^\infty dk_N \left[F\rho_N F^\dagger-F^{*}\rho_{\bar N} F^T\right]\nonumber\\
& -\frac{\gamma_\nu^d (T)}{2 \cosh\mu_\alpha} \int_0^\infty \frac{dk_L}{T^2}\rho_{\rm eq}(k_L)\int_0^\infty dk_N \left[F\rho_N F^\dagger-F^{*}\rho_{\bar N} F^T\right] \,.
\label{eq_mu}
\end{align}
Here we have taken the active leptons to be in thermal equilibrium, allowing us to trade their equations for  equations of the chemical potentials $\mu_e,\mu_\mu,\mu_\tau$.
\begin{equation}\label{Eqrho}
\rho_{L}=N_D \, \rho_{\rm eq}(k) \, A,\,\,\,\,\,\,\rho_{\bar L}=N_D \, \rho_{\rm eq}(k) \, A^{-1},\,\,\,\,\,\rho_{\rm eq}=e^{-\frac{k}{T}} \,,
\end{equation}
with $A=\text{diag}(e^{\mu_e},e^{\mu_\mu},e^{\mu_\tau})$ representing a matrix of chemical potentials, $\rho_\text{eq}$  the equilibrium abundance of the mode with wavenumber $k$, determined by the temperature of the thermal bath $T$, and $N_D = 2$.

The first term on the rhs of Eq.~\eqref{eq_N} describes the oscillations of the heavy neutrinos in the presence of the effective Hamiltonian $H_N$, containing the free propagation and the effective potential induced by the medium effects. The following two terms describe the decay of the sterile states and the final two terms account for their production.  Both of these processes contain diagrams sensitive to the asymmetry in the active sector, leading to the terms proportional to $(A^{\pm 1} - I)$.  
The corresponding equation for the anti-particles $\rho_{\bar N}$  is straightforwardly obtained from Eq.~(\ref{eq_N}) by replacing $N \leftrightarrow \bar N$, $F \leftrightarrow F^{*}$ and $A \leftrightarrow A^{-1}$. For simplicity we will only show the equations for $\rho_N$ in the following.
 Equation~\eqref{eq_mu} contains the decay and production of the active states, which in turn depend on the abundance and momentum of the sterile states. 
The functions $\gamma$ encoding the decay and production rates are defined as:
\begin{equation}
\gamma_N^d(k)=\frac{N_D N_C h_t^2}{64 \pi^3} \frac{T^2}{k},\,\,\,\,\,\,\gamma_\nu^d(k)=\frac{1}{N_D}\gamma_N^d(k) \,,
\end{equation}
with $N_C = 3$ and the top Yukawa coupling $h_t \simeq 1$. Here it is assumed that these rates are dominated by scattering processes involving the top quark. As was pointed out in Refs.~\cite{Anisimov:2010gy, Besak:2012qm}, scattering processes involving EW gauge bosons and multiple scatterings mediated by soft gauge bosons also contribute at the leading order, inducing an enhancement of the decay/production rates by about a factor 3 around the EW scale. Such processes can also contribute, at higher order, to the CP asymmetry~\cite{Anisimov:2010gy}, however, their corresponding contribution has, see e.g.~\cite{Frossard:2012pc}, not been calculated yet for the scenario under consideration. As already commented in e.g.~\cite{Canetti:2012kh} their impact on the scenario under investigation cannot be straightforwardly determined.
Hence for a consistent treatment, we restrict ourselves to the top quark scattering processes. Analogously to \cite{Canetti:2012kh} we have regarded the uncertainty introduced at this point to be of similar order as the other uncertainties in the treatment, and we do not expect it to significantly modify our final results. A calculation of the missing CP-violating rates is an important step towards a precise treatment of leptogenesis through neutrino oscillations, but is beyond the scope of the present work.

 The abundances of the various species are given by
\begin{equation}
\label{eq:YN0}
Y_{N,L}=\frac{1}{s}\int \frac{d^3 k}{{\left(2 \pi\right)}^3} \, \rho_{N,L}(k) \,,
\end{equation}
where $s=\frac{2 \pi^2 g_{s}}{45}T^3$ denotes the entropy density of the thermal bath.
The system of integro-differential equations~(\ref{eq_N})-(\ref{eq_mu}) can be solved by specifying the masses of the heavy neutrinos $M_{1,2}$ and their Yukawa couplings $F_{\alpha i}$. The abundance of the heavy neutrino species and the asymmetry in the neutrino spectrum are given by:
\begin{align}
Y_{N,i}=\frac{1}{s}\int \frac{d^3 k}{{\left(2 \pi\right)}^3} \, {\left[\rho_{N}(k)\right]}_{ii} \,,\,\,\,\,i=1,2\nonumber\\
Y_{\Delta N,i}=\frac{1}{s}\int \frac{d^3 k}{{\left(2 \pi\right)}^3} \, {\left[\Delta\rho_{N}(k)\right]}_{ii}\,,\,\,\,\,\Delta \rho=\rho_N-\rho_{\bar N}
\end{align}

while the asymmetry in the leptonic flavour can be determined from the chemical potentials as:
\begin{equation}
Y_{\Delta L_\alpha}=\frac{45 N_D}{\pi^4 g_{s}} \sinh \mu_\alpha\,,\,\,\,\,\alpha=e,\mu,\tau
\end{equation}
According the conservation of the total (active plus sterile) lepton number, the baryon abundance $Y_B$ is given by:
\begin{equation} 
Y_B=-\frac{28}{79}\sum_{\alpha}Y_{\Delta L_\alpha}=\frac{28}{79} \sum_i Y_{\Delta N_i} \,.
\end{equation}
The properties of the system~(\ref{eq_N})-(\ref{eq_mu}) and of its solutions have been extensively studied in~\cite{Asaka:2011wq}. A useful simplification is to assume that the momentum distribution of the heavy neutrinos is proportional to the equilibrium one (this is equivalent to state the the heavy neutrinos are in kinetic equilibrium), i.e.:
\begin{equation}
\rho_{N,\bar N}=R_{N,\bar N}(t)  \, \rho_{\rm eq}(k)\ .
\label{eq:RN}
\end{equation}
With this substitution we can trace the evolution of the abundances of the heavy neutrinos through the only time dependent functions $R_{N,\bar N}$. The system of Boltzmann equations is then casted as:
\begin{align}
\label{eq:full_system}
 \frac{dR_N}{dt}= & -i\left[H(k_N),R_N\right] 
- \frac{3}{2}\gamma^d_N(k_N) \left\{F^\dagger F,R_N-I\right\}+2 \gamma_N^d(k_N) \left(F^\dagger\left(A-I\right)F\right) \nonumber\\
& -\frac{1}{2}\gamma^d_N(k_N)\left \{\left( F^\dagger\left(A^{-1}-I\right)F\right),R_N\right \}\nonumber\\
 \frac{d \mu_\alpha}{dt}= & -\frac{3}{2} \gamma_\nu^d(T){\left(FF^\dagger\right)}_{\alpha \alpha} \tanh\mu_\alpha 
 -\frac{\gamma_\nu^d(T)}{4}\left(1+\tanh\mu_\alpha\right)\left(F\left(R_{\bar N}^T-I\right)F^\dagger\right)_{\alpha \alpha} \\ \nonumber
&  +\frac{\gamma_\nu^d(T)}{4}\left(1-\tanh\mu_\alpha\right)(F^{*}\left(R_{N}^T-I\right)F^T)_{\alpha \alpha} 
+\frac{\gamma_\nu^d(T)}{2 \cosh\mu_\alpha}\left[F R_N F^\dagger-F^{*}R_{\bar N}F^T\right]_{\alpha \alpha}
\end{align}
As discussed in~\cite{Asaka:2011wq}, in very good approximation the system can be solved by reducing it to a system of ordinary equations for a single mode $k_*$, equivalent to the one presented~\cite{Asaka:2005pn}, by a suitable replacement of the type $k_* \sim T$. 
Notice that the choice of $k_{*}$ must maintain the system self-consistent, i.e.\ it should preserve lepton number. This condition can be stated as:
\begin{equation}
\Tr{\frac{dR_N}{dt}|_{k=k_{*}}-\frac{dR_{\bar N}}{dt}|_{k=k_{*}}+N_D \frac{d A}{dt}-N_D\frac{d A^{-1}}{dt}}=0
\end{equation}
and can be satisfied only for $k_{*}=2 \, T$, rather than for $k_{*}=3 \, T$, corresponding to the conventional thermal average.\footnote{In the notation used in this appendix this implies $\gamma(t) = \gamma(T) = \gamma(k_*/2) = 2 \, \gamma(k_*) $. }
The system~(\ref{eq:full_system}), with the substitution $k \rightarrow k_{*}=2 T$ is the one used in our study. Notice that, although very similar, the system~(\ref{eq:full_system}) does not exactly coincide with the one presented in~\cite{Asaka:2005pn}. In particular the coefficient of the third term of the right-hand side of the equation for $R_N$ differs by a factor 2/3. This is an important point since this term represent the connection term between the active and sterile sector which is mostly responsible of the generation of the lepton asymmetry.

As stated in the main text, despite the simplification discussed, an extensive numerical analysis is still very difficult. For this reason we have limited the numerical study to some relevant benchmarks, as reported e.g.\ in Figures~\ref{fig:bench_natural}, \ref{fig:bench_numsm} and \ref{fig:bench_thermal}  and have adopted, for the study of the parameter space, an analytical solution which is valid in the so-called weak wash-out regime. This analytical solution is derived following the procedure proposed in~\cite{Asaka:2005pn,Asaka:2010kk}. The final expression differs, however, by a $O(1)$ factor with respect to these references due to the different starting system, as mentioned above.

\subsection{Analytical solution in the weak washout regime}

An analytical expression of $Y_B$ can be obtained by solving Eq.~(\ref{eq:full_system}) perturbatively for small values of $\mu_\alpha$ and $F$.
Let us first consider the leading order in $\mu_\alpha$, i.e.\ we set tanh$\mu_\alpha \rightarrow 0$, cosh$\mu_\alpha \rightarrow 1$, $A - I \rightarrow 0$ and $A^{-1} - I \rightarrow 0$. The initial conditions are $R_{N,\bar N}(0)=0$, $\mu_{\alpha}=0$. 
The first step is to solve the equations for $R_{N,\bar N}$. First of all, one can perform the following transformation~\cite{Shaposhnikov:2008pf}:
\begin{equation}
\label{eq:transformation}
R_N=E(t) \tilde{R}_N E^{\dagger}(t) \,,
\end{equation}
with
\begin{equation}
E(t)=\exp\left[-i \int_0^t dt^{'} \Delta E \right],\quad \Delta E=\text{diag}(E_1,E_2)\ ,
\end{equation}
where $E_i$ denotes the energies of the two heavy neutrinos. This transformation encodes the oscillations processes in the sterile neutrino production term.  In this way we obtain
\begin{align}
\label{eq:diff2}
& \frac{d\tilde{R}_N}{dt}=-i\left[\tilde{H},\tilde{R}_N\right]-\frac{3}{2}\left\{\tilde{\Gamma}^d_N,\tilde{R}_N-I\right\} \,,
\end{align}
where we have dropped the terms proportional to $(A - I)$ and $(A^{-1} - I)$ and have defined
\begin{equation}
\tilde{\Gamma}{_N^d}(t)=E^\dagger(t)  \Gamma_N^d (t)  E(t)\,, \quad \Gamma_N^d=\frac{1}{2}\gamma_N^d(T) F^\dagger F \,.
\end{equation}
Physically, this corresponds to ignoring the back reaction of the asymmetry in the active sector on the production of the sterile neutrinos. For the size of asymmetries in the active sector which we are phenomenologically interested in, this is a very good approximation. The asymmetry in the active sector will however become important in the next order of our perturbative expansion, which we will need when determining the asymmetry in the sterile sector, as we will see below.

In the weak washout regime characterised by $R_N \ll 1$, we can solve Eq.~\eqref{eq:diff2} by dropping all the terms proportional to $R_N$,
\begin{equation}
\label{eq:first_step}
{\tilde{\textcolor{black}{R}}}_N =3\int_0^t dt_1 E^\dagger(t_1)  \Gamma_N^d (t_1)  E(t_1)\ .
\end{equation}
Let us now move to the equation for the chemical potential. At leading order in $\mu_\alpha$ and after inserting Eq.~\eqref{eq:first_step}, we find
\begin{align}
& \mu_\alpha= \frac{3}{4}\int_0^t dt_1 \gamma_\nu^d (t_1)  \int_0^{t_1} dt_2 \gamma_N^d(t_2) \left(\left[F E(t_1) E(t_2)^\dagger F^\dagger F  E(t_2)E(t_1)^\dagger  F^\dagger\right] \right.\nonumber\\
& \left. -  \left[F^{*}  E(t_1) E(t_2)^\dagger  F^T F^{*} E(t_2)E(t_1)^\dagger  F^T\right] \right)_{\alpha \alpha}\nonumber\\
& - \frac{3}{8}\int_0^t dt_1 \gamma_\nu^d (t_1)  \int_0^{t_1} dt_2 \gamma_N^d(t_2) \left(\left[F {\left[E(t_1) E(t_2)^\dagger  F^T F^{*} E(t_2)E(t_1)^\dagger\right]}^T  F^\dagger\right] \right.\nonumber\\
& \left. -  \left[F^{*}  {\left[E(t_1) E(t_2)^\dagger  F^\dagger F E(t_2)E(t_1)^\dagger\right]}^T  F^T\right] \right)_{\alpha \alpha} \, .
\end{align} 
After some manipulation, exploiting in particular
\begin{align}
&  E(t_1) E^\dagger (t_2)_{ij} = \text{diag}\left[\exp(i \int_{t_1}^{t_2} E_i)\right]  \,, \\
 & F_{\alpha i} (F^\dagger F)_{ij} (F^\dagger)_{j \alpha} - F^*_{\alpha i} (F^T F^*)_{ij} (F^T)_{j \alpha} = 2 \, \text{Im} ( F_{\alpha i} (F^\dagger F)_{ij} (F^\dagger)_{j \alpha})\,,
\end{align}
this expression can be simplified to
\begin{equation}
\label{eq:mualpha_asaka}
\mu_\alpha= {\frac{9}{2}} \, \delta_{\alpha}\int_0^t dt_1 \gamma_\nu^d (t_1) \int_0^{t_1} dt_2  \gamma_N^d (t_2) \sin\left(\int_{t_2}^{t_1} dt_3 E_2(t_3)-E_3(t_3)\right) \,,
\end{equation}
with
\begin{equation}
\delta_{\alpha} \equiv \sum_{i >j} \text{Im}\left[F_{\alpha i} \left(F^{\dagger} F\right)_{ij} F^{\dagger}_{j\alpha}\right] \,.
\end{equation}
This result denotes the leading order asymmetry in the individual
flavours of the active sector induced by the sterile neutrino
oscillations. This asymmetry in turn generates an effective potential
for the sterile neutrino states, inducing an asymmetry in the sterile
sector, as we will discuss below. The backreaction of this asymmetry
in the sterile flavours will finally generate a net asymmetry (at next
order in the perturbative expansion) in the active sector.

Introducing
\begin{align}
& \int_{t_2}^{t_1} dt_3 (E_1(t_3)-E_2(t_3))=z(T_1)-z(T_2) \,, \nonumber\\
& z(T)=\int_0^t \frac{\Delta M^2_{12}}{2 T}=-\int_{T_0}^T \frac{M_0}{T^3} \frac{\Delta M^2_{12}}{4 T}=\frac{M_0 \Delta M^2_{12}}{12 T^3} \,,
\end{align}
the remaining integral can be computed by  changing the variables to $x_i \equiv \frac{T_L}{T_i}$ (with $dt_i=\frac{M_0}{T_L^2} {x_i} dx_i$), where
\begin{equation}\label{eq:TL}
T_L \equiv {\left(\frac{1}{12}M_0 \Delta M^2_{12}\right)}^{1/3}
\end{equation}
will turn out to be the characteristic temperature of the leptogenesis process.
From
\begin{align}
\gamma_N^d(t_i)=\frac{N_D N_C h_t^2}{64 \pi^3}T_i\,, \quad 
\gamma_\nu^d(t_i)=\frac{N_C h_t^2}{64 \pi^3}T_i \,, \quad \frac{N_C h_t^2}{64 \pi^3}=\frac{\sin\phi}{8} \,,
\end{align}
where $\sin\phi \simeq 0.012$ is defined in~\cite{Akhmedov:1998qx}\footnote{{To give a physical intuition, $\sin \phi$ roughly corresponds to the ratio of decay rate over effective potential for the sterile states, or correspondingly to the ratio of the imaginary over the real part of the one-loop diagram $N L \rightarrow N L$.}}, we find
\begin{align}
\label{eq:secondstep}
& \mu_\alpha={\frac{9}{64}} \sin^2\phi \frac{M_0^2}{T_L^2}\delta_\alpha J_{32}\left(\frac{T_L}{T}\right)\,, \\ 
& J_{32}(x)=\int_0^x dx_1 \int_0^{x_1}dx_2  \, \sin\left(x_1^3-x_2^3\right)\ .
\end{align}
The function $J_{32}$ has a very interesting behaviour. 
At early times, i.e. $x \ll 1$, $J_{32}(x)=\frac{3}{20}x^5$, while, after a sharp transition at $x \simeq 1$, it becomes constant. The asymptotic value for $x \gtrsim 1$ is given by:
\begin{equation}
J_{32}(x)=\frac{2^{1/3} \, \pi^{3/2}}{9 \, \Gamma(5/6)} \, .
\end{equation}
Given this behaviour, it is safe to assume that the lepton asymmetry, encoded in the chemical potential $\mu_{\alpha}$, is mostly generated at the temperature $T_L$.

The last step is to compute the asymmetry in the sterile sector. At leading order we have to compute:
\begin{align}
\frac{d \left(\Delta R\right)_{ii}}{dt}& = \gamma_N^d(t) \left[F^\dagger A F-F^T A^{-1} F^{*}\right]_{ii}
& =2 \gamma_N^d(t) \left[F^\dagger \sinh \mu_{\alpha} F\right]_{ii} \approx 2 \,\gamma_N^d(t) \left[F^\dagger \mu_\alpha F\right]_{ii}
\end{align}
Performing a direct integration this yields
\begin{equation}
\label{eq:third_step}
\left(\Delta R\right)_{ii} (T)=\frac{3\, 2^{2/3} \pi^{3/2}}{64 \,3^{1/3} \Gamma(5/6)}\sin^3 \phi\frac{M_0}{T} \frac{M_0^{4/3}}{\Delta M_{12}^{4/3}} \left(F^\dagger \delta_\alpha F\right)_{ii}
\end{equation} 
where we have profited from the asymptotic behaviour of the function $J_{32}$ to analytically solve the integral,  since the asymmetry in the sterile sector is generated mainly at $T < T_L$.
The asymmetry stored in the sterile sector is obtained as the trace of Eq.~(\ref{eq:third_step}). Since the total lepton number is conserved (recall that in the parameter space of interest the Majorana mass terms are much smaller than the temperature of the thermal bath), the same asymmetry but with an opposite sign is contained in the active flavours. SM sphaleron processes couple only to the active flavours, converting the asymmetry stored there into a baryon asymmetry,
\begin{equation}
 Y_{\Delta B} =  - \frac{28}{79} Y_{\Delta \alpha} = \frac{28}{79} Y_{\Delta N} = \frac{28}{79} Y_{N0} (\Delta R_{11}(T_{\rm W}) + \Delta R_{22}(T_{\rm W}))\,,
\end{equation}
with $Y_{N0} \simeq 0.0022$ denoting the equilibrium abundance, cf.\ Eq.~\eqref{eq:YN0}. Evaluating Eq.~\eqref{eq:third_step} at $T = T_{\rm W}$ demonstrates the strong enhancement $M_0/T_{\rm W}$ of the asymmetry, due to the separation of time-scales $ T_{\text W} < T_L \ll M_0$. 
We remark that at each step of the solution increasing powers of $\sin\phi$ and of the Yukawas are present, rendering the analytical procedure reliable.

%% file: tables0720.tex
\section{Numerical benchmark points \label{app_benchmarks}}

\subsection{Benchmarks in the weak wash-out regime}

\begin{itemize}
\item First benchmark (``perturbative'' model, Fig.~\ref{fig:bench_natural}):
\begin{align}
& M=1.5\, \mbox{GeV},\,\,\,\,\,\,\Delta m= 133\, \mbox{eV}\nonumber\\
& Y^{\rm eff}=\left(
\begin{array}{cc}
-3.35 \times 10^{-8} -i\, 1.27 \times 10^{-8} & -1.38 \times 10^{-8} + i\, 3.20 \times 10^{-8} \\
-2.89 \times 10^{-8} + i\, 5.89 \times 10^{-8} & 6.74 \times 10^{-8} + i\, 2.57 \times 10^{-8} \\
 2.30 \times 10^{-8} + i\, 6.99 \times 10^{-8} & 7.87 \times 10^{-8} - i\, 2.04 \times 10^{-8} 
\end{array}
\right)
\end{align}

\item Second benchmark (``generic'' model, Fig.~\ref{fig:bench_numsm}):
\begin{align}
& M=15\, \mbox{GeV},\,\,\,\,\,\,\Delta m= 163\, \mbox{eV}\nonumber\\
& Y^{\rm eff}=\left(
\begin{array}{cc}
4.91 \times 10^{-9} - i\, 3.67 \times 10^{-8} &  1.59 \times 10^{-8} - i\, 1.99 \times 10^-8 \\
6.23 \times 10^{-9} -  i\, 5.74 \times 10^{-8} & 1.13 \times 10^{-7} + i\, 1.09 \times 10^{-9} \\
-1.28 \times 10^{-8} + i\, 1.63 \times 10^{-8} & 1.10 \times 10^{-7} - i\, 2.24 \times 10^{-9}
\end{array}
\right)
\end{align}

\item Third benchmark (perturbative regime with large $Y^\text{eff}$, Fig.~\ref{fig:bench_thermal}):
\begin{align}
& M=3 \, \mbox{GeV},\,\,\,\,\,\,\Delta m= 9\, \mbox{keV}\nonumber\\
& Y^{\rm eff}=\left(
\begin{array}{cc}
-1.27 \times 10^{-8} -i\, 1.96 \times 10^{-8} & 1.87 \times 10^{-8} - i\, 6.78 \times 10^{-9}\\
-3.92 \times 10^{-8} + i\, 8.04 \times 10^{-8} & -9.30 \times 10^{-8} - i\, 3.50 \times 10^{-8}\\
 3.12 \times 10^{-8} + i\, 1.20 \times 10^{-7} & -1.31 \times 10^{-7} + i\, 2.78 \times 10^{-8}
\end{array}
\right)
\end{align}
\end{itemize}

\subsection{Benchmarks in the strong wash-out regime}

\begin{itemize}
\item First benchmark (Fig.~\ref{fig:benchhy1}):
\begin{align}
& M=5.5 \, \mbox{GeV},\,\,\,\,\,\,\Delta m= 5.5\, \mbox{keV}\nonumber\\
& Y^{\rm eff}=\left(
\begin{array}{cc}
 5.78 \times 10^{-8} + i\, 1.39 \times 10^{-7} & -1.37 \times 10^{-7} + i\, 5.92 \times 10^{-8}\\
 -1.79 \times 10^{-8} - i\, 1.90 \times 10^{-7} & 1.98 \times 10^{-7} - i\, 1.76 \times 10^{-8}\\
 -4.54 \times 10^{-9} - i\, 4.58 \times 10^{-7} & 4.63 \times 10^{-7} - i\, 4.45 \times 10^{-9}
\end{array}
\right)
\end{align}

\item Second benchmark (Fig.~\ref{fig:benchhy2}):
\begin{align}
& M=1.5 \, \mbox{GeV},\,\,\,\,\,\,\Delta m= 8.5\, \mbox{keV}\nonumber\\
& Y^{\rm eff}=\left(
\begin{array}{cc}
-1.65 \times 10^{-7} -i\, 1.26 \times 10^{-7} & 1.26 \times 10^{-7} - i\, 1.65 \times 10^{-7} \\
-8.65 \times 10^{-8} + i\, 2.59 \times 10^{-7} & -2.61 \times 10^{-7} -i\, 8.62 \times 10^{-8} \\
 9.37 \times 10^{-8} + i\, 5.16 \times 10^{-7} & -5.18 \times 10^{-7} + i\, 9.34 \times 10^{-8}
\end{array}
\right)
\end{align}

\item Third benchmark (Fig.~\ref{fig:benchhy3}):
\begin{align}
& M=4.5 \, \mbox{GeV},\,\,\,\,\,\,\Delta m= 2\, \mbox{keV}\nonumber\\
& Y^{\rm eff}=\left(
\begin{array}{cc}
 7.61\times 10^{-7} +i\, 6.89 \times 10^{-7} & -6.87 \times 10^{-7} + i\, 7.62 \times 10^{-7}\\
 4.00 \times 10^{-7} +i\, 2.79 \times 10^{-6} & -2.79 \times 10^{-6} +i\, 4.00 \times 10^{-7}\\
 -2.39 \times 10^{-7} +i\, 1.60 \times 10^{-6} & -1.60 \times 10^{-6} -i\, 2.39 \times 10^{-7}
\end{array}
\right)
\end{align}

\item First ``flavoured'' benchmark (left panels of Fig.~\ref{fig:bench_flavored}):
\begin{align}\label{eq:benchSW1}
& M=1 \, \mbox{GeV},\,\,\,\,\,\,\Delta m= 8.5\, \mbox{keV}\nonumber\\
& Y^{\rm eff}=\left(
\begin{array}{cc}
 -1.51 \times 10^{-7} - i\, 1.30 \times 10^{-7} & -1.30 \times 10^{-7} + i\, 1.51 \times 10^{-7}\\
 -6.69 \times 10^{-8} - 7.07 \times 10^{-7} & -7.07 \times 10^{-7} + i\, 6.68 \times 10^{-8}\\
 2.57 \times 10^{-8} - 3.92 \times 10^{-7} & -3.93 \times 10^{-7} - i\,2.57 \times 10^{-8}
\end{array}
\right)
\end{align}

\item Second ``flavoured'' benchmark (right panels of Fig.~\ref{fig:bench_flavored}):
\begin{align}\label{eq:benchSW2}
& M=1.4 \, \mbox{GeV},\,\,\,\,\,\,\Delta m= 1.6\, \mbox{keV}\nonumber\\
& Y^{\rm eff}=\left(
\begin{array}{cc}
1.20 \times 10^{-7} +i\, 1.08 \times 10^{-7} & 1.08 \times 10^{-8} - i\, 1.20 \times 10^{-8} \\
 1.28 \times 10^{-8} - i\, 3.60 \times 10^{-7} & -3.61 \times 10^{-7} - i\, 1.28 \times 10^{-8}\\
-4.41 \times 10^{-8} - i\, 8.29 \times 10^{-7} &  -8.30 \times 10^{-6} +i\, 4.40 \times 10^{-8}
\end{array}
\right)
\end{align}

\end{itemize}